\documentclass[11pt, a4paper]{article}
\usepackage[top=20truemm, bottom=20truemm, left=20truemm, right=20truemm]{geometry}
\usepackage{amsmath, amssymb, bm, fancyhdr, natbib, newtxtext, subfigure, upgreek, xcolor}
\usepackage[pdftex]{graphicx}
\setcitestyle{authoryear, open={(},close={)}}
\DeclareSymbolFont{eulerup}{U}{zeur}{m}{n}
\DeclareMathSymbol{\upartial}{\mathalpha}{eulerup}{"40}
\DeclareMathSymbol{\ud}{\mathalpha}{eulerup}{`d}
\newcommand{\upi}{\uppi}
\newenvironment{keywords}{\small\textbf{Keywords:}}{}
\newenvironment{note}{\small\color{red}}{}

\usepackage{lineno}
\nolinenumbers

\begin{document}

%\jvol{00} \jnum{00} \jyear{2012} \jmonth{February}

\markboth{R. Nakashima and S. Yoshida}{2D ideal MHD waves with a continuous spectrum}

%\articletype{GUIDE}

\title{Two-dimensional ideal magnetohydrodynamic waves on a rotating sphere under a non-Malkus field: I. Continuous spectrum and its ray-theoretical interpretation}

\author{
Ryosuke Nakashima${\dag}$$^{\ast}$\thanks{$^\ast$Corresponding author. Email: r.nakashima.geophysics@gmail.com
\vspace{6pt}} and Shigeo Yoshida${\ddag}$\\
\vspace{6pt}  ${\dag}$Faculty of Science, Kyushu University, Fukuoka, Japan\\
\vspace{6pt}  ${\ddag}$Department of Earth and Planetary Sciences, Faculty of Science, Kyushu University, Fukuoka, Japan\\
%\vspace{6pt}\received{v4.4 released October 2012}
}

\maketitle
\begin{note}
This is an Accepted Manuscript of an article published by Taylor \& Francis in \textit{Geophysical \& Astrophysical Fluid Dynamics} (\url{https://www.tandfonline.com/journals/ggaf20}) on 12 August 2024, available at: \url{https://doi.org/10.1080/03091929.2024.2384388}. It is deposited under the terms of the  Creative Commons Attribution License (\url{https://creativecommons.org/licenses/by/4.0/}). This license enables reusers to distribute, remix, adapt, and build upon the material in any medium or format, so long as attribution is given to the creator. The license allows for commercial use.
\end{note}
 
\begin{abstract}
Two-dimensional ideal incompressible magnetohydrodynamic (MHD) linear waves at the surface of a rotating sphere are studied as a model to imitate the outermost layer of the Earth's core or the solar tachocline. This thin conducting layer is permeated by a toroidal magnetic field the magnitude of which depends only on the latitude. The Malkus background field, which is proportional to the sine of the colatitude, provides two well-known groups of branches; on one branch, retrograde Alfv\'en waves gradually become fast magnetic Rossby (MR) waves as the field amplitude decreases, and on the other, prograde Alfv\'en waves undergo a gradual transition into slow MR waves. In the case of non-Malkus fields, we demonstrate that the associated eigenvalue problems can yield a continuous spectrum instead of Alfv\'en and slow MR discrete modes. The critical latitudes attributed to the Alfv\'en resonance eliminate these discrete eigenvalues and produce an infinite number of singular eigenmodes. The theory of slowly varying wave trains in an inhomogeneous magnetic field shows that a wave packet related to this continuous spectrum propagates toward a critical latitude corresponding to the wave and is eventually absorbed there. The expected behaviour whereby the retrograde propagating packets pertaining to the continuous spectrum approach the latitudes from the equatorial side and the prograde ones approach from the polar side is consistent with the profiles of their eigenfunctions derived using our numerical calculations. Further in-depth discussions of the Alfv\'en continuum would develop the theory of the ``wave--mean field interaction'' in the MHD system and the understanding of the dynamics in such thin layers.
\end{abstract}

\begin{keywords}
Magnetic Rossby waves; Continuous modes; Stably stratified layer; Geomagnetic variations;
\end{keywords}

% = = = = = = = = = = = = = = = = = = = = = = = = = = = = = = %
%                                                             %
%                         Section 1                           %
%                                                             %
% = = = = = = = = = = = = = = = = = = = = = = = = = = = = = = %

\section{Introduction}\label{SEC_introduction}
%%%%%
Geomagnetic and geodetic variations may be partially accounted for by magnetohydrodynamic (MHD) waves within the Earth's outer core \citep[e.g.][]{Triana2021, Gillet2022, Hori2022}. For example, \citet{doi:10.1098/rsta.1966.0026} suggested that the westward drift of the geomagnetic field may originate from slow magnetic Rossby (MR) waves in the liquid core. \citet{braginsky1970torsional} ascribed both the $60$-year length-of-day and geomagnetic variations to the torsional oscillations. If this type of attribution is substantiated, the comparison between observations and the theory results in some inferences on the relevant physical quantities. \citet{Zatman1997} accepted Braginsky's \citeyearpar{braginsky1970torsional} explanation and inferred that the magnitude of the cylindrical radial field is approximately $0.2\,\mathrm{mT}$. This suggestion was challenged by \citet{Gillet2010}, who associated the torsional oscillations (or torsional waves) with $6$-year length-of-day signals to conclude that its magnitude is approximately $2\,\mathrm{mT}$.\par
%%%%%
A stably stratified layer at the top of the outer core has been proposed on various grounds. Simple thermal \citep{10.1111/j.1365-246X.1982.tb06972.x} and compositional \citep{doi:10.1080/03091928408210077} stratification were first considered, and the recent proposed physical mechanisms of stratification have become more sophisticated. For the thermal stratification, a subadiabatic temperature gradient due to the high thermal conductivity of the core \citep{Pozzo2012, Zhange2119001119} has been invoked, and for the compositional stratification, barodiffusion and chemical interactions between the core and mantle \citep{buffett2010stratification, GUBBINS201321, brodholt2017composition, davies2018partitioning} and a remnant of the Moon-creating impact \citep{Landeau2016} were proposed. The seismological evidence reported for such a layer has been a controversial subject \citep[e.g.][]{Helffrich2010, KANESHIMA2018234, doi:10.1126/sciadv.aar2538, 10.1093/gji/ggaa368}. Regional stratification -- instead of global stratification -- owing to the core-mantle boundary (CMB) heterogeneity was also proposed \citep{Mound2019}. The obscure properties of the layer could be inferred from the identification of the sources of geomagnetic and geodetic signatures from the core.\par
%%%%%
Magnetic-Archimedes-Coriolis (MAC) and MR waves in such a stably stratified layer at the top of the core have often been invoked as possible causes of geomagnetic fluctuations \citep{19931517, Braginsky1998, BRAGINSKY199921, Buffett2014, https://doi.org/10.1002/2015GL064067, 10.1093/gji/ggv552, KNEZEK20181, https://doi.org/10.1029/2021GL094692}. If a stratified layer exists at the top of the core, the vigorous convection prevailing in the bulk of the core overshoots the interface between the bulk and the layer \citep{TAKEHIRO2001357, 10.1093/gji/ggaa250} and can excite MHD waves that travel in the layer \citep{10.1093/gji/ggx088, PhysRevFluids.2.094804, 10.1093/gji/ggx492,10.1093/gji/ggab343}.\par
\begin{figure}
	\begin{center}
	\begin{minipage}{50mm}
		\subfigure[Malkus field ($\mathcal{B}=1$).]{
		\resizebox*{50mm}{!}{\includegraphics{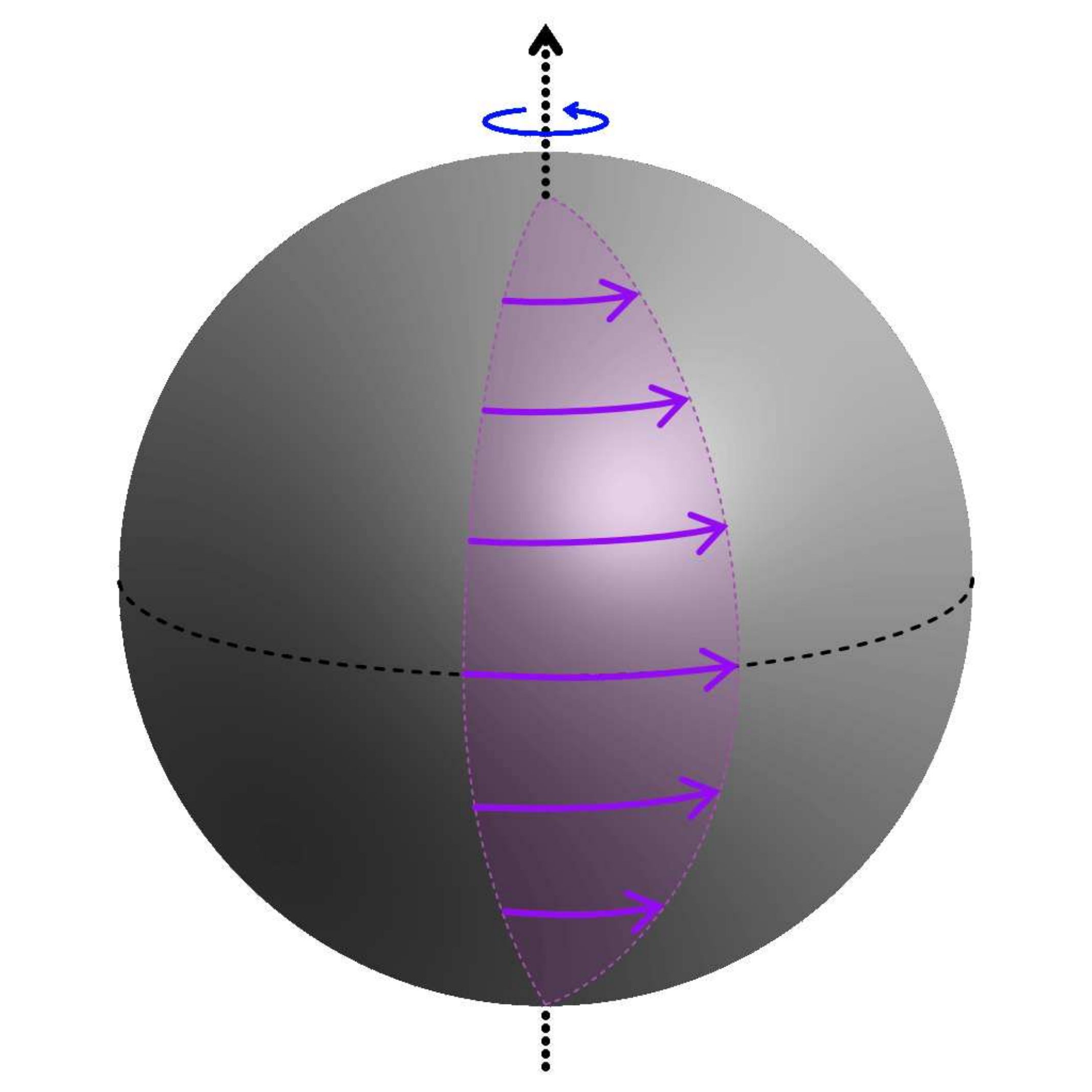}}}%
	\end{minipage}
	\begin{minipage}{50mm}
		\subfigure[$\mathcal{B}=\cos\theta$.]{
		\resizebox*{50mm}{!}{\includegraphics{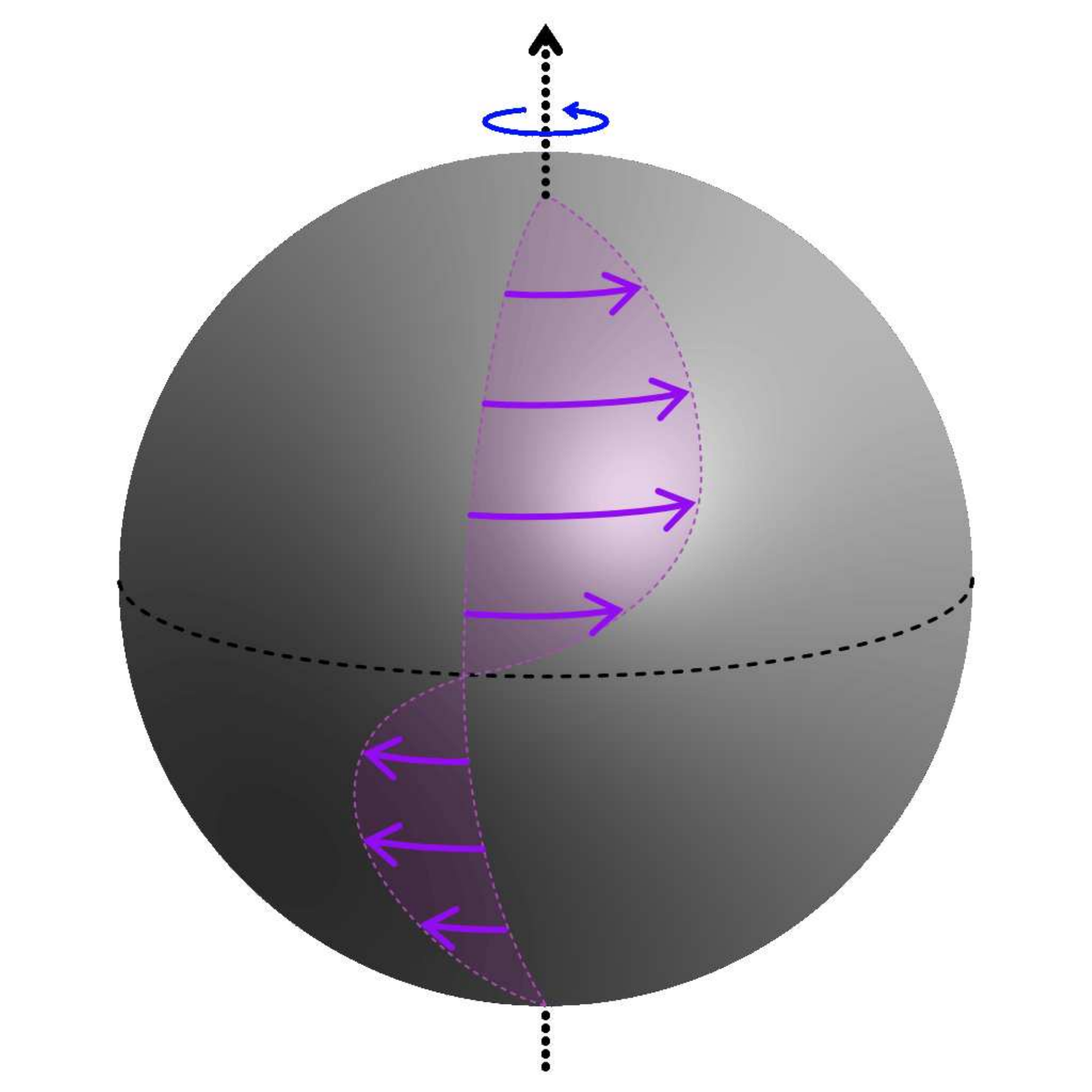}}}%
	\end{minipage}
		\caption{Schematic illustrations of the profiles of the imposed toroidal magnetic fields $B_{0\phi}\propto\mathcal{B}\sin\theta$.}
		\label{FIG_profile}
	\end{center}
	\end{figure}

%%%%%
Two-dimensional (2D) ideal incompressible MHD linear waves within a thin conducting fluid layer over a rotating sphere are focused on in the present study. A non-uniform toroidal ($\phi$-directional, $\phi$ being the longitude) magnetic field is imposed on the fluid film as a background field, and the layer rotates rigidly with the sphere. To investigate the suggestion of \citet{doi:10.1098/rsta.1966.0026}, a similar problem was first considered by \citet{doi:10.1098/rspa.1967.0129}, in which the azimuthal main field is spatially uniform. Such a toroidal field does not vanish even at the north and south poles ($\theta=0$ and $\upi$, where $\theta$ is the colatitude). One should therefore express the basic field as $B_{0\phi}(\theta)=B_0\mathcal{B}\sin\theta$ with a constant $B_0$ and a bounded continuous function $\mathcal{B}(\cos\theta)$ (e.g. \citealt{Gilman_1997}; see Figure \ref{FIG_profile}). The simplest profile $B_{0\phi}=B_0\sin\theta$ in this expression, or $\mathcal{B}=1$, is equivalent to the well-known Malkus field (e.g. \citealt{malkus_1967, FINLAY2008403}; see Figure \ref{FIG_profile}(a)). This elementary 2D model is useful for qualitatively understanding the wave dynamics in the thin layer at the top of the core.\par
\begin{figure}
\begin{center}
	\resizebox*{150mm}{!}{\includegraphics{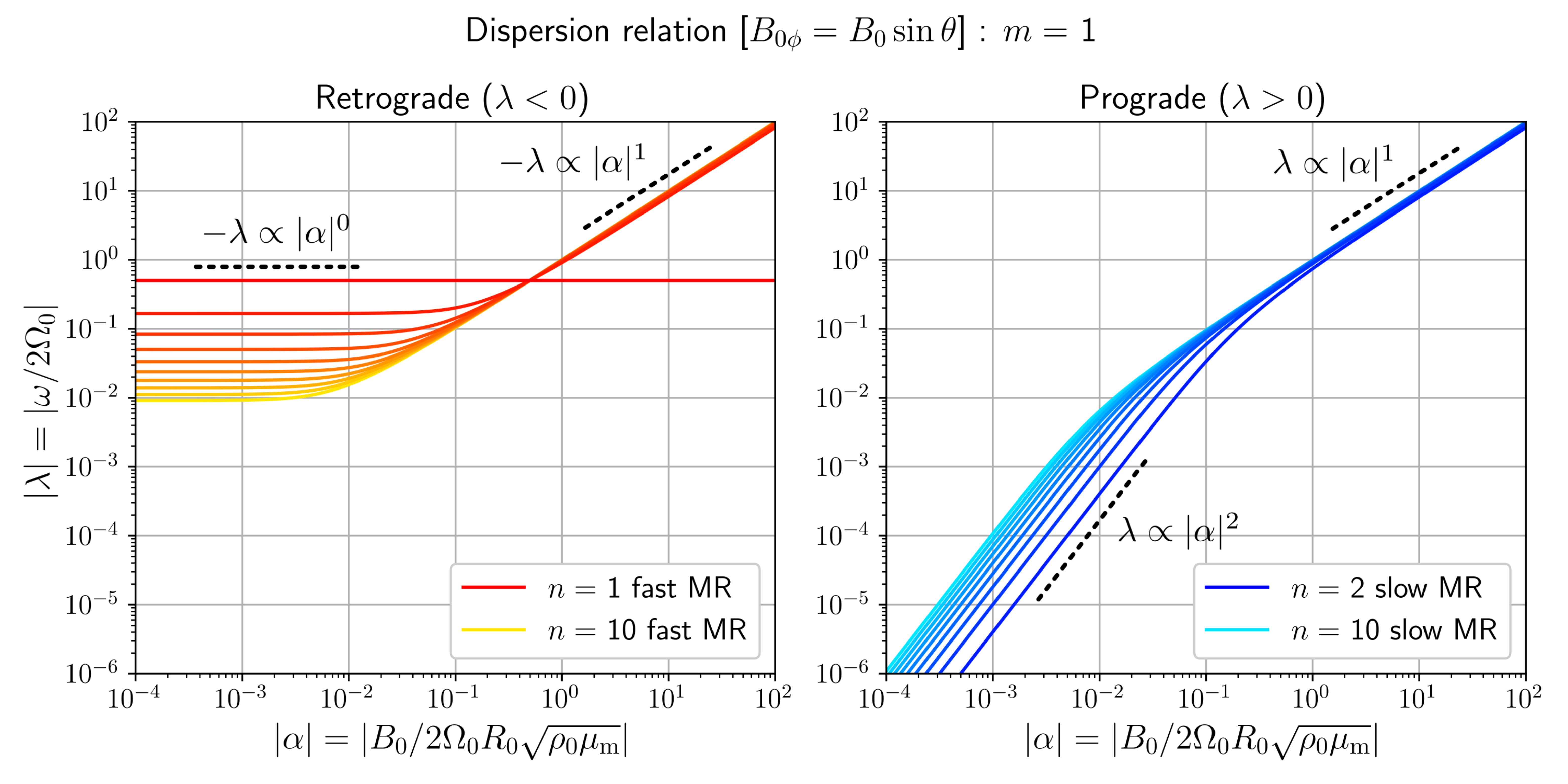}}%
	\caption{Nondimensional angular frequency $\lambda$ calculated from \eqref{EQ_dispersion_Malkus} as a function of the absolute value $|\alpha|$ of the Lehnert number when the zonal wavenumber $m=1$. The left and right panels present retrograde propagating waves ($\lambda<0$) and prograde ones ($\lambda>0$). Fast MR waves for $n=1,2,\ldots,10$ are indicated by the warm-coloured curves, while slow MR waves for $n=2,\ldots,10$ are indicated by the cool-coloured curves. As $|\alpha|\to\infty$, these modes approach the lines $|\lambda|\propto|\alpha|^1$ of Alfv\'en waves, with the exception of the case of $m=n=1$.}
	\label{FIG_dispersion_Malkus}
\end{center}
\end{figure}

%%%%%
The dispersion relation of 2D waves under the Malkus background field should be reviewed before moving on to the main focus of this study. Under the Malkus field, the local Alfv\'en wave velocities are $V_{\mathrm{A}\phi}(\theta)=(B_0/\sqrt{\rho_0\mu_\mathrm{m}})\sin\theta$, which is proportional to $\sin\theta$, where $\rho_0$ is a uniform density, and $\mu_\mathrm{m}$ is the constant magnetic permeability. If local Alfv\'en waves were excited at the same longitude but at different latitudes simultaneously, they would travel with the same longitudinal angular velocity in a manner similar to rigid body rotation, as the length of the circle of colatitude $\theta$ is proportional to $\sin\theta$, as shown in Figure \ref{FIG_profile}(a). This indicates that the Malkus field at the surface of a sphere behaves like a uniform horizontal magnetic field and reduces our perturbation equations to a regular Sturm--Liouville problem because interior poles vanish \citep{doi:10.1063/1.525100}, as shown in Section \ref{SEC_mathematical}. If the Malkus field permeates a rigidly rotating thin layer on top of a sphere having a radius $R_0$, the dispersion relation for the 2D ideal incompressible MHD wave \citep{Zaqarashvili2007, doi:10.1080/03091929.2017.1301937} is given by
\begin{equation}
\lambda\,=\,\frac{-m\pm m\sqrt{1+4\alpha^2n(n+1)[n(n+1)-2]}}{2n(n+1)}\,,
\label{EQ_dispersion_Malkus}
\end{equation}
where $\lambda\equiv\omega/2\varOmega_0$ is the angular frequency that is nondimensionalised by twice the constant (signed) rotation rate $\varOmega_0$. In the above expression, $m$ denotes the (positive) zonal wavenumber, $\alpha\equiv B_0/2\varOmega_0R_0\sqrt{\rho_0\mu_\mathrm{m}}$ is the (signed) Lehnert number, and $n$ is the degree of the associated Legendre polynomial $\mathrm{P}_n^m(\cos\theta)$. Although the Lehnert number is often defined as $Le\equiv 2|\alpha|$ in geophysics literature, we follow the precedents mentioned above. The double sign in \eqref{EQ_dispersion_Malkus} corresponds to prograde and retrograde propagating Alfv\'en waves $\lambda\simeq\pm m|\alpha|\sqrt{1-2/n(n+1)}$ for a large value of $|\alpha|$, and the two classes of MR waves as $|\alpha|\to0$: the slow mode $\lambda\simeq m\alpha^2[n(n+1)-2]$ peculiar to the MHD system and the fast mode $\lambda\simeq-m/n(n+1)$. The fast mode is closely related to the well-known (hydrodynamic) Rossby waves in meteorology and oceanography \citep[e.g.][]{doi:10.1098/rsta.1968.0003}. Figure \ref{FIG_dispersion_Malkus} shows that the exponent of $|\alpha|$ in a branch of $\lambda$, or $(\upartial \log|\lambda|/\upartial \log|\alpha|)$, can be used to distinguish the three types of waves derived from \eqref{EQ_dispersion_Malkus}: Alfv\'en ($\lambda\propto\pm|\alpha|^1$), fast MR ($\lambda\propto-|\alpha|^0$), and slow MR ($\lambda\propto|\alpha|^2$) waves.\par
%%%%%
The simplest equatorially antisymmetric non-Malkus field $B_{0\phi}=B_0\sin\theta\cos\theta$, or $\mathcal{B}=\cos\theta$, should be more appropriate for the Earth's core than the pure Malkus field (Figure \ref{FIG_profile}(b)). In the current study, we focus our attention on the case of this basic field profile because MHD waves under a non-Malkus field are poorly understood. It should be noted that the field is eastward in the northern hemisphere when $B_0>0$ and westward when $B_0<0$. It is important that $|\alpha|$ is significantly less than unity in the Earth's core. If the field strength $|B_0|$ inside the core is several millitesla \citep{Gillet2010}, one can estimate that $|\alpha|\approx10^{-4}$ with $\varOmega_0\approx+0.729\times10^{-4}\,\mathrm{s^{-1}}$, $R_0\approx3480\,\mathrm{km}$, $\rho_0\approx10^4\,\mathrm{kg\,m^{-3}}$, and $\mu_\mathrm{m}\approx1.26\times10^{-6}\,\mathrm{H\,m^{-1}}$.\par
%%%%%
Non-Malkus fields complicate our linear wave problem because regular singular points appear in the governing equations, as demonstrated in Section \ref{SEC_mathematical}. The singular colatitudes $\theta=\theta_\mathrm{c}$ result from the Alfv\'en resonance, at which the azimuthal phase speed $|\omega|/(m/R_0\sin\theta_\mathrm{c})$ of a wave is equal to the local Alfv\'en speed $|V_{\mathrm{A}\phi}(\theta_\mathrm{c})|=(|B_0\mathcal{B}(\cos\theta_\mathrm{c})|/\sqrt{\rho_0\mu_\mathrm{m}})\sin\theta_\mathrm{c}$ \citep[e.g.][]{doi:10.1063/1.1694148, goedbloed_poedts_2004}. For a given angular frequency $\lambda$, such latitudes exist if $\min(\mathcal{B}^2)\leq\lambda^2/m^2\alpha^2\leq \max(\mathcal{B}^2)$. For example, such singular latitudes ordinarily exist for low-frequency neutral modes under a non-Malkus configuration expressed by a function $\mathcal{B}$ that has at least one zero in $0<\theta<\upi$. Even if such points exist for a given angular frequency $\lambda$, however, its eigenmode can be constructed when its eigenfunction is permitted to have singular profiles only at the points, and such singular modes are required for completeness \citep{VANKAMPEN1955949, CASE1959349, BARSTON1964282}. This means that its spectrum should include a continuous range, $\min(\mathcal{B}^2)\leq\lambda^2/m^2\alpha^2\leq \max(\mathcal{B}^2)$. An eigenmode belonging to this continuous spectrum is hereinafter referred to as a ``continuous eigenmode'' or ``continuous mode''. In contrast to the eigenmode of a discrete spectrum, focusing on a single continuous eigenmode is futile since its eigenfunction is singular at its corresponding critical latitudes. Nevertheless, the integration (or superposition) with respect to $\lambda$ in which the integrand is the eigenfunctions weighted by a coefficient depending only on $\lambda$, which may be considered as the inverse Fourier transform, can constitute a physically relevant solution. It should be noted that it is not always true that the Fourier transform of a well-behaved function is not pathological. Solutions composed only of continuous eigenmodes do not exhibit behaviour typical of collective oscillations by normal discrete eigenvalues but represent the transient growth of initial disturbances \citep{TheInitialGrowthofDisturbancesinaBaroclinicFlow} and non-diffusive attenuation \citep[e.g.][]{ADAM1986263}. The latter involves phase mixing and Landau damping, which cause algebraic decay \citep{doi:10.1063/1.1706010, Balmforth1995} and exponential decay \citep{landau, doi:10.1063/1.1692936, sedlacek_1971, Tataronis1973}, respectively. These unintuitive aspects due to the advent of continuous spectra have long been discussed in various research areas: inviscid shear flows, collisionless plasmas, ideal MHD systems, 2D vortices and differentially rotating disks, self-gravitating systems, and the Kuramoto model \citep[e.g.][]{ADAM1986263, balmforth1995normal, STROGATZ20001, BARRE2015723}.\par
%%%%%
Related instability problems may pertain to the dynamics of the solar tachocline underlying the convection zone. Because unstable modes can be easily identified even in the presence of continuous spectra, this circumstance is significantly different from that of the study of linear waves. \citet{Gilman_1997, Fox_1999, Gilman_1999} and the subsequent research \citep{Dikpati_1999, Gilman_2000, Zaqarashvili_2010} investigated the ``joint instability'' or ``magnetic Rossby wave instability'' arising in the tachocline, which can occur in the MHD system accompanied by latitudinal differential rotation. According to these studies, some non-axisymmetric infinitesimal perturbations are likely to destabilise the coexistence of the solar-like angular velocity profile deduced from helioseismic observations and a variety of plausible toroidal field configurations such as broad profiles written as $B_{0\phi}=(B_0+B_1\cos^2\theta)\sin\theta\cos\theta$ (where $B_1$ is also constant) or latitudinally localised field bands expressed by Gaussian functions. These global unstable modes may play an important role in the persistence of such a thin shear layer via the latitudinal transport of the angular momentum \citep{1992A&A...265..106S} and place an upper limit on the strength of a toroidal field stored within the layer through the $\omega$-effect \citep{https://doi.org/10.1002/asna.200710882, Arlt_2007_Stability}. The nonlinear evolution of the modes has been developed by \citet{Cally2001} and \citet{Cally_2003, cally_dikpati_gilman_2004}, who reported the novel ``clamshell'' and ``tipping'' patterns. Furthermore, viscosity and magnetic diffusion were also introduced in the radial \citep{Dikpati_2004} and horizontal \citep{doi:10.1080/03091920500372084} directions in an anisotropic manner.\par
%%%%%
Even beyond the strict 2D model, stability analyses of the tachocline have been conducted. The MHD shallow water equations \citep{Gilman_2000_Magnetohydrodynamic} and the three-dimensional (3D) thin shell model or ``MHD hydrostatic primitive'' equations \citep{Miesch2004} were newly proposed for evaluating the impacts of subadiabatic stratification and the weak vertical displacement of fluid particles. \citet{Gilman_2002} and \citet{Dikpati_2003} demonstrated that the combinations of the differential rotation and toroidal fields in the shallow model easily become unstable again and that the growing perturbations have non-zero kinetic helicities, which are related to the $\alpha$-effect \citep{moffatt1978magnetic}. Furthermore, unstable MR waves in the layer may have caused some of the periodicities in the solar activity (\citealt{Zaqarashvili_2010, Zaqarashvili_2010_Quasi-biennial, Zaqarashvili_2015, Dikpati2017, Gachechiladze_2019}; see \citealt{Zaqarashvili2021}, for a review). For instance, nonlinear development in the MHD shallow water system \citep{Dikpati2017, Dikpati_2018_Phase, Dikpati_2018} indicated that MR waves exchange angular momentum with mean fields and that their wave patterns deformed by consequent reconstructions of the mean profiles can trigger nonlinear quasi-periodic oscillations. In addition, searches were performed not only for non-axisymmetric unstable modes \citep{10.1046/j.1365-8711.2003.06236.x, Gilman_2007, Kitchatinov_2008}, but also for axisymmetric ones \citep{10.1111/j.1365-2966.2008.13934.x, Dikpati_2009} in extensive studies of the 3D thin shell model. These studies include nonlinear simulations of these growth modes \citep{Miesch_2007, Hollerbach2009} and linear stability analyses that took into consideration the vertical profiles of the differential rotation and the background toroidal field \citep{https://doi.org/10.1002/asna.200710882, Arlt_2007_Stability}.\par
%%%%%
Critical lines (or levels, latitudes, etc.) and their concomitant continuous modes have important bearings even on unstable modes, which are beyond the scope of this study. Although the eigenfrequencies of unstable modes are complex numbers, their eigenfunctions are affected by the positions of critical points, as demonstrated by, e.g. \citet{Fox_1999} and \citet{wang_gilbert_mason_2022}. Moreover, a neutral mode sometimes interacts with continuous modes when the branch of the neutral one overlaps with the continuous spectrum. This results in the appearance of a pair of unstable and decay modes \citep{Iga_1999, taniguchi_ishiwatari_2006} or non-modal growth \citep{doi:10.1063/5.0011351}. On the other hand, continuous spectra sometimes cover and hide the branches of such interacting neutral modes \citep{iga_2013}. Therefore, we also intend to detail the critical latitudes and neutral continuous eigenmodes in advance to understand the linear stability in similar problems \citep[e.g.][]{wang_gilbert_mason_2022_analytical}, although our current problem does not comprise such unstable modes.\par
%%%%%%
In contrast to the background magnetic field that we focus on in this paper, the dominant field may be radial in the outermost Earth's core owing to the small conductivity in the mantle \citep[e.g.][]{KNEZEK20181}. Nevertheless, critical latitudes can remain important. This is because the latitudes often reside in equations of similar eigenvalue problems when a radial background field depending on $\theta$ passes through a thin layer \citep[e.g. the equations in][]{10.1093/gji/ggz233}. Although the critical latitudes are sometimes neglected in approximations that are apparently appropriate for each target of study \citep{Zaqarashvili_2018, 10.1093/gji/ggz233}, it is unclear whether the simplifications that drop the Alfv\'en resonance are always valid or not. Thus, it is necessary to scrutinise the influences of the critical latitudes on eigenmodes. Although a more realistic model should have a combination of a toroidal and poloidal imposed field with a finite layer thickness \citep[e.g.][]{doi:10.1080/03091920903439746,https://doi.org/10.1029/2020GL090803,doi:10.1098/rspa.2022.0108}, our simplified model would help us to understand such more complex, realistic ones.\par
%%%%%
Non-ideality and nonlinearity are also intimately related to the critical points. A small viscosity and magnetic diffusion can transform a singular point on the real axis on the ``fictional'' complex plane of the involved spatial coordinate into a complex turning point because the diffusion terms include higher-order derivatives \citep[e.g.][]{drazin_reid_1981, Shivamoggi1992}. Since the eigenfunctions that we intend to determine are functions on the real axis in the complex plane, they become non-singular with the removal of the singular points from the real axis. This appears to indicate that it is necessary to recover these ignored damping terms. However, very weak diffusions only result in thin boundary layers around the critical points (and near any walls), and the eigenfunctions would still be similar to the profiles of continuous eigenmodes obtained from the non-diffusive limit sufficiently outside the layers. The ideal model can, therefore, help us understand a weakly diffusive system \citep[][]{Rincon_2003_Oscillations,Reese_2004_Oscillations}. Although decaying normal modes due to measurable dissipations \citep{Steinolfson1985ResistiveWD, Gizon_2020} and nonlinear boundary layers \citep[e.g.][]{ATheoryofStationaryLongWavesPartIIIQuasiNormalModesinaSingularWaveguide, doi:10.1146/annurev.fl.18.010186.002201} may also be important to our problem, they are beyond the scope of this article.\par
%%%%%
This paper is organised as follows. In Section \ref{SEC_mathematical}, we shall derive the governing equations for our problem, and then present a method for seeking eigenmodes numerically based on the associated Legendre polynomial expansion in Section \ref{SUBSEC_numerical} after presenting some general results in Sections \ref{SUBSEC_existence}, \ref{SUBSEC_frobenius}, and \ref{SUBSEC_necessary}. Section \ref{SEC_numerical} presents numerical solutions for the case where the background field is expressed as the simplest equatorially antisymmetric non-Malkus field $\mathcal{B}=\cos\theta$. In Section \ref{SUBSEC_eigenfunctions} and Appendix \ref{SUBSEC_comparison}, the structures of the obtained eigenfunctions are examined outside and near critical latitudes, respectively. In Section \ref{SEC_discussion}, we also conduct numerical integrations of ray-tracing equations at large wavenumbers \citep[e.g.][]{10.1093/gji/ggx143, 10.3389/fspas.2022.856912} to interpret our eigenvalue problem from a different perspective. This approach tracks the paths of wave packets migrating with their group velocities. It should be noted that this is different from the calculations of the Stokes drift by \citet{Dikpati_2020}, who computed trajectories of fluid particles advected by an oscillatory flow induced by MR waves. Finally, we present the conclusions of this study in Section \ref{SEC_conclusion}.

% = = = = = = = = = = = = = = = = = = = = = = = = = = = = = = %
%                                                             %
%                         Section 2                           %
%                                                             %
% = = = = = = = = = = = = = = = = = = = = = = = = = = = = = = %

\section{Mathematical formulation}\label{SEC_mathematical}
%%%%%
We now begin with a description of the governing equations for the 2D ideal incompressible MHD in a thin layer on a rotating sphere. Here, we use the spherical coordinate system $(r,\theta,\phi)$. Since the vertical flow should be weak within the thin stratified layer, we approximate that the velocity field $\bm{u}=(0, u_\theta, u_\phi)$ relative to the rotating frame of reference is horizontal for simplicity. Furthermore, the magnetic field $\bm{B}=(0, B_\theta, B_\phi)$ is assumed to have only horizontal components as our purpose is primarily to lay the foundation for more realistic models. To maintain consistency with this simplified setting, we select the azimuthal field $B_{0\phi}=B_0\mathcal{B}\sin\theta$ as the imposed field, as mentioned previously in the introduction section. In contrast to the toroidal field, radial ones normally bring vertical derivatives into their governing equations, and the operator is incompatible with a 2D model, which is computationally undemanding. Our 2D model can also be straightforwardly extended into the ``MHD shallow water'' system \citep{Zaqarashvili2007, Zaqarashvili_2009, Heng_2009, Zaqarashvili_2011, doi:10.1080/03091929.2017.1301937}. Moreover, \citet{10.1093/gji/ggaa260} reported that small vertical motions under a strong stratification can enhance toroidal fields because of the ``Malkus constraint''. This suggestion supports our choice as a reasonable main field.\par
%%%%%
In the horizontal 2D model, we only have to consider the horizontal components of the momentum and induction equations. The horizontal momentum and the horizontal induction equations \citep[e.g.][]{Gilman_1997} are written as
\begin{subequations}\label{EQ_governing_pre}
\begin{linenomath}\begin{align}
\frac{\mathrm{D}u_\theta}{\mathrm{D}t}\,-\,\frac{u_\phi^2\cot\theta}{R_0}\,&-\,2\varOmega_0\cos\theta u_\phi\notag\\
\,&=\,-\frac{1}{R_0}\frac{\upartial}{\upartial\theta}\left(\frac{\varPi}{\rho_0}\right)\,+\,\frac{(\bm{B}\bm{\cdot}\bm{\nabla}_\mathrm{H})B_\theta}{\rho_0\mu_\mathrm{m}}\,-\,\frac{B_\phi^2\cot\theta}{R_0\rho_0\mu_\mathrm{m}}\,,
\label{EQ_momentum_theta}\\
\frac{\mathrm{D}u_\phi}{\mathrm{D}t}\,+\,\frac{u_\phi u_\theta\cot\theta}{R_0}\,&+\,2\varOmega_0\cos\theta u_\theta\notag\\
\,&=\,-\frac{1}{R_0\sin\theta}\frac{\upartial}{\upartial\phi}\left(\frac{\varPi}{\rho_0}\right)\,+\,\frac{(\bm{B}\bm{\cdot}\bm{\nabla}_\mathrm{H})B_\phi}{\rho_0\mu_\mathrm{m}}\,+\,\frac{B_\phi B_\theta\cot\theta}{R_0\rho_0\mu_\mathrm{m}}\,,
\label{EQ_momentum_phi}\\
\frac{\mathrm{D}B_\theta}{\mathrm{D}t}\,&=\,(\bm{B}\bm{\cdot}\bm{\nabla}_\mathrm{H})u_\theta\,,
\label{EQ_induction_theta}\\
\frac{\mathrm{D}B_\phi}{\mathrm{D}t}\,+\,\frac{u_\phi B_\theta\cot\theta}{R_0}\,&=\,(\bm{B}\bm{\cdot}\bm{\nabla}_\mathrm{H})u_\phi\,+\,\frac{B_\phi u_\theta\cot\theta}{R_0}\,,
\label{EQ_induction_phi}
\end{align}\end{linenomath}
with the solenoidal conditions
\begin{linenomath}\begin{align}
\frac{\upartial(u_\theta\sin\theta)}{\upartial\theta}\,+\,\frac{\upartial u_\phi}{\upartial\phi}\,&=\,0\,,
\label{EQ_solenoidal_u}\\
\frac{\upartial(B_\theta\sin\theta)}{\upartial\theta}\,+\,\frac{\upartial B_\phi}{\upartial\phi}\,&=\,0\,,
\label{EQ_solenoidal_b}
\end{align}\end{linenomath}
\end{subequations}
where $\varPi$ is the reduced pressure, $\bm{\nabla}_\mathrm{H}\equiv(\hat{\bm{e}}_\theta/R_0)(\upartial/\upartial\theta)+(\hat{\bm{e}}_\phi/R_0\sin\theta)(\upartial/\upartial\phi)$ is the horizontal nabla operator, and the material derivative is expressed as $(\mathrm{D}/\mathrm{D}t)\equiv(\upartial/\upartial t)+\bm{u}\bm{\cdot}\bm{\nabla}_\mathrm{H}$. The 2D vorticity and 2D uncurled induction equations, obtained from \eqref{EQ_governing_pre}, on the spherical surface $r=R_0$ are given as
\begin{subequations}\label{EQ_governing}
\begin{linenomath}\begin{align}
\frac{\mathrm{D}\zeta}{\mathrm{D}t}\,-\,\frac{2\varOmega_0\sin\theta}{R_0}u_\theta\,&=\,\frac{(\bm{B}\bm{\cdot}\bm{\nabla}_\mathrm{H})(\mu_\mathrm{m}J)}{\rho_0\mu_\mathrm{m}}\,,
\label{EQ_vorticity}\\
\frac{\mathrm{D}A}{\mathrm{D}t}\,&=\,0\,.
\label{EQ_potential}
\end{align}\end{linenomath}
\end{subequations}
In the above equations, the radial components of the vorticity $\zeta$ and electrical current $J$ are defined as
\begin{subequations}
\begin{linenomath}\begin{align}
\zeta\,&\equiv\,\frac{1}{R_0\sin\theta}\left[\frac{\upartial(u_\phi\sin\theta)}{\upartial\theta}-\frac{\upartial u_\theta}{\upartial\phi}\right]\,=\,-\nabla_\mathrm{H}^2\psi\,,
\label{DEF_vorticity}\\
J\,&\equiv\,\frac{1}{\mu_\mathrm{m}}\frac{1}{R_0\sin\theta}\left[\frac{\upartial(B_\phi\sin\theta)}{\upartial\theta}-\frac{\upartial B_\theta}{\upartial\phi}\right]\,=\,-\mu_\mathrm{m}^{-1}\nabla_\mathrm{H}^2A\,,
\label{DEF_current}
\end{align}\end{linenomath}
\end{subequations}
where we introduce the stream function $\psi$ and the magnetic vector potential $\bm{A}=(A,0,0)$. Owing to the solenoidal conditions of the fields $\bm{u}$ and $\bm{B}$, the scalars $\psi$ and $A$ are related to the two vectors as
\begin{subequations}
\begin{linenomath}\begin{align}
u_\theta\,=\,\frac{1}{R_0\sin\theta}\frac{\upartial\psi}{\upartial\phi}\,,&\qquad u_\phi\,=\,-\frac{1}{R_0}\frac{\upartial\psi}{\upartial\theta}\,,
\label{DEF_stream}\\
B_\theta\,=\,\frac{1}{R_0\sin\theta}\frac{\upartial A}{\upartial\phi}\,,&\qquad B_\phi\,=\,-\frac{1}{R_0}\frac{\upartial A}{\upartial\theta}\,.
\label{DEF_potential}
\end{align}\end{linenomath}
\end{subequations}
Using these expressions, the governing equations \eqref{EQ_governing} are written in terms of $\psi$ and $A$ \citep[e.g.][]{doi:10.1098/rspa.2020.0174} as
\begin{subequations}\label{EQ_governing_stream_potential}
\begin{linenomath}\begin{align}
\frac{\upartial(\nabla_\mathrm{H}^2\psi)}{\upartial t}\,+\,\mathcal{J}(\nabla_\mathrm{H}^2\psi, \psi)\,+\,\frac{2\varOmega_0}{R_0^2}\frac{\upartial\psi}{\upartial\phi}\,&=\,\frac{1}{\rho_0\mu_\mathrm{m}}\mathcal{J}(\nabla_\mathrm{H}^2A, A)\,,
\label{EQ_vorticity_stream_potential}\\
\frac{\upartial A}{\upartial t}\,+\,\mathcal{J}(A, \psi)\,&=\,0\,,
\label{EQ_potential_stream_potential}
\end{align}\end{linenomath}
where the operator $\mathcal{J}(f, g)$ for any two scalar functions $f$ and $g$ is defined as
\begin{equation}
\mathcal{J}(f, g)\,\equiv\,\frac{1}{R_0^2\sin\theta}\left(\frac{\upartial f}{\upartial\theta}\frac{\upartial g}{\upartial\phi}-\frac{\upartial f}{\upartial\phi}\frac{\upartial g}{\upartial\theta}\right)\,.
\label{DEF_jacobian}
\end{equation}
\end{subequations}
\par
%%%%%
In what follows, waves having a small amplitude are considered in this system. We introduce a small positive parameter $\varepsilon$ ($\ll 1$) representing the amplitude of the waves to rewrite $\psi$ and $A$ as
\begin{equation}
\psi\,=\,\varepsilon\psi_1\,+\,\mathrm{O}(\varepsilon^2)\,,\qquad A\,=\,A_0(\theta)\,+\,\varepsilon a_1\,+\,\mathrm{O}(\varepsilon^2)\,.
\label{EQ_linearization}
\end{equation}
The basic state is assumed to be rigid body rotation $\psi_0\equiv0$ with a latitude-dependent toroidal field $B_{0\phi}(\theta)$. With the expression of $B_{0\phi}$ given in the previous section, we obtain $(\ud A_0/\ud\theta)=-R_0B_0\mathcal{B}\sin\theta$. It should be noted that a smooth background field at the poles necessitates that the function $A_0$ satisfies $A_0=\sum_{n=1}^\infty T_n\mathrm{P}_n$, where $T_n$ ($n=1,2,\ldots$) are constants, and $\mathrm{P}_n$ is the Legendre polynomial of degree $n$. \eqref{EQ_governing_stream_potential} is then written as
\begin{subequations}\label{EQ_linear}
\begin{linenomath}\begin{align}
\left(\frac{\upartial}{\upartial t}\nabla_\mathrm{H}^2+\frac{2\varOmega_0}{R_0^2}\frac{\upartial}{\upartial\phi}\right)\psi_1\,&=\,\frac{1}{R_0\rho_0\mu_\mathrm{m}\sin\theta}\left[B_{0\phi}\nabla_\mathrm{H}^2-\frac{1}{R_0}\frac{\ud(\mu_\mathrm{m}J_0)}{\ud\theta}\right]\frac{\upartial a_1}{\upartial\phi}\,+\,\mathrm{O}(\varepsilon)\,,
\label{EQ_vorticity_linear}\\
\frac{\upartial a_1}{\upartial t}\,&=\,\frac{B_{0\phi}}{R_0\sin\theta}\frac{\upartial \psi_1}{\upartial\phi}\,+\,\mathrm{O}(\varepsilon)\,,
\label{EQ_potential_linear}
\end{align}\end{linenomath}
\end{subequations}
where $J_0(\theta)=\mu_\mathrm{m}^{-1}(1/R_0\sin\theta)[\ud(B_{0\phi}\sin\theta)/\ud\theta]$ is the background electrical current.\par
%%%%%
The normal mode approach is a common procedure in the study of linear waves. For a given azimuthal wavenumber $m$ and an angular frequency $\omega$ -- determined subsequently using a dispersion relation -- which becomes \eqref{EQ_dispersion_Malkus} for the Malkus field $\mathcal{B}=1$ and is sought numerically in Section \ref{SEC_numerical} when $\mathcal{B}=\cos\theta$, we postulate that $\psi_1(\theta,\phi,t)\equiv\mathrm{Re}[\tilde{\psi}(\mu;m,\omega)\mathrm{e}^{\mathrm{i}\varphi(\phi,t)}]=\tilde{\psi}\mathrm{e}^{\mathrm{i}\varphi}/2+\mathrm{c.c.}$ and $a_1\equiv\mathrm{Re}[\tilde{a}\mathrm{e}^{\mathrm{i}\varphi}]$, where $\mu=\cos\theta$, and $\varphi\equiv m\phi-\omega t$ is the phase of the waves. We use $\mathrm{c.c.}$ for the complex conjugate of the preceding terms. With the nondimensional time $\tau\equiv2\varOmega_0 t$, one also has $\varphi=m\phi-\lambda\tau$. Upon substituting this ansatz into \eqref{EQ_linear}, we obtain
\begin{subequations}\label{EQ_nondim}
\begin{linenomath}\begin{align}
(-\lambda\nabla_\mathrm{h}^2+m)\tilde{\psi}\,&=\,m|\alpha|\left\{\mathcal{B}\nabla_\mathrm{h}^2-\frac{\ud^2[\mathcal{B}(1-\mu^2)]}{\ud\mu^2}\right\}\left[\frac{\operatorname{sgn}(\alpha)\tilde{a}}{\sqrt{\rho_0\mu_\mathrm{m}}}\right]\,,
\label{EQ_vorticity_nondim}\\
-\lambda\left[\frac{\operatorname{sgn}(\alpha)\tilde{a}}{\sqrt{\rho_0\mu_\mathrm{m}}}\right]\,&=\,m|\alpha|\mathcal{B}\tilde{\psi}\,,
\label{EQ_potential_nondim}
\end{align}\end{linenomath}
\end{subequations}
where $\nabla_\mathrm{h}^2\equiv R_0^2\nabla_\mathrm{H}^2$ is the dimensionless horizontal Laplacian. It should be noted that $\operatorname{sgn}(\alpha)\tilde{a}/\sqrt{\rho_0\mu_\mathrm{m}}$ has the same dimensions as $\tilde{\psi}$. Hereinafter, our interest is limited to the case where $m\neq0$. The above equations are easily transformed into a single ordinary differential equation of the form
\begin{equation}
\frac{\ud}{\ud\mu}\left[\varLambda(1-\mu^2)\frac{\ud\tilde{\psi}}{\ud\mu}\right]\,-\,\left\{\frac{m^2\varLambda}{1-\mu^2}+m\left[\lambda+2m\alpha^2\mathcal{B}\frac{\ud(\mathcal{B}\mu)}{\ud\mu}\right]\right\}\tilde{\psi}\,=\,0\,,
\label{EQ_differential}
\end{equation}
where the factor $\varLambda(\mu)\equiv\lambda^2-m^2\alpha^2\mathcal{B}^2$ is crucial to our problem. If there exist real values $\mu$ that satisfy $\varLambda(\mu)=0$ within the interval $-1<\mu<1$, which are hereinafter denoted as $\mu_\mathrm{c}$, those points are interior poles depending on $\lambda$. The poles yield continuous spectra, as stated in the introduction section. On the other hand, the Malkus field $\mathcal{B}=1$ obviously produces no singular points, with the exception of the endpoints $\mu=\pm1$. On dividing \eqref{EQ_differential} by the factor $\varLambda$ ($\neq0$), one can obtain its dispersion relation \eqref{EQ_dispersion_Malkus} without any hurdles because \eqref{EQ_differential} is then reduced to the associated Legendre differential equation. It should be noted that, if $\lambda$ and $\tilde{\psi}$ are an eigenvalue and its corresponding eigenfunction of \eqref{EQ_differential}, respectively, the same holds true for their complex conjugates.\par

% = = = = = = = = = = = = = = = = = = = = = = = = = = = = = = %

\subsection{Existence of continuous spectra}\label{SUBSEC_existence}
%%%%%
The existence of continuous spectra can be justified when the function $\varLambda$ has zeros in $-1<\mu<1$. Since \eqref{EQ_differential} is a second-order differential equation, it should have two linearly independent solutions. Let $\tilde{\psi}_\mathrm{I}$ be a non-singular solution, and let the other be tentatively expressed in the heuristic form $\tilde{\varPsi}_\mathrm{I\!I}=\tilde{\psi}_\mathrm{I}\int^\mu f(\mu_*)\ud\mu_*$ with a function $f(\mu)$. Then, one has a non-zero Wronskian $\tilde{\psi}_\mathrm{I}(\ud\tilde{\varPsi}_\mathrm{I\!I}/\ud\mu)-(\ud\tilde{\psi}_\mathrm{I}/\ud\mu)\tilde{\varPsi}_\mathrm{I\!I}=\tilde{\psi}_\mathrm{I}^2f$, guaranteeing that $\tilde{\psi}_\mathrm{I}$ and $\tilde{\varPsi}_\mathrm{I\!I}$ are independent unless $f=0$. On substituting the form of $\tilde{\varPsi}_\mathrm{I\!I}$ into \eqref{EQ_differential}, we obtain
\begin{equation}
\frac{\ud\ln[\varLambda(1-\mu^2)f]}{\ud\mu}\,+\,2\frac{\ud\ln\tilde{\psi}}{\ud\mu}\,=\,0\,,
\end{equation}
from which we obtain $\varLambda(1-\mu^2)f\tilde{\psi}_\mathrm{I}^{2}=\text{const.}\,$. Therefore the solution should have the form
\begin{equation}
\tilde{\psi}\,=\,C_\mathrm{I}\tilde{\psi}_\mathrm{I}\,+\,C_\mathrm{I\!I}\tilde{\varPsi}_\mathrm{I\!I}\,,\qquad\tilde{\varPsi}_\mathrm{I\!I}(\mu)\,=\,\tilde{\psi}_\mathrm{I}(\mu)\int^\mu\frac{\ud\mu_*}{(1-\mu_*^2)\varLambda(\mu_*)\tilde{\psi}_\mathrm{I}^2(\mu_*)}\,,
\label{EQ_solution}
\end{equation}
with both $C_\mathrm{I}$ and $C_\mathrm{I\!I}$ being constants. Let us examine how the integral changes if the interval of integration in \eqref{EQ_solution} passes over zeros of $\varLambda$ (we assume that $\tilde{\psi}_\mathrm{I}^2\neq0$ within the interval). With an arbitrary starting point $\mu_0$ of the integration interval (in the following equation, we suppose that $\mu_0<\operatorname{min}(\mu, \mu_\mathrm{c})$ and that at most one singular point exists between $\mu_0$ and $\mu$), we can express the second solution of the linearly independent set as the improper integral
\begin{equation}
\tilde{\varPsi}_\mathrm{I\!I}(\mu)\,=\,
\begin{cases}
\displaystyle \tilde{\psi}_\mathrm{I}(\mu)\int_{\mu_0}^\mu\frac{\ud\mu_*}{(1-\mu_*^2)\varLambda(\mu_*)\tilde{\psi}_\mathrm{I}^2(\mu_*)}\,, & \mu<\mu_\mathrm{c}\,,\\[0.5cm]
\displaystyle \tilde{\psi}_\mathrm{I}(\mu)\left[\lim_{\varDelta_1\to+0}\int_{\mu_0}^{\mu_\mathrm{c}-\varDelta_1}\frac{\ud\mu_*}{(1-\mu_*^2)\varLambda(\mu_*)\tilde{\psi}_\mathrm{I}^2(\mu_*)}\right.\\
\displaystyle \hspace{3cm}\,+\,\left.\lim_{\varDelta_2\to+0}\int_{\mu_\mathrm{c}+\varDelta_2}^\mu\frac{\ud\mu_*}{(1-\mu_*^2)\varLambda(\mu_*)\tilde{\psi}_\mathrm{I}^2(\mu_*)}\right]\,, & \mu_\mathrm{c}<\mu\,.
\end{cases}
\label{EQ_plemelj}
\end{equation}
We may then deduce from \eqref{EQ_plemelj} that \eqref{EQ_solution} is also written as
\begin{subequations}
\begin{equation}
\tilde{\psi}\,=\,C_\mathrm{I}\tilde{\psi}_\mathrm{I}\,+\,C_\mathrm{I\!I}\tilde{\psi}_\mathrm{I\!I}\,+\,\tilde{\psi}_\mathrm{I}\sum_{i}C_{\mathrm{I\!I\!I}, i}\mathrm{H}(\mu-\mu_{\mathrm{c}, i})\,,
\label{EQ_solution_all}
\end{equation}
where the integer $i$ is the index of the singular latitudes, and $\mathrm{H}(\mu)$ is the step function. In addition, we have introduced a new second solution
\begin{equation}
\tilde{\psi}_\mathrm{I\!I}(\mu)\,\equiv\,\tilde{\psi}_\mathrm{I}(\mu)\,\mathcal{P}\!\!\int^\mu\frac{\ud\mu_*}{(1-\mu_*^2)\varLambda(\mu_*)\tilde{\psi}_\mathrm{I}^2(\mu_*)}\,,
\label{DEF_solution_all}
\end{equation}
where $\mathcal{P}$ denotes the Cauchy principal value. In the vicinity of $\mu=\mu_\mathrm{c}$, we know that $\mathcal{B}^2=\mathcal{B}_\mathrm{c}^2+(\mathcal{B}_\mathrm{c}^2)'(\mu-\mu_\mathrm{c})+(\mathcal{B}_\mathrm{c}^2)''(\mu-\mu_\mathrm{c})^2/2+(\mathcal{B}_\mathrm{c}^2)'''(\mu-\mu_\mathrm{c})^3/6+\mathrm{O}(|\mu-\mu_\mathrm{c}|^4)$, where $\mathcal{B}_\mathrm{c}^2\equiv\mathcal{B}^2(\mu_\mathrm{c})=\lambda^2/m^2\alpha^2$, $(\mathcal{B}_\mathrm{c}^2)'\equiv(\ud\mathcal{B}^2/\ud\mu)|_{\mu=\mu_\mathrm{c}}$, and so on. In this paper, we narrow down a target only to the case where $(\mathcal{B}_\mathrm{c}^2)'$ does not vanish. Using this series expansion, we obtain the magnitude $C_{\mathrm{I\!I\!I}, i}$ of the discontinuities at the latitudes in the form
\begin{equation}
C_{\mathrm{I\!I\!I}, i}\,\equiv\,-\frac{C_\mathrm{I\!I}I_i}{(1-\mu_{\mathrm{c}, i}^{2})m^2\alpha^2(\mathcal{B}_{\mathrm{c}, i}^2)'\tilde{\psi}_\mathrm{I}^2(\mu_{\mathrm{c}, i})}
\label{EQ_plemelj_2}
\end{equation}
with the integral $I_i$ given by
\begin{linenomath}\begin{align}
I_i\,&\equiv\,\lim_{\varDelta, \varDelta_{1, i}\to+0}\int_{\mu_{\mathrm{c}, i}-\varDelta}^{\mu_{\mathrm{c}, i}-\varDelta_{1, i}}\frac{\ud\mu_*}{\mu_*-\mu_{\mathrm{c}, i}}\,+\,\lim_{\varDelta, \varDelta_{2, i}\to+0}\int_{\mu_{\mathrm{c}, i}+\varDelta_{2, i}}^{\mu_{\mathrm{c}, i}+\varDelta}\frac{\ud\mu_*}{\mu_*-\mu_{\mathrm{c}, i}}\notag\\
\,&=\,\lim_{\varDelta_{1, i}, \varDelta_{2, i}\to+0}\ln\frac{\varDelta_{1, i}}{\varDelta_{2, i}}\,.
\label{DEF_detour}
\end{align}\end{linenomath}
\end{subequations}
This integral can be an arbitrary number, depending on how we take the limit $\varDelta_{1, i}, \varDelta_{2, i}\to+0$. In fact, \citet{VANKAMPEN1955949} suggested that, in the problem of plasma oscillations, the counterpart of $I$ may be considered as an arbitrary parameter, which can be determined by a normalisation condition of the counterpart of $\tilde{\psi}$, or the distribution function of plasma. As observed in \eqref{EQ_solution_all}, the existence of more than two linearly independent solutions despite \eqref{EQ_differential} being a second-order differential equation can result in an excess of arbitrary coefficients which should be adjusted to satisfy boundary conditions. This excessive freedom results in continuous spectra.\par
%%%%%
Another explanation for the appearance of continuous spectra is derived from the condition that should be satisfied by the solution of \eqref{EQ_differential} in the vicinity of the singular point. On integrating \eqref{EQ_differential} with respect to $\mu$ over the narrow range sandwiched between $\mu_\mathrm{c}-\varDelta_1$ and $\mu_\mathrm{c}+\varDelta_2$ with $\varDelta_1$, $\varDelta_2\to+0$, we obtain
\begin{subequations}
\begin{equation}
\lim_{\varDelta_1\to+0}\varDelta_1\left.\frac{\ud\tilde{\psi}}{\ud\mu}\right|_{\mu=\mu_\mathrm{c}-\varDelta_1}\,+\,\lim_{\varDelta_2\to+0}\varDelta_2\left.\frac{\ud\tilde{\psi}}{\ud\mu}\right|_{\mu=\mu_\mathrm{c}+\varDelta_2}\,\to\,0\,,
\label{EQ_sandwich}
\end{equation}
provided that the condition
\begin{equation}
\lim_{\varDelta_1, \varDelta_2\to+0}\int_{\mu_\mathrm{c}-\varDelta_1}^{\mu_\mathrm{c}+\varDelta_2}|\tilde{\psi}|\ud\mu\,\to\,0
\label{EQ_sandwich_condition}
\end{equation}
\end{subequations}
is fulfilled. The fact that $\tilde{\psi}_\mathrm{I}$ is surely a solution for \eqref{EQ_differential} and that $\tilde{\psi}_\mathrm{I}\mathrm{H}(\mu-\mu_\mathrm{c})$ always satisfies \eqref{EQ_sandwich} shows that the third term in \eqref{EQ_solution_all} is a weak solution for \eqref{EQ_differential}, and $C_\mathrm{I\!I\!I}$ should then be an undetermined parameter.\par
% = = = = = = = = = = = = = = = = = = = = = = = = = = = = = = %

\subsection{Frobenius method}\label{SUBSEC_frobenius}
%%%%%
An alternative survey of the structure of the solution at the critical latitudes is conducted using the Frobenius method, which is a standard approach for providing a power series solution about a regular singular point of an ordinary differential equation \citep[e.g.][]{braun1975differential}. Let us suppose that a power series of the form $\tilde{\psi}^{(\mathrm{c})}\equiv\sum_{k=0}^\infty a_k(\mu-\mu_\mathrm{c})^{k+\varrho}$ ($a_0\neq0$) is a solution for \eqref{EQ_differential} around a critical latitude $\mu_\mathrm{c}$, where $\varrho$ is a number to be determined by substituting the form into basic equations. The equation that determines $\varrho$ is referred to as the indicial equation. On substituting this assumption for $\tilde{\psi}$ in \eqref{EQ_differential}, we obtain the equation $\varrho^2=0$ from its leading order term. As a result, two linearly independent solutions near the latitude are obtained on the basis of the method for dealing with a repeated root. The method is based on the fact that, for the temporary expression
\begin{subequations}
\begin{equation}
\breve{\psi}^{(\mathrm{c})}(a_0,\varrho)\,\equiv\,a_0(\mu-\mu_\mathrm{c})^\varrho\,+\,\sum_{k=1}^\infty \breve{a}_k(a_0,\varrho)(\mu-\mu_\mathrm{c})^{k+\varrho}
\end{equation}
with the sequence $\breve{a}_k(a_0, \varrho)$ ($k\geq1$) selected to satisfy \eqref{EQ_differential} apart from the term proportional to $(\mu-\mu_\mathrm{c})^{\varrho-1}$, we can rewrite the linear differential equation \eqref{EQ_differential} as
\begin{equation}\label{EQ_linear_operator}
\mathcal{M}[\breve{\psi}^{(\mathrm{c})}(a_0,\varrho)]\,=\,-m^2\alpha^2(\mathcal{B}_\mathrm{c}^2)'(1-\mu_\mathrm{c}^2)a_0\varrho^2(\mu-\mu_\mathrm{c})^{\varrho-1}\,,
\end{equation}
\end{subequations}
where $\mathcal{M}$ is the linear differential operator on the left side of \eqref{EQ_differential}. On setting $\varrho=0$ and $a_0=1$ in \eqref{EQ_linear_operator} and its $\varrho$ derivative, we obtain $\tilde{\psi}_\mathrm{I}^{(\mathrm{c})}=\breve{\psi}^{(\mathrm{c})}(\varrho=0)$ and $\tilde{\psi}_\mathrm{I\!I}^{(\mathrm{c})}=(\upartial \breve{\psi}^{(\mathrm{c})}/\upartial\varrho)|_{\varrho=0}$, which give
\begin{subequations}\label{DEF_Frobenius}
\begin{linenomath}\begin{align}
\tilde{\psi}_\mathrm{I}^{(\mathrm{c})}\,&\equiv\,1\,+\,\sum_{k=1}^\infty a_k(\mu-\mu_\mathrm{c})^k\,,\\
\tilde{\psi}_\mathrm{I\!I}^{(\mathrm{c})}\,&\equiv\,\tilde{\psi}_\mathrm{I}^{(\mathrm{c})}\ln|\mu-\mu_\mathrm{c}|\,+\,\sum_{k=1}^\infty b_k(\mu-\mu_\mathrm{c})^k\,,
\label{DEF_Frobenius_b}
\end{align}\end{linenomath}
\end{subequations}
where we used the formula $[\upartial (\mu-\mu_\mathrm{c})^{k+\varrho}/\upartial\varrho]=(\mu-\mu_\mathrm{c})^{k+\varrho}\ln|\mu-\mu_\mathrm{c}|$. It should be noted that \eqref{DEF_Frobenius_b} has a logarithmic singularity. Some of the expansion coefficients are determined from $\breve{a}_k(a_0, \varrho)$ with slightly tedious but standard manipulations such as $a_1=D_1$, $a_2=(D_1^2+2D_1D_2+D_3)/4$, $b_1=-2D_1+D_2$, and $b_2=(-3D_1^2-2D_1D_2+2D_2^2-D_3+2D_4)/4$, where
\begin{subequations}\label{DEF_coefficient}
\begin{linenomath}\begin{align}
D_1\,&\equiv\,-\frac{\lambda/m\alpha^2+2\mathcal{B}_\mathrm{c}^2+(\mathcal{B}_\mathrm{c}^2)'\mu_\mathrm{c}}{(\mathcal{B}_\mathrm{c}^2)'(1-\mu_\mathrm{c}^2)}\,,\\
D_2\,&\equiv\,\frac{2(\mathcal{B}_\mathrm{c}^2)'\mu_\mathrm{c}-(\mathcal{B}_\mathrm{c}^2)''(1-\mu_\mathrm{c}^2)/2}{(\mathcal{B}_\mathrm{c}^2)'(1-\mu_\mathrm{c}^2)}\,,\\
D_3\,&\equiv\,\frac{m^2(\mathcal{B}_\mathrm{c}^2)'/(1-\mu_\mathrm{c}^2)-3(\mathcal{B}_\mathrm{c}^2)'-(\mathcal{B}_\mathrm{c}^2)''\mu_\mathrm{c}}{(\mathcal{B}_\mathrm{c}^2)'(1-\mu_\mathrm{c}^2)}\,,\\
D_4\,&\equiv\,\frac{(\mathcal{B}_\mathrm{c}^2)'+(\mathcal{B}_\mathrm{c}^2)''\mu_\mathrm{c}-(\mathcal{B}_\mathrm{c}^2)'''(1-\mu_\mathrm{c}^2)/6}{(\mathcal{B}_\mathrm{c}^2)'(1-\mu_\mathrm{c}^2)}\,.
\end{align}\end{linenomath}
\end{subequations}
If we substitute the first Frobenius solution $\tilde{\psi}_\mathrm{I}^{(\mathrm{c})}$ for $\tilde{\psi}_\mathrm{I}$ into the integral expression \eqref{DEF_solution_all} to calculate the second solution up to its second-order terms, the resulting expression certainly agrees with the second Frobenius solution $\tilde{\psi}_\mathrm{I\!I}^{(\mathrm{c})}$ up to the same order (strictly speaking, $\tilde{\psi}_\mathrm{I\!I}\simeq-(\tilde{\psi}_\mathrm{I\!I}^{(\mathrm{c})}+\text{const.}\times\tilde{\psi}_\mathrm{I}^{(\mathrm{c})})/(1-\mu_\mathrm{c}^2)m^2\alpha^2(\mathcal{B}_\mathrm{c}^2)'$ around the latitude). Accordingly, we can associate $\tilde{\psi}_\mathrm{I}$ with $\tilde{\psi}_\mathrm{I}^{(\mathrm{c})}$, and $\tilde{\psi}_\mathrm{I\!I}$ with $\tilde{\psi}_\mathrm{I\!I}^{(\mathrm{c})}$, in the vicinity of $\mu=\mu_\mathrm{c}$. The series solutions \eqref{DEF_Frobenius} also justify \eqref{EQ_sandwich_condition}, since the term that becomes the largest contributor to the integral is $\int_{\mu_\mathrm{c}-\varDelta_1}^{\mu_\mathrm{c}+\varDelta_2}\ln|\mu-\mu_\mathrm{c}|\ud\mu$. The comparison between linear combinations of these linearly independent solutions and our numerical solutions is presented in Appendix \ref{SUBSEC_comparison}.\par
% = = = = = = = = = = = = = = = = = = = = = = = = = = = = = = %

\subsection{Necessary conditions for instability}\label{SUBSEC_necessary}
%%%%%
The necessary conditions for instability often receive interest from many hydrodynamicists. We prove an inequality that gives an eigenvalue bound, of the form of the semicircle theorem, in Appendix \ref{proofs}. It is written as
\begin{subequations}\label{EQ_semicircle_bound}
\begin{linenomath}\begin{align}
\left[\frac{\mathrm{Re}(\lambda)}{m}+\alpha^2\max\left(2\mathcal{B}\frac{\ud(\mathcal{B}\mu)}{\ud\mu}\right)\right]^2&+\left(\frac{\mathrm{Im}(\lambda)}{m}\right)^2\notag\\
\,&\le\,\alpha^4\left[\max\left(2\mathcal{B}\frac{\ud(\mathcal{B}\mu)}{\ud\mu}\right)\right]^2\,-\,\alpha^2\min(\mathcal{B}^2)\,,
\label{EQ_semicircle}
\end{align}\end{linenomath}
only if $\mathrm{Im}(\lambda)\neq0$. Moreover, we also find another bound for the case where $\mathrm{Im}(\lambda)\neq0$ in the form
\begin{equation}
-\frac{1}{2m(m+1)}\leq\frac{\mathrm{Re}(\lambda)}{m}\leq 0\,.
\label{EQ_bound}
\end{equation}
\end{subequations}
These relations indicate that if unstable modes exist, they must propagate in the retrograde direction, and the value $\max\{2\mathcal{B}[\ud(\mathcal{B}\mu)/\ud\mu]\}$ regarding the gradient of an imposed magnetic field must be positive. \citet{doi:10.1098/rspa.2000.0725, mak_griffiths_hughes_2016}, and \citet{wang_gilbert_mason_2022, wang_gilbert_mason_2022_analytical} derived similar theorems for the MHD or MHD shallow water systems with a background shear flow. The theorem \eqref{EQ_semicircle} indicates that the magnetic shear ascribed to the spherical geometry may have a destabilising effect. We were not, however, able to identify unstable modes that are likely to be physically meaningful when $\mathcal{B}=\mu$, as described in Section \ref{SEC_numerical}.

% = = = = = = = = = = = = = = = = = = = = = = = = = = = = = = %

\subsection{Numerical method}\label{SUBSEC_numerical}
%%%%%
A numerical method for solving our eigenvalue problem is described here. In this article, we focus on the simplest equatorially antisymmetric non-Malkus field $\mathcal{B}=\mu$. This choice prompts us to use the associated Legendre polynomial expansion, as \eqref{EQ_differential} exactly becomes the associated Legendre differential equation if $\mathcal{B}=1$, and the recurrence formulae of these polynomials are useful in the present situation. For a fixed $m$, the polynomials $\mathrm{P}_n^m$ ($n\geq m$) constitute a basis for function expansion in the Galerkin discretisation on a spherical surface. Thus, we obtain
\begin{subequations}
\begin{equation}
\tilde{\psi}(\mu)\,\equiv\,\sum_{n=m}^{N_\mathrm{t}}\tilde{\psi}^{[n]}\mathcal{N}_n^m\mathrm{P}_n^m(\mu)\,,\qquad\frac{\operatorname{sgn}(\alpha)\tilde{a}}{\sqrt{\rho_0\mu_\mathrm{m}}}\,\equiv\,\sum_{n=m}^{N_\mathrm{t}}\tilde{a}^{[n]}\mathcal{N}_n^m\mathrm{P}_n^m\,,
\label{DEF_expansion}
\end{equation}
where $N_\mathrm{t}$ denotes the truncation degree, and the normalising factor is written as
\begin{equation}
\mathcal{N}_n^m\,\equiv\,(-1)^m\sqrt{\frac{2n+1}{2}\frac{(n-m)!}{(n+m)!}}\,.
\label{DEF_normalizing}
\end{equation}
\end{subequations}
On assuming \eqref{DEF_expansion} to be an approximate solution for \eqref{EQ_nondim} and using useful relations of the forms
\begin{subequations}\label{EQ_recurrence}
\begin{linenomath}\begin{align}
\nabla_\mathrm{h}^2\mathrm{P}_n^m\,&=\,-n(n+1)\mathrm{P}_n^m\,,\\
\mu\mathrm{P}_n^m\,&=\,\frac{n+m}{2n+1}\mathrm{P}_{n-1}^m\,+\,\frac{n-m+1}{2n+1}\mathrm{P}_{n+1}^m\,,
\end{align}\end{linenomath}
\end{subequations}
and the orthogonality relation $\int_{-1}^1\mathrm{P}_n^m\mathrm{P}_{\nu}^m\ud\mu=\delta_{n,\nu}/\mathcal{N}_n^m\mathcal{N}_{\nu}^m$ with the Kronecker delta $\delta_{n,\nu}$, we obtain the following simultaneous equations for all the integers $n$ that satisfy $m\leq n\leq N_\mathrm{t}$:
\begin{subequations}\label{EQ_matrix}
\begin{linenomath}\begin{align}
\left[\lambda+\frac{m}{n(n+1)}\right]\tilde{\psi}^{[n]}\,&=\,-m|\alpha|\left[\frac{(n-3)(n+2)}{n(n+1)}k_n^m\tilde{a}^{[n-1]}\right.\notag\\
&\qquad\qquad\qquad\,+\,\left.\frac{(n-1)(n+4)}{n(n+1)}k_{n+1}^m\tilde{a}^{[n+1]}\right]\,,\\
\lambda\tilde{a}^{[n]}\,&=\,-m|\alpha|\left(k_n^m\tilde{\psi}^{[n-1]}\,+\,k_{n+1}^m\tilde{\psi}^{[n+1]}\right)\,,
\end{align}\end{linenomath}
where the number sequence
\begin{equation}
k_n^m\,=\,\sqrt{\frac{(n-m)(n+m)}{(2n-1)(2n+1)}}\qquad\text{for}\quad n=m, m+1, \ldots, N_\mathrm{t}\,,
\label{DEF_submatrix}
\end{equation}
\end{subequations}
is introduced for convenience.\par
%%%%%
The system of linear equations \eqref{EQ_matrix} is equivalent to the eigenvalue problem for the corresponding $2(N_\mathrm{t}-m+1)\times2(N_\mathrm{t}-m+1)$ matrix, and the arrays of the expansion coefficients $\tilde{\psi}^{[n]}$ and $\tilde{a}^{[n]}$ are its eigenvectors. We performed numerical calculations for solving this eigenvalue problem using our Python code, which is based on the \texttt{numpy.linalg.eig} function of the NumPy library. In these calculations, the truncation number $N_\mathrm{t}$ was set as $2000$. As can be observed from \eqref{EQ_matrix}, the subset of $\tilde{\psi}^{[n]}$ with $n$ being odd numbers pertains only to the subset of $\tilde{a}^{[n]}$ with $n$ being even, and the same is true of the relationship between $\tilde{\psi}^{[n]}$ with $n$ being even and $\tilde{a}^{[n]}$ with $n$ being odd. Based on this dichotomy, we refer to eigenmodes for which $\tilde{\psi}^{[n]}$'s are non-zero only when $n-m$ is even (i.e., $\tilde{a}^{[n]}$'s do not vanish only when $n-m$ is odd, and $u_\theta$ and $b_\phi$ are equatorially symmetric, and $u_\phi$ and $b_\theta$ are equatorially antisymmetric) as the sinuous modes \citep[cf.][]{doi:10.1080/03091929.2017.1301937}. Conversely, eigenmodes for which $\tilde{\psi}^{[n]}$'s become non-zero only when $n-m$ is odd (or $u_\theta$ is antisymmetric, and $u_\phi$ is symmetric about the equator) are hereinafter referred to as the varicose modes.\par
%%%%%
The validity of eigenmodes obtained numerically is diagnosed from the aspect of convergence. This is realised by evaluating
\begin{equation}
\sum_{n=m}^{\lfloor N_\mathrm{t}/2\rfloor}\left|\tilde{\psi}^{[n]}\right|^2\,>\,10^2\sum_{n=\lfloor N_\mathrm{t}/2\rfloor+1}^{N_\mathrm{t}}\left|\tilde{\psi}^{[n]}\right|^2\quad\text{and}\quad\sum_{n=m}^{\lfloor N_\mathrm{t}/2\rfloor}\left|\tilde{a}^{[n]}\right|^2\,>\,10^2\sum_{n=\lfloor N_\mathrm{t}/2\rfloor+1}^{N_\mathrm{t}}\left|\tilde{a}^{[n]}\right|^2,
\label{EQ_convergence}
\end{equation}
where $\lfloor x\rfloor$ is the integer part of $x$. Only eigenmodes that pass the screening via this validation are studied and illustrated in the results section.\par
\begin{figure}
\begin{center}
	\resizebox*{75mm}{!}{\includegraphics{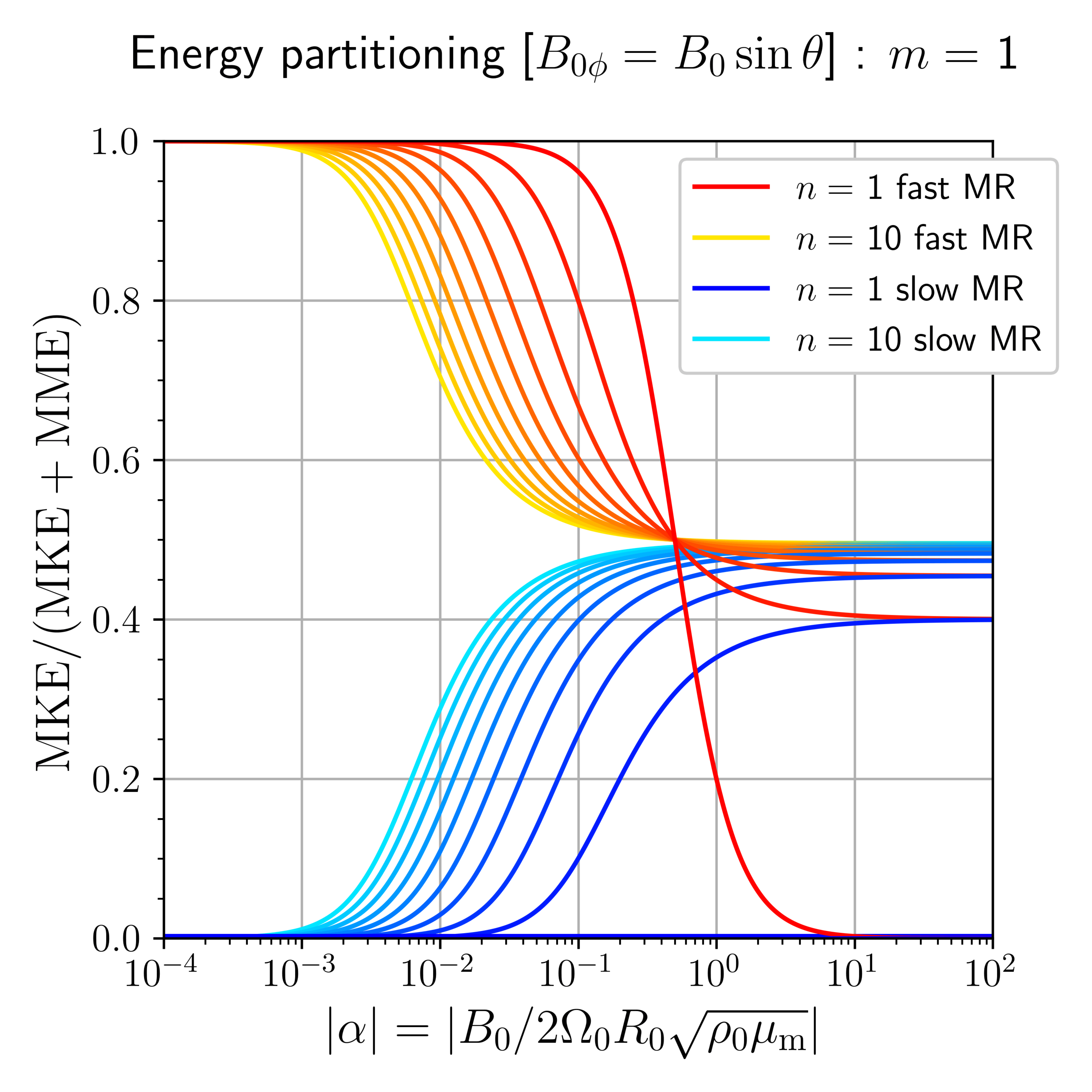}}%
	\caption{Ratio of the mean kinetic energy $\text{MKE}$ to the mean total energy $\text{MKE}+\text{MME}$ against the absolute value $|\alpha|$ of the Lehnert number when the zonal wavenumber $m=1$ and the Malkus field is imposed. This plot is obtained from the relation $\text{MKE}/(\text{MKE}+\text{MME})=\lambda^2/(\lambda^2+m^2\alpha^2)$, where the nondimensional angular frequency $\lambda$ is calculated using \eqref{EQ_dispersion_Malkus}. As $|\alpha|\to0$, either $\text{MKE}$ or $\text{MME}$ approaches zero, depending on whether the eigenmode is a slow or fast MR wave. The colours of the curves correspond to those in Figure \ref{FIG_dispersion_Malkus}.}
	\label{FIG_energy_Malkus}
\end{center}
\end{figure}

%%%%%
The normalisation of the amplitudes of eigenmodes is valuable for understanding their characteristics by comparing physical quantities such as energies. The normalisation was realised by letting the mean total energies $\mathrm{MKE}+\mathrm{MME}$ of the perturbations be $(\rho_0/8R_0^2)\mathrm{e}^{2\mathrm{Im}(\omega)t}$, where the mean kinetic and mean magnetic energies of an eigenmode are expressed as
\begin{subequations}\label{EQ_normalization}
\begin{linenomath}\begin{align}
\mathrm{MKE}\,&\equiv\,\frac{1}{4\upi}\int_0^{\upi}\ud\theta\int_{0}^{2\upi}\sin\theta\ud\phi\frac{\rho_0|\bm{u}_1|^2}{2}\,=\,\frac{\rho_0}{8R_0^2}\mathrm{e}^{2\mathrm{Im}(\omega)t}\sum_{n=m}^{N_\mathrm{t}}n(n+1)\left|\tilde{\psi}^{[n]}\right|^2\,,\\
\mathrm{MME}\,&\equiv\,\frac{1}{4\upi}\int_0^{\upi}\ud\theta\int_{0}^{2\upi}\sin\theta\ud\phi\frac{|\bm{b}_1|^2}{2\mu_\mathrm{m}}\,=\,\frac{\rho_0}{8R_0^2}\mathrm{e}^{2\mathrm{Im}(\omega)t}\sum_{n=m}^{N_\mathrm{t}}n(n+1)\left|\tilde{a}^{[n]}\right|^2\,,
\end{align}\end{linenomath}
\end{subequations}
respectively, with $\bm{u}=\varepsilon\bm{u}_1+\mathrm{O}(\varepsilon^2)$ and $\bm{B}=\bm{B}_0+\varepsilon\bm{b}_1+\mathrm{O}(\varepsilon^2)$. The energy partitioning between $\text{MKE}$ and $\text{MME}$ allows us to examine the force balance of the eigenmodes and to classify the types of waves. For instance, Figure \ref{FIG_energy_Malkus} presents the energy partitions for various eigenmodes under the Malkus field and shows that $\text{MKE}$s dominate the mean total energies in the case of the fast MR waves, $\text{MME}$s are predominant over $\text{MKE}$s in the case of the slow MR waves, and Alfv\'en waves exhibit almost equipartition between the two for a large $|\alpha|$. This normalisation of eigenvectors is applied to all the figures of the profiles of eigenmodes displayed in the following sections. The associated Legendre polynomials $\mathrm{P}_n^m$ employed to construct the eigenfunctions from their corresponding eigenvectors are given by the \texttt{scipy.special.lpmv} function of the SciPy library.

% = = = = = = = = = = = = = = = = = = = = = = = = = = = = = = %
%                                                             %
%                         Section 3                           %
%                                                             %
% = = = = = = = = = = = = = = = = = = = = = = = = = = = = = = %

\section{Numerical results}\label{SEC_numerical}
\begin{figure}
\begin{center}
	\resizebox*{150mm}{!}{\includegraphics{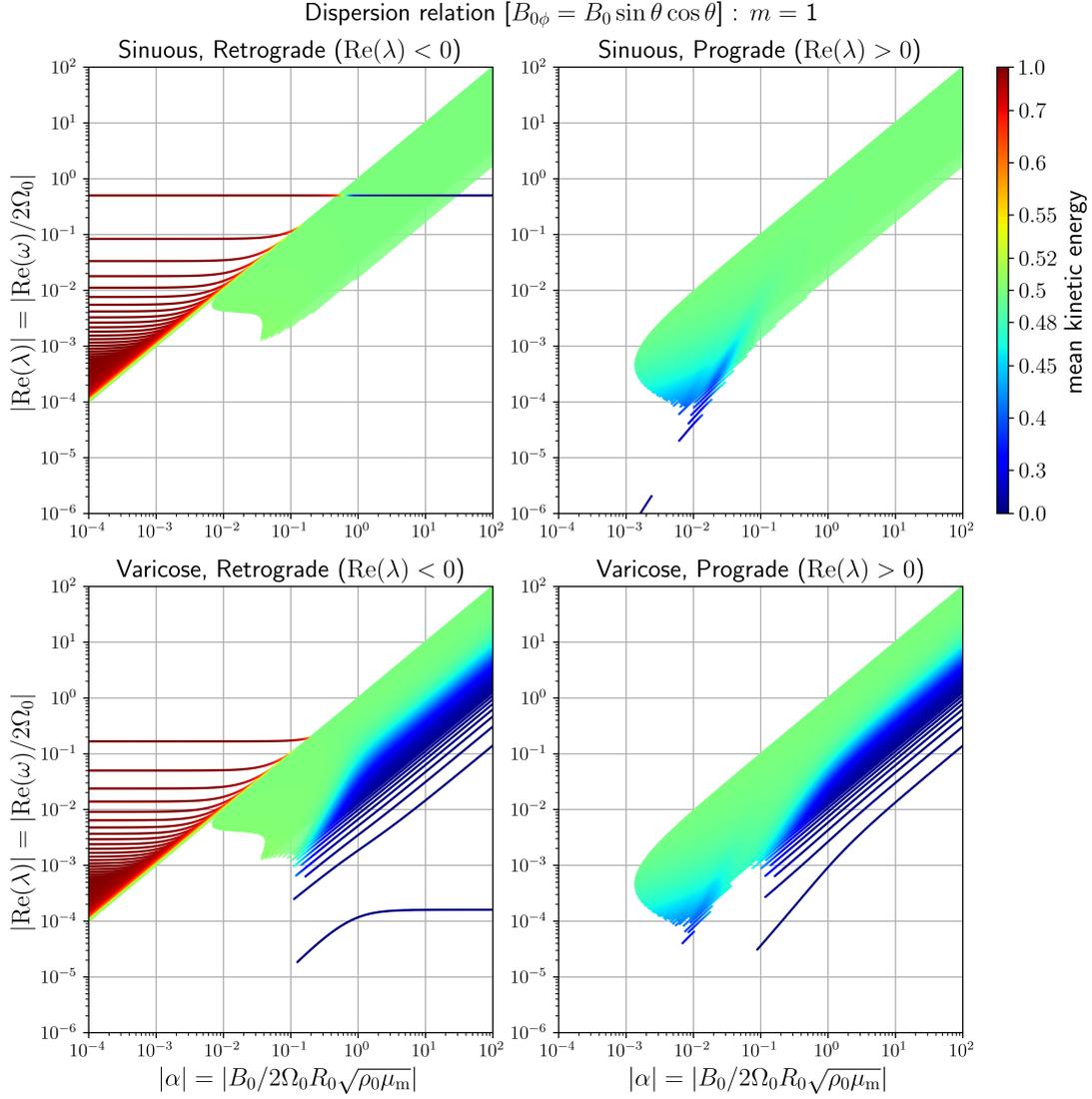}}%
	\caption{Eigenfrequencies versus the Lehnert number when the zonal wavenumber $m=1$ and the simplest equatorially antisymmetric non-Malkus field $\mathcal{B}=\mu$ pervades the system. The ordinates of each panel represent the real part $\mathrm{Re}(\lambda)$ of the dimensionless angular frequency, and the abscissas present the absolute value $|\alpha|$ of the Lehnert number. Retrograde modes ($\mathrm{Re}(\lambda)<0$) and prograde modes ($\mathrm{Re}(\lambda)>0$) are presented in the left and right columns, respectively. The upper and lower rows illustrate the sinuous and varicose modes, respectively. The colours of the markers represent the ratio of the mean kinetic energy $\text{MKE}$ to the mean total energy $\text{MKE} + \text{MME}$.}
	\label{FIG_dispersion_sincos_m1}
\end{center}
\end{figure}

%%%%%
We start this section by presenting the dispersion relation for our current problem. Figure \ref{FIG_dispersion_sincos_m1} shows the real parts of the eigenfrequencies obtained numerically when $m=1$ as functions of $|\alpha|$ (we find that their imaginary parts vanish, that is $\mathrm{Re}(\lambda)=\lambda$, for all the eigenmodes except for unreliable eigenmodes, which are discussed briefly later). The figure has four panels. The left and right columns present the retrograde and prograde modes, respectively, and the upper and lower rows present the sinuous and varicose modes, respectively. It should be noted that, in the sinuous modes, $\tilde{\psi}$ is symmetric and $\tilde{a}$ is antisymmetric about the equator, while $\tilde{\psi}$ is equatorially antisymmetric and $\tilde{a}$ is equatorially symmetric in the varicose modes, as defined in Section \ref{SUBSEC_numerical}. Each colour in the scatter plots represents the fraction of the mean kinetic energy within the mean total energy of an eigenmode corresponding to a point on the diagrams. It should be noted that we used a nonlinear colour scale that is created using the arctangent function to highlight whether an eigenmode is similar to the Alfv\'en wave ($\text{MKE}\approx\text{MME}$; the colour of its marker is greenish) or not. To make it easier to find markers of modes dissimilar to the Alfv\'en wave (we select the range $\text{MKE}<0.49$ or $0.51<\text{MKE}$), we furthermore set their size to be larger than that of the markers representing the Alfv\'en wave. Based on the information presented in Figure \ref{FIG_energy_Malkus}, even in the present situation, we would be justified in thinking of an eigenmode having a reddish marker as a mode similar to the fast MR wave and a bluish one as a mode similar to the slow MR wave.\par
\begin{figure}
\begin{center}
	\resizebox*{75mm}{!}{\includegraphics{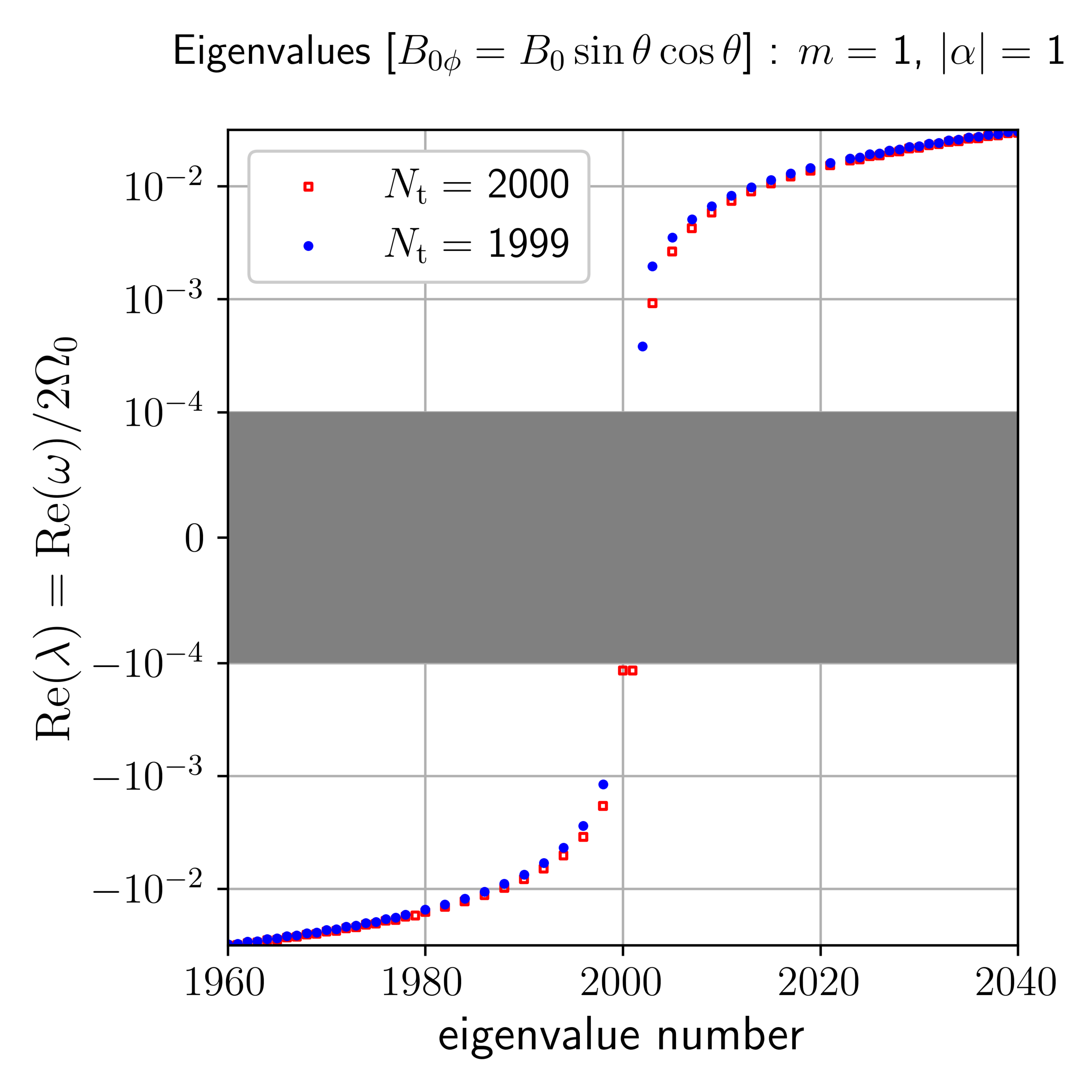}}%
	\caption{Dependence of the real parts $\mathrm{Re}(\lambda)$ of the eigenvalues on the truncation degree $N_\mathrm{t}$ when the zonal wavenumber $m=1$, the absolute value of the Lehnert number $|\alpha|=1$, and the background field is the simplest equatorially antisymmetric non-Malkus one ($\mathcal{B}=\mu$). The horizontal axis is the eigenvalue number when eigenvalues are arranged in ascending order. The red open squares and the blue circles represent the case where $N_\mathrm{t}=2000$ and $1999$, respectively.}
	\label{dependdegree}
\end{center}
\end{figure}

\begin{figure}
\begin{center}
	\resizebox*{150mm}{!}{\includegraphics{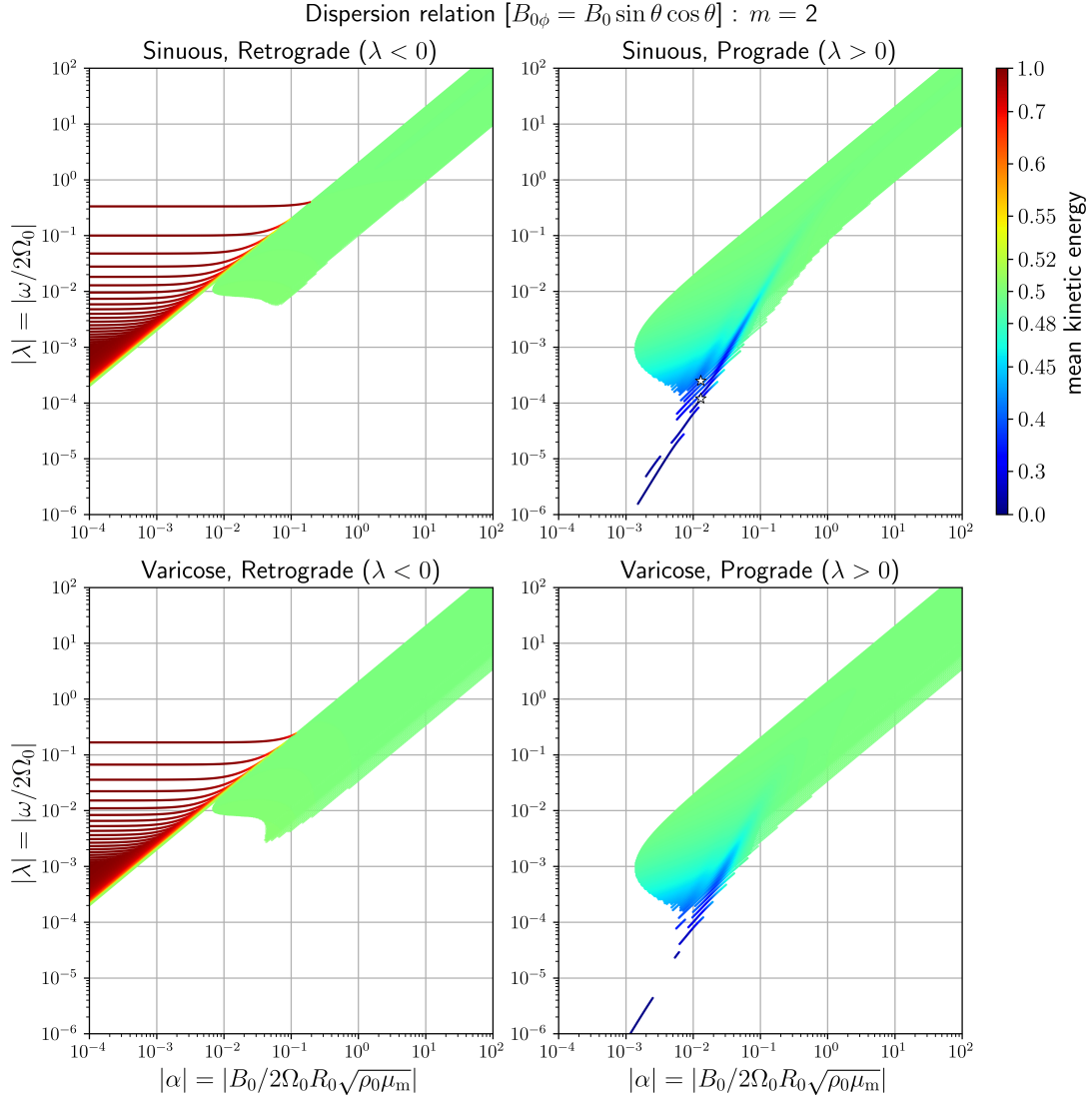}}%
	\caption{Same as Figure \ref{FIG_dispersion_sincos_m1}, but for $m=2$. The two white asterisks of the upper right panel correspond to the two eigenmodes depicted in Figure \ref{FIG_eigenfunction_sincos_m1a0013}.}
	\label{FIG_dispersion_sincos_m2}
\end{center}
\end{figure}

%%%%%
In all the panels, we can observe the bands crowded with eigenmodes just below the lines $|\lambda|=m|\alpha|$. The kinetic and magnetic energies of the majority of the eigenmodes in the bands are partitioned almost equally. We conjecture that these bands should be identified with the continuous spectrum due to the Alfv\'en resonance, which is hereinafter referred to as the Alfv\'en continuous spectrum or Alfv\'en continuum, although our numerical method yields only approximate discrete modes even when the system has a continuous spectrum. Even though the spectrum should encompass the expected range $m|\alpha|\operatorname{min}(\mu)=-m|\alpha|\leq\lambda\leq m|\alpha|=m|\alpha|\operatorname{max}(\mu)$ without any gaps as explained in the previous sections, our eigenmodes satisfying \eqref{EQ_convergence} do not have eigenvalues smaller than some levels in terms of absolute values. This is because a fine structure in its eigenfunction appears around the equator, as the critical latitudes $\mu_\mathrm{c}=\pm\lambda/m|\alpha|$ approach the equator as $\lambda\to0$ for a given $m$ and a given $|\alpha|$, in addition to the vertical axes of the panels presented on a logarithmic scale. It is necessary to perform the calculation with a higher truncation degree for \eqref{DEF_expansion} to express such a fine structure. Moreover, small values of $|\alpha|$ ($\ll1$) reduce the typical meridional wavelengths of perturbations, as shown in Figure \ref{FIG_alleigenfunction_sincos_m1a001}. For the aforementioned reason, the bands are cut off below certain values of $|\alpha|$. If numerical calculations were performed with an infinite degree, the obtained eigenvalues would cover the entire range below the line $|\lambda|=m|\alpha|$. We also performed calculations with truncation degrees less than that shown in Figure \ref{FIG_dispersion_sincos_m1}, for example, $N_\mathrm{t}=1000$ (not shown). The outlines of these dispersion diagrams appear almost unchanged aside from the difference in the widths of the bands; the higher the truncation degree, the wider the band. In the lower panels of Figure \ref{FIG_dispersion_sincos_m1}, some branches of the varicose modes look like discrete eigenvalues that lie below the bands. For these eigenmodes, $\text{MME}$s are dominant. In particular, the lowermost two eigenvalues, which are represented by overlapping curves (since they are a complex conjugate pair, as mentioned in Section \ref{SEC_mathematical}) in the lower left panel, have non-zero imaginary parts (not shown), and their values are consistent with \eqref{EQ_semicircle} and \eqref{EQ_bound}. However, we consider the branches including the unstable modes as unreliable eigenvalues, or a part of the Alfv\'en continuous modes, because the calculations (see Figure \ref{dependdegree}) reveal that these eigenvalues depend strongly upon $N_\mathrm{t}$ as opposed to normal discrete modes \citep[cf.][]{doi:10.1063/1.5116633}. These Alfv\'en continuous modes for which $\mathrm{MME}$s dominate over $\mathrm{MKE}$s, which are found only in the $m=1$ varicose modes (see also Figure \ref{FIG_dispersion_sincos_m2}), could possibly be related to the ``clamshell instability'' \citep[e.g.][see also the last paragraph of Appendix \ref{SEC_discrete}]{wang_gilbert_mason_2022_analytical}.\par
%%%%%
Discrete branches equivalent to slow MR waves found in Figure \ref{EQ_dispersion_Malkus} disappear from Figure \ref{FIG_dispersion_sincos_m1} as a result of the modification of the main field. Instead, the blue markers, for which the fractions of \text{MME}s of their eigenmodes are close to unity as in the case of slow MR waves, are distributed within the Alfv\'en continuum. These markers form blue upward wedges having the approximate slope $\lambda\propto|\alpha|^{2}$ at $|\alpha|\approx10^{-2}$ and $\lambda\approx10^{-4}$ in the right panels of Figure \ref{FIG_dispersion_sincos_m1}. Therefore, we suggest that the discrete modes of slow MR waves turn into continuous ones under a non-Malkus field. This is similar to the case of equatorial Rossby waves in the study of \citet{taniguchi_ishiwatari_2006}, who investigated eigenmodes in a linear shear flow on an equatorial $\beta$-plane.\par
%%%%%%
Outside the continuous spectrum, that is, above the lines $|\lambda|=m|\alpha|$ in the diagrams, fast MR waves remain discrete eigenmodes even in the non-Malkus field. Their semi-analytical solutions can be obtained from eigenvalues of the spheroidal differential equation to which \eqref{EQ_differential} is reduced when $\mathcal{B}=\mu$ and $m^2\alpha^2/\lambda^2$ is small (see Appendix \ref{SEC_fast}). Moreover, we find that the lowermost wavenumber branch of the $m=1$ sinuous modes of fast MR waves can penetrate the band of the continuous spectrum without interaction (see the upper left panel of Figure \ref{FIG_dispersion_sincos_m1}). This is because the Lorentz force does not act on this eigenmode. This mode is explained in detail in Appendix \ref{SEC_discrete}.\par
%%%%%%
Figure \ref{FIG_dispersion_sincos_m2} depicts the eigenfrequencies for $m=2$. They roughly epitomise the diagrams when $m\geq3$ (not shown). Their outlines do not change much from those of Figure \ref{FIG_dispersion_sincos_m1} with the exception of the absence of the branch of the fast MR waves that penetrates the continuous spectrum. As in the $m=1$ case, more conspicuous blue upward wedges exist around $10^{-2}\leq|\alpha|\leq10^{-1}$ and $10^{-4}\leq\lambda\leq10^{-2}$ in the right panels of Figure \ref{FIG_dispersion_sincos_m2}.

% = = = = = = = = = = = = = = = = = = = = = = = = = = = = = = %

\subsection{Eigenfunctions of the Alfv\'en continuous modes}\label{SUBSEC_eigenfunctions}
\begin{figure}
\begin{center}
\begin{minipage}{75mm}
	\subfigure[Meridional structures.]{
	\resizebox*{75mm}{!}{\includegraphics{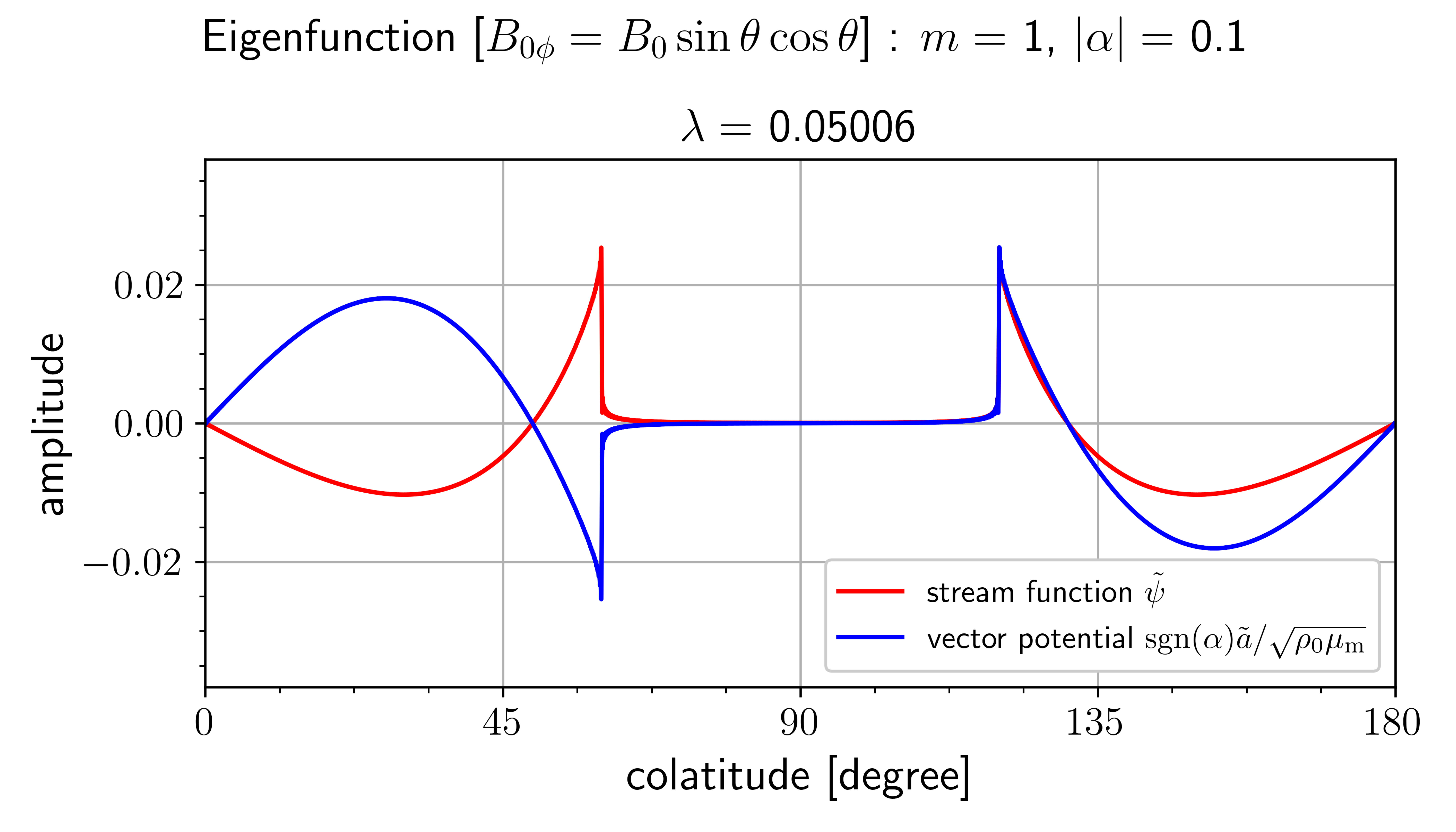}}}%
\end{minipage}
\begin{minipage}{75mm}
	\subfigure[Global contour maps.]{
	\resizebox*{75mm}{!}{\includegraphics{./eigenfunction_sincos_m1a01_map.pdf}}}%
\end{minipage}
	\caption{Eigenfunction of the sinuous mode with the nondimensional angular frequency $\lambda\approx0.05006$ when the zonal wavenumber $m=1$, the absolute value of the Lehnert number $|\alpha|=0.1$, and the simplest equatorially antisymmetric non-Malkus field $\mathcal{B}=\mu$ is imposed. The critical colatitudes $\theta_\mathrm{c}\approx59.96^\circ$ and $120.04^\circ$. (a) Amplitudes of the stream function $\tilde{\psi}$ (red line) and the scaled magnetic vector potential $\operatorname{sgn}(\alpha)\tilde{a}/\sqrt{\rho_0\mu_\mathrm{m}}$ (blue line) as functions of the colatitude. (b) Contour maps of the stream function $\psi_1$ (left panel) and the scaled magnetic vector potential $\operatorname{sgn}(\alpha)a_1/\sqrt{\rho_0\mu_\mathrm{m}}$ (right panel) in the Mollweide projection.}
	\label{FIG_eigenfunction_sincos_m1a01}
\end{center}
\end{figure}

\begin{figure}
\begin{center}
\begin{minipage}{75mm}
	\subfigure[Sinuous mode with $\lambda\approx0.00012$.]{
	\resizebox*{75mm}{!}{\includegraphics{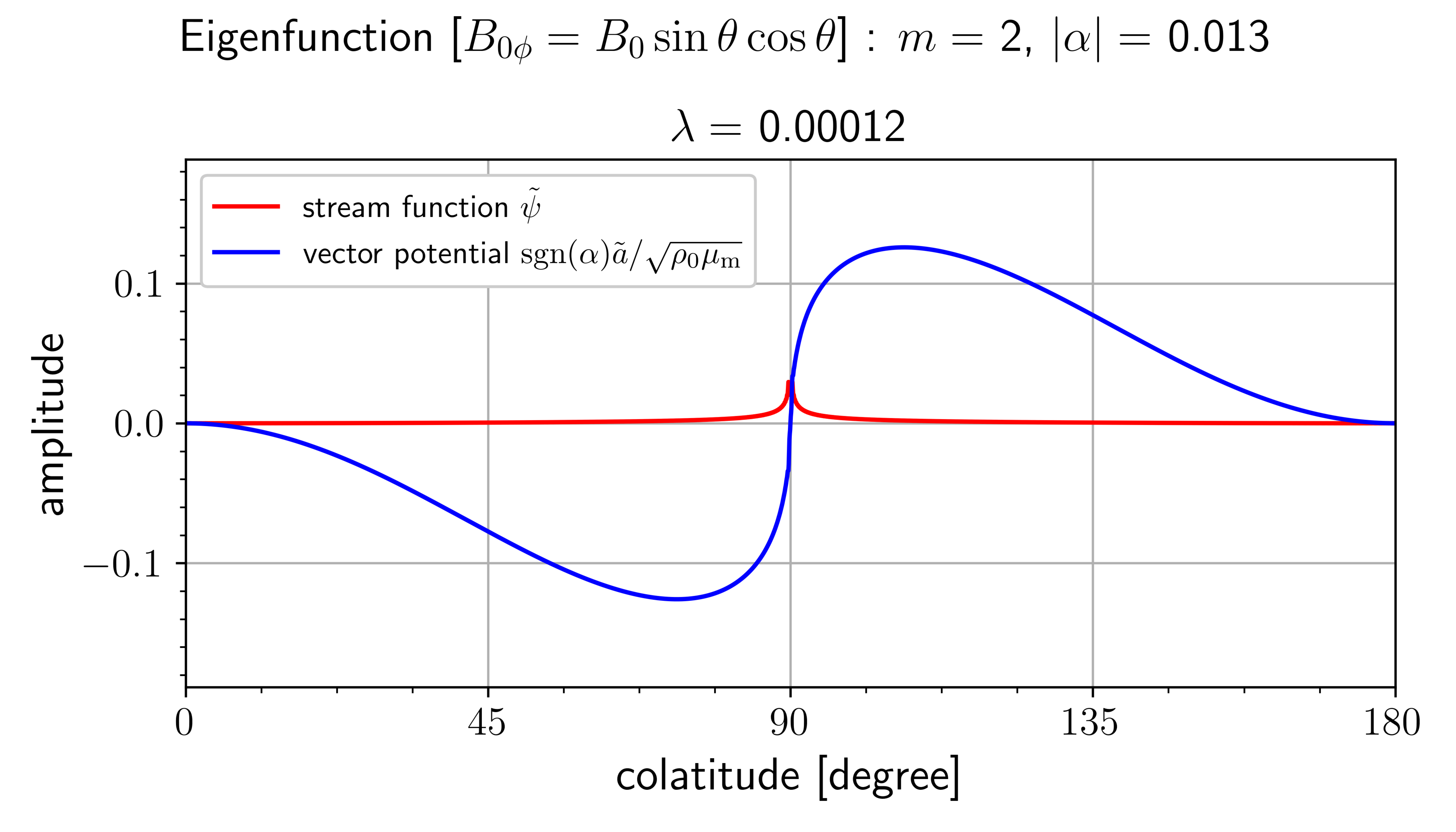}}}%
\end{minipage}
\begin{minipage}{75mm}
	\subfigure[Sinuous mode with $\lambda\approx0.00025$.]{
	\resizebox*{75mm}{!}{\includegraphics{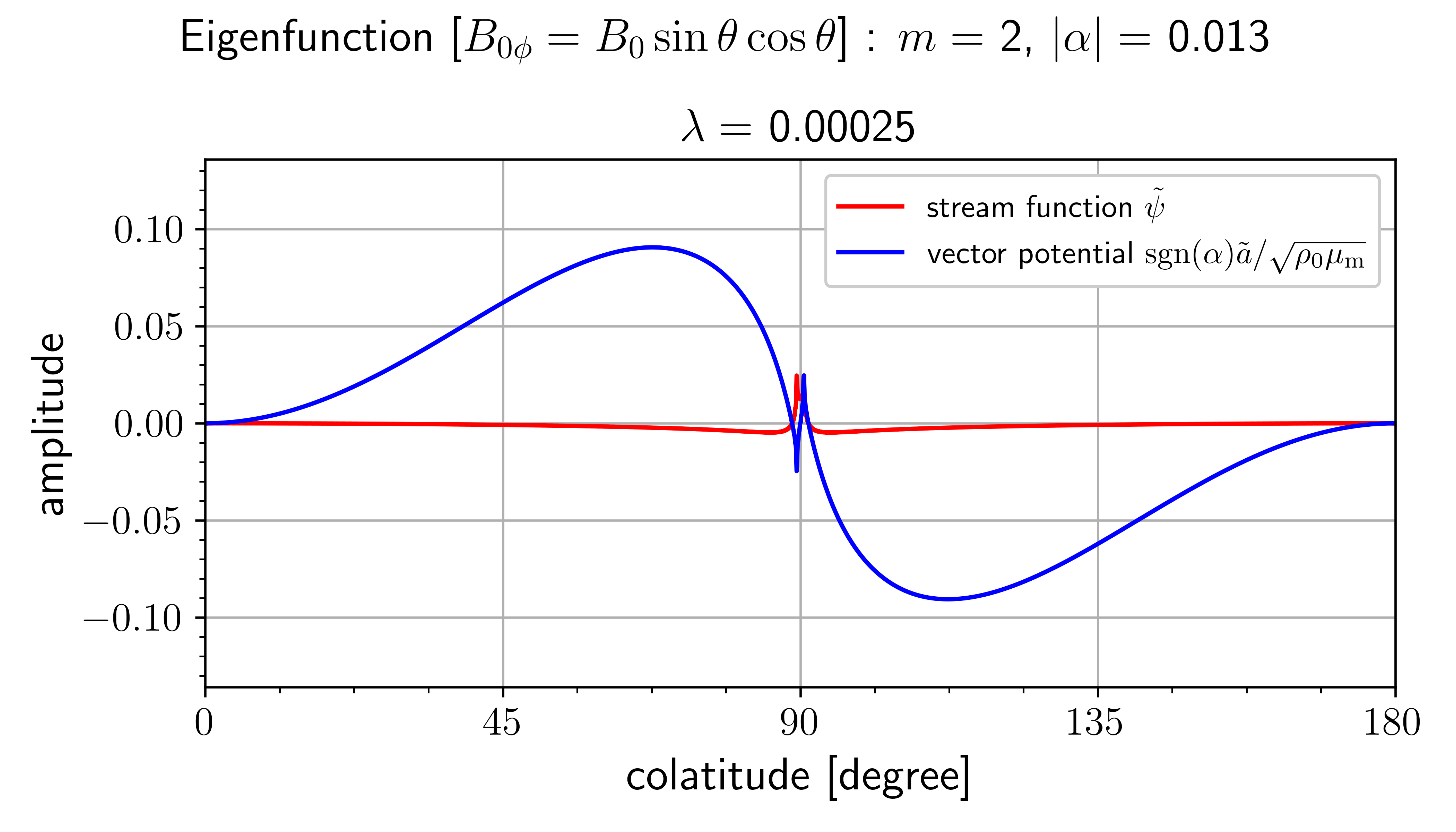}}}%
\end{minipage}
	\caption{Eigenfunctions of two magnetic-energy dominant modes for $m=2$ and $|\alpha|=0.013$. The axes and colours are identical to those in Figure \ref{FIG_eigenfunction_sincos_m1a01}(a).}
	\label{FIG_eigenfunction_sincos_m1a0013}
\end{center}
\end{figure}

%%%%%
We investigate the eigenfunctions of the Alfv\'en continuous modes in this subsection and Appendix \ref{SUBSEC_comparison}. Figure \ref{FIG_eigenfunction_sincos_m1a01} presents the typical structures of the perturbations in the stream function $\psi_1$ and the magnetic vector potential $a_1$ for the continuous modes. The dependences of their amplitudes $\tilde{\psi}$ and $\tilde{a}$ on the colatitude are illustrated in Figure \ref{FIG_eigenfunction_sincos_m1a01}(a) and used for preparing the contour maps of Figure \ref{FIG_eigenfunction_sincos_m1a01}(b). From these figures, we note that spiky singular structures appear at the critical latitudes of the eigenmode. As shown in Section \ref{SUBSEC_frobenius}, the eigenfunction has a logarithmic singularity or a step function singularity or both. In Appendix \ref{SUBSEC_comparison}, we provide the ratios of their contributions to the eigenfunctions around their corresponding critical latitudes. We extract two eigenmodes that are magnetic energy dominant from each of the two noticeable blue upward wedges in the upper right panel of Figure \ref{FIG_dispersion_sincos_m2}, and their eigenfunctions are plotted in Figure \ref{FIG_eigenfunction_sincos_m1a0013}.\par
\begin{figure}
\begin{center}
	\resizebox*{150mm}{!}{\includegraphics{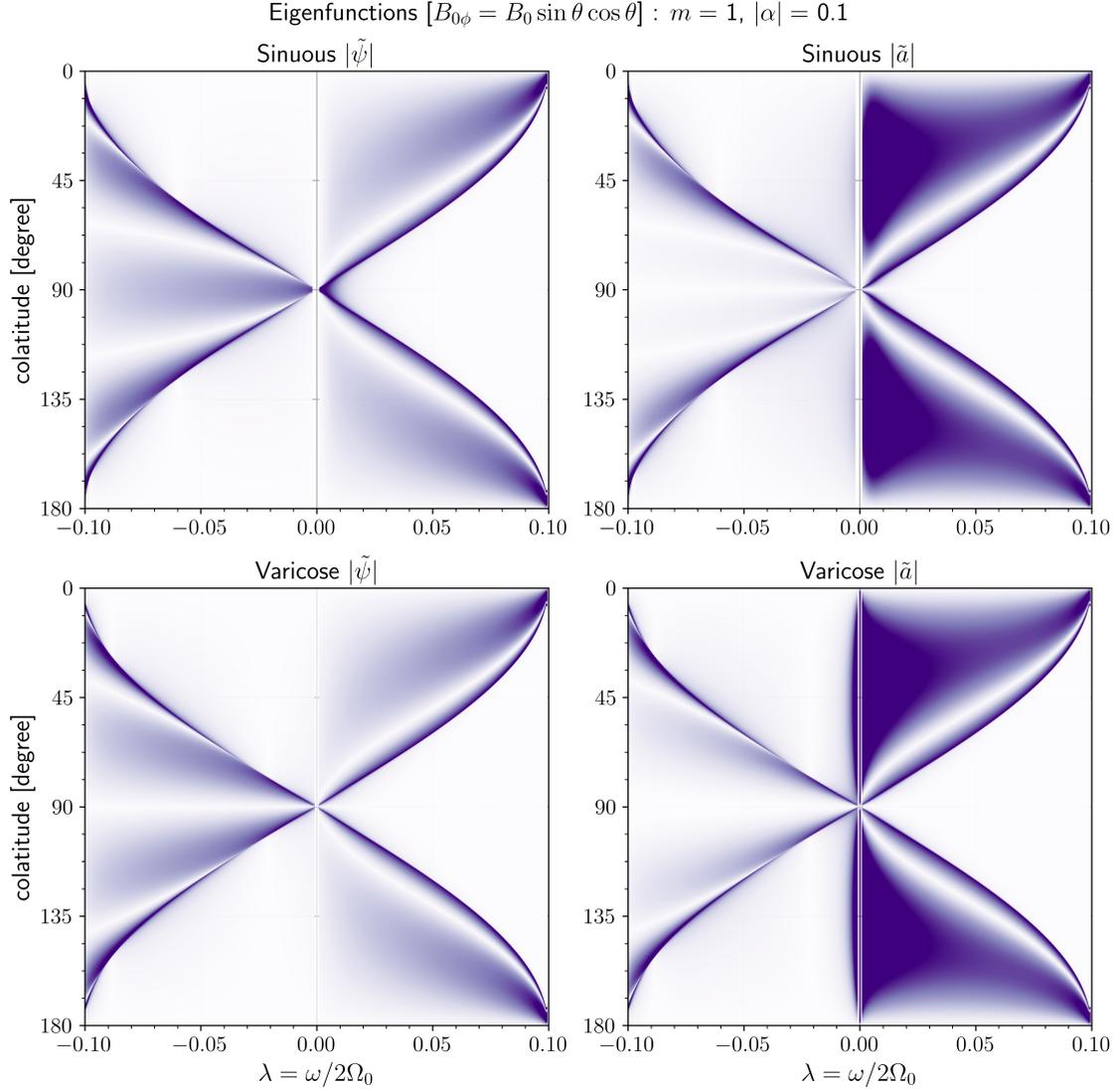}}%
	\caption{Amplitudes of the eigenfunctions of all the obtained continuous modes when the zonal wavenumber $m=1$, the absolute value of the Lehnert number $|\alpha|=0.1$, and the basic field is the simplest equatorially antisymmetric non-Malkus one ($\mathcal{B}=\mu$). The four panels are divided into the stream function and magnetic vector potential in the left and right columns, respectively, and sinuous and varicose modes in the upper and lower rows, respectively. The vertical axes of each panel correspond to the colatitude, and the horizontal axes represent the nondimensional angular frequency $\lambda$. The darker the colours, the higher the absolute values $|\tilde{\psi}|$ and $|\tilde{a}|$ of the amplitudes of the stream function and the magnetic vector potential.}
	\label{FIG_alleigenfunction_sincos_m1a01}
\end{center}
\end{figure}

\begin{figure}
\begin{center}
	\resizebox*{150mm}{!}{\includegraphics{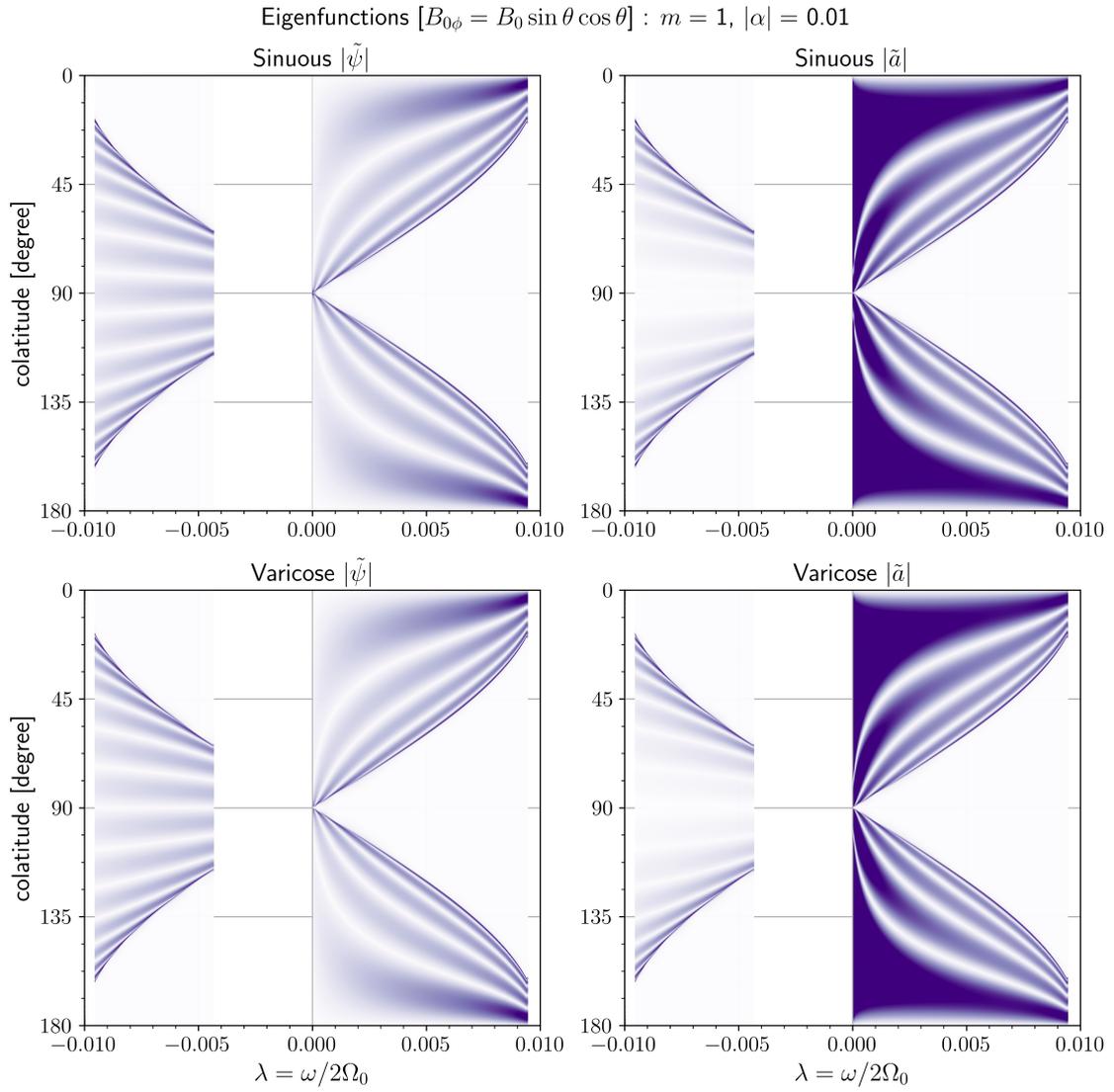}}%
	\caption{Same as Figure \ref{FIG_alleigenfunction_sincos_m1a01}, but for $|\alpha|=0.01$.}
	\label{FIG_alleigenfunction_sincos_m1a001}
\end{center}
\end{figure}

\begin{figure}
\begin{center}
	\resizebox*{150mm}{!}{\includegraphics{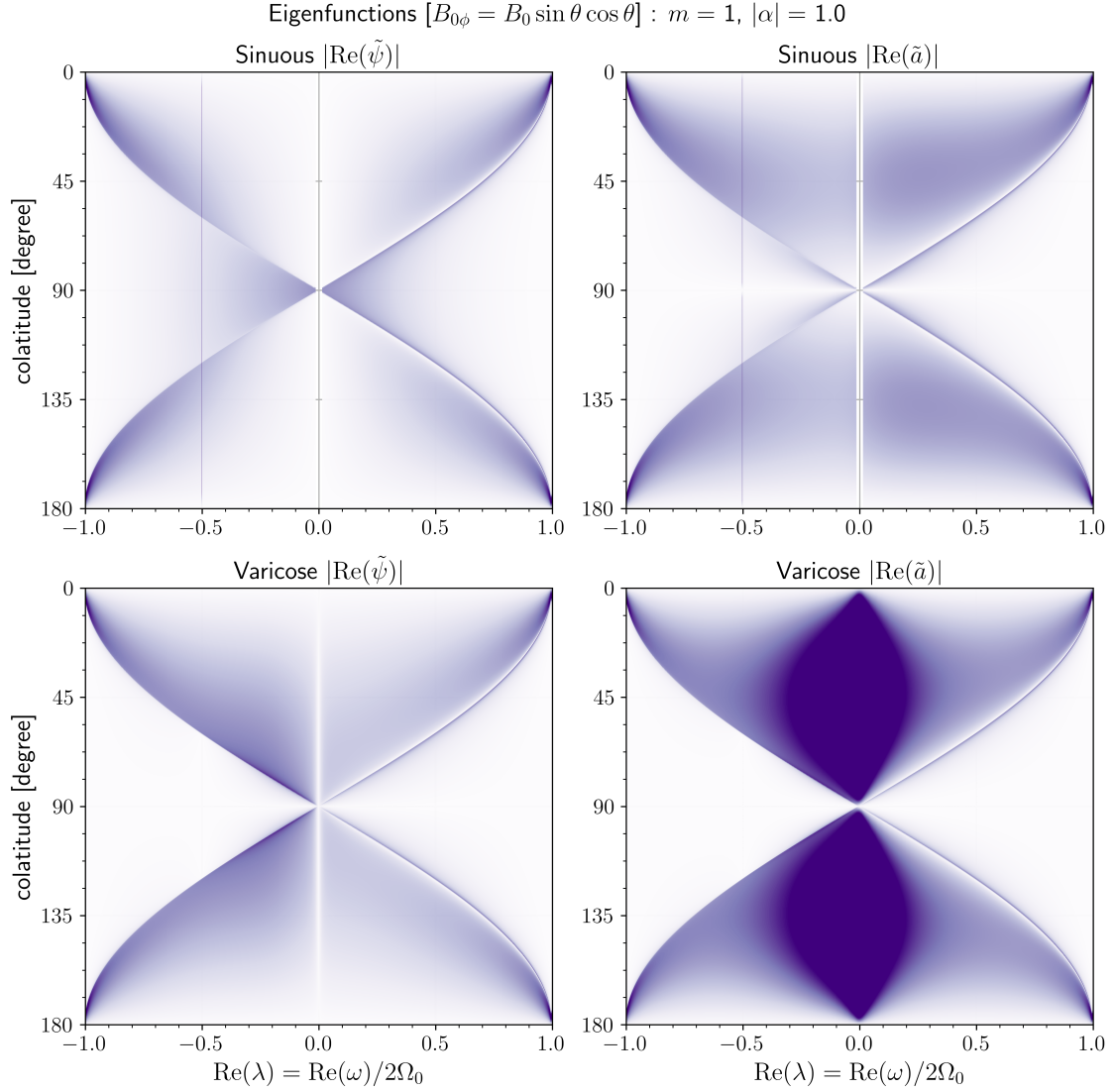}}%
	\caption{Same as Figure \ref{FIG_alleigenfunction_sincos_m1a01}, but for $m=1$ and $|\alpha|=1$.}
	\label{FIG_alleigenfunction_sincos_m2a1}
\end{center}
\end{figure}

To obtain the whole picture of the eigenfunctions of the continuous modes, we exhibit those of all the obtained continuous modes. Figures \ref{FIG_alleigenfunction_sincos_m1a01} and \ref{FIG_alleigenfunction_sincos_m1a001} present the heatmaps of the absolute values, or $|\tilde{\psi}|$ and $|\tilde{a}|$, of their amplitudes as functions of $\lambda$ and the colatitude when $|\alpha|=0.1$ and $0.01$, respectively, and $m=1$. The left columns in these figures correspond to $|\tilde{\psi}|$, and $|\tilde{a}|$ is depicted in their right columns. The sinuous and varicose modes are presented in the upper and lower rows, respectively. The colour darkens as the absolute value increases. The maps indicate that the retrograde continuous modes are evanescent on the polar side of the critical latitudes, while the prograde ones are evanescent on the equatorial side. However, only the case in which $|\alpha|$ is sufficiently smaller than unity displays this behaviour (see Figure \ref{FIG_alleigenfunction_sincos_m2a1}). Furthermore, the comparison of the eigenmodes having the same value of $\lambda/m|\alpha|$ in Figures \ref{FIG_alleigenfunction_sincos_m1a01} and \ref{FIG_alleigenfunction_sincos_m1a001} shows that the smaller the value of $|\alpha|$ is, the smaller the typical north--south wavelengths of their amplitudes. In Section \ref{SEC_discussion}, we will therefore examine the behaviour of the wave packets possessing large wavenumbers at small values of $|\alpha|$, which may be applied to the Earth's core conditions, based on the ray theory to better understand our numerical results.\par
For moderate or large values of $|\alpha|$, less striking differences exist in the evanescent property between the retrograde and prograde modes which possess the same absolute value $|\lambda|$ of their angular frequency. This statement is based on Figure \ref{FIG_alleigenfunction_sincos_m2a1}, which shows the heatmaps for $m=1$ and $|\alpha|=1$, and other experiments with several values of $m$ and $|\alpha|$ (not shown). Therefore, the contrast in the property between the retrograde and prograde modes as demonstrated in Figures \ref{FIG_alleigenfunction_sincos_m1a01} and \ref{FIG_alleigenfunction_sincos_m1a001} can be attributed to the planetary $\beta$ effect, that is, the effect of rotation. The ray-tracing analysis and local dispersion relation, which we are going to discuss in Section \ref{SEC_discussion} offer a similar explanation. In addition, the fast MR mode buried in the continuous modes at $\lambda=-1/2$ is discernible in the upper panels of Figure \ref{FIG_alleigenfunction_sincos_m2a1} (see also Appendix \ref{SEC_discrete}). We attempted to use these plots to discover buried discrete eigenmodes other than this fast MR mode, although no such eigenmodes were found.\par
\begin{figure}
\begin{center}
\begin{minipage}{75mm}
	\subfigure[$\mathcal{B}=\mu$.]{
	\resizebox*{75mm}{!}{\includegraphics{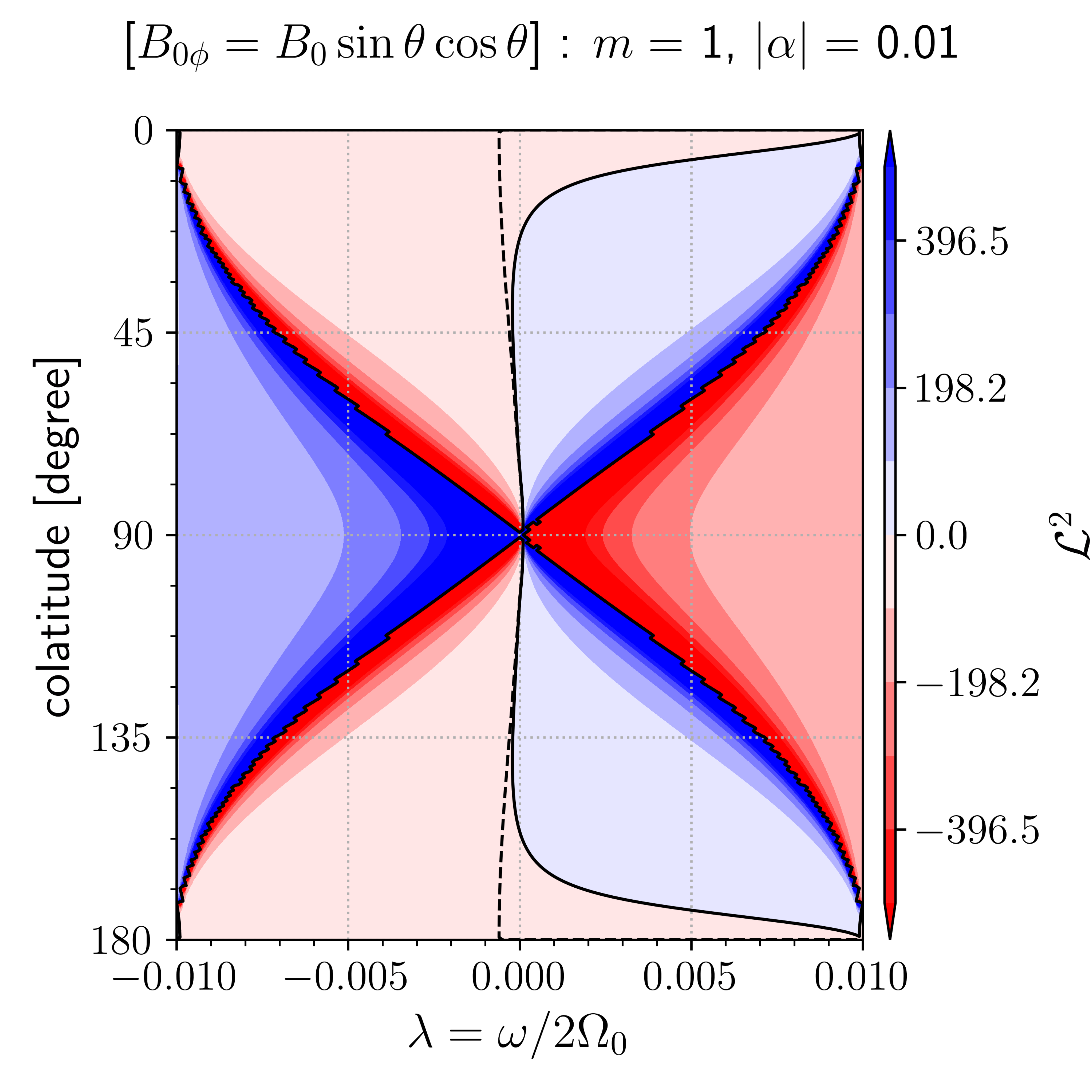}}}%
\end{minipage}
\begin{minipage}{75mm}
	\subfigure[$\mathcal{B}=\mu\sqrt{1-\mu^2}$.]{
	\resizebox*{75mm}{!}{\includegraphics{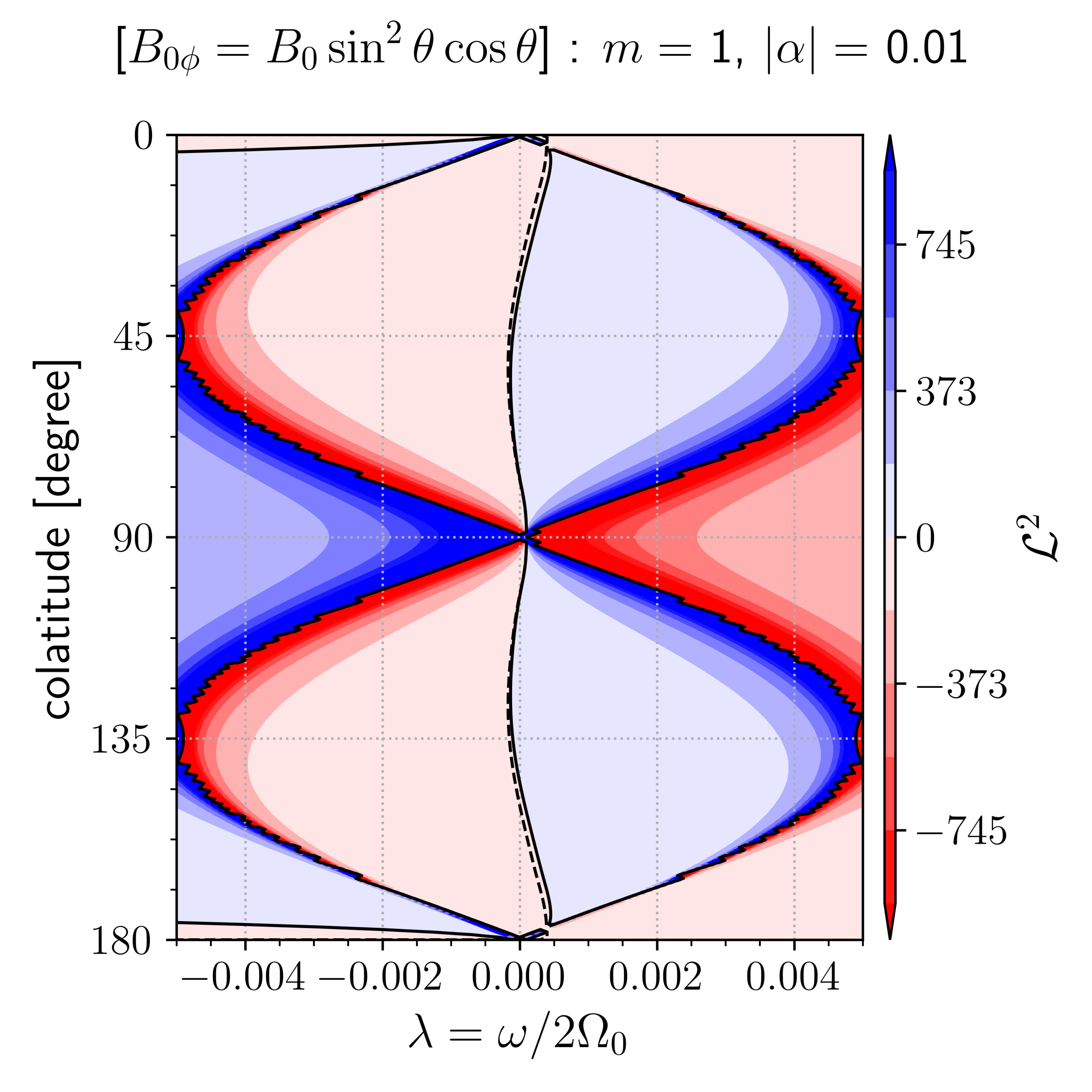}}}%
\end{minipage}
	\caption{Dependence of the function $\mathcal{L}^2$ given by \eqref{DEF_alternative_differential_potential} on the nondimensional angular frequency $\lambda$ and the colatitude when the zonal wavenumber $m=1$, the absolute value of the Lehnert number $|\alpha|=0.01$, and the basic fields are (a) $\mathcal{B}=\mu$ and (b) $\mathcal{B}=\mu\sqrt{1-\mu^2}$. The solid and dashed curves present the contour lines corresponding to $\mathcal{L}^2=0$ and $-1$, respectively.}
	\label{FIG_waveeq}
\end{center}
\end{figure}

%%%%%
The evanescent property can be determined using the function
\begin{subequations}
\begin{equation}
\mathcal{L}^2(\mu; m, \lambda)\,\equiv\,-m^2\,-\,\frac{m(1-\mu^2)}{\varLambda}\left[\lambda+2m\alpha^2\mathcal{B}\frac{\ud(\mathcal{B}\mu)}{\ud\mu}\right]\,-\,\frac{1-\mu^2}{2\sqrt{\varLambda}}\frac{\ud}{\ud\mu}\left(\frac{1-\mu^2}{\sqrt{\varLambda}}\frac{\ud\varLambda}{\ud\mu}\right)\,,
\label{DEF_alternative_differential_potential}
\end{equation}
which appears in an alternative form of the differential equation \eqref{EQ_differential}
\begin{equation}
\frac{\ud}{\ud\mu}\left[(1-\mu^2)\frac{\ud(\tilde{\psi}\sqrt{\varLambda})}{\ud\mu}\right]\,+\,\frac{\mathcal{L}^2}{1-\mu^2}(\tilde{\psi}\sqrt{\varLambda})\,=\,0\,.
\label{EQ_alternative_differential}
\end{equation}
\end{subequations}
The Mercator projection transformation $y=(1/2)\ln[(1+\mu)/(1-\mu)]$ yields a differential equation of the harmonic oscillation,
\begin{subequations}
\begin{equation}
\frac{\ud^2(\tilde{\psi}\sqrt{\varLambda})}{\ud y^2}\,+\,\mathcal{L}^2(\tilde{\psi}\sqrt{\varLambda})\,=\,0\,,
\end{equation}
which shows that the sign of $\mathcal{L}^2$ determines the evanescent property of $\tilde{\psi}\sqrt{\varLambda}$ at the latitude. We can also rewrite \eqref{EQ_alternative_differential} into the form
\begin{equation}
\frac{\ud^2[\tilde{\psi}\sqrt{(1-\mu^2)\varLambda}]}{\ud \mu^2}\,+\,\frac{\mathcal{L}^2+1}{(1-\mu^2)^2}[\tilde{\psi}\sqrt{(1-\mu^2)\varLambda}]\,=\,0\,,
\end{equation}
\end{subequations}
which, similarly, shows that the value of $\mathcal{L}^2$ indicates the evanescent property. The equivalent of this differential equation was derived by \citet{Fox_1999}, although the form of $\mathcal{L}^2$ differs from theirs because their equation is based on a non-rotating frame; the partial derivatives with respect to $t$ in our equations have to be replaced by $(\upartial/\upartial t)+\varOmega_0(\upartial/\upartial \phi)$ in the non-rotating frame. The left panel in Figure \ref{FIG_waveeq} presents contour plots of $\mathcal{L}^2$ as a function of $\lambda$ and the colatitude when $|\alpha|=0.01$ and $\mathcal{B}=\mu$. We observe that the area where $\mathcal{L}^2>0$ (or $\mathcal{L}^2>-1$) in the panel certainly agrees with the wavy regions in Figure \ref{FIG_alleigenfunction_sincos_m1a001}. We now consider the case where $|\alpha|$ is small. On noting that $\lambda=\mathrm{O}(|\alpha|)$ and $|\varLambda|=\mathrm{O}(|\alpha|^2)$ for the continuous modes, we obtain
\begin{equation}
\mathcal{L}^2\,=\,-\frac{m(1-\mu^2)}{\varLambda}\lambda\,+\,\mathrm{O}(|\varLambda|^{-2}|\alpha|^4)
\label{DEF_alternative_differential_potential_approximate}
\end{equation}
unless the latitudinal position $\mu$ is very close to a critical latitude ($|\varLambda|\gg\mathrm{O}(|\alpha|^3)$), since $(\ud\varLambda/\ud\mu)=\mathrm{O}(|\alpha|^2)$ and $(\ud^2\varLambda/\ud\mu^2)=\mathrm{O}(|\alpha|^2)$, as can be seen from the definition of the function $\varLambda$. When $\mathcal{B}=\mu$ (and $|\mu^2-\lambda^2/m^2\alpha^2|\gg\mathrm{O}(|\alpha|)$), the oscillatory condition $\mathcal{L}^2>0$ on the equatorial side ($\mu^2<\lambda^2/m^2\alpha^2$) of the critical latitudes therefore requires that $\lambda<0$ (the retrograde continuous modes), and $\mathcal{L}^2>0$ on the polar side ($\mu^2>\lambda^2/m^2\alpha^2$) for the prograde modes ($\lambda>0$). This explains the contrasting evanescent behaviour between the retrograde and prograde continuous modes for a small value of $|\alpha|$.\par

% = = = = = = = = = = = = = = = = = = = = = = = = = = = = = = %
%                                                             %
%                         Section 4                           %
%                                                             %
% = = = = = = = = = = = = = = = = = = = = = = = = = = = = = = %

\section{Interpretation in terms of ray theory and discussion}\label{SEC_discussion}
%%%%%
To better understand the continuous modes and their eigenfunctions, we reduce the system investigated thus far to a more restricted situation where $|\alpha|\ll1$, which is appropriate for the Earth's core. We apply the ray theory, in which an inhomogeneous background field varies with a spatial scale much larger than typical wavelengths. In addition, we track the path of a wave packet migrating with its group velocity. It should be noted that we now use $\theta$ as a dependent variable instead of $\mu$ to define a north--south wavenumber in terms of $\theta$ for the ray theory and consider $\mathcal{B}$ as the function of $\theta$ rather than $\mu$. We introduce the local coordinates $(\varTheta, \varPhi)$, which suitably measure the spatial scale size of the typical wavelength of a wave train. The introduction of small parameters aids in incorporating such a setting into the governing equations. In Section \ref{SUBSEC_eigenfunctions}, we found that the typical meridional wavelengths decrease as the value of $|\alpha|$ decreases. Thus, it would be reasonable to select $|\alpha|$ as the parameter, if the value of $|\alpha|$ is sufficiently smaller than unity. The local coordinates are then stretched in the forms
\begin{subequations}\label{DEF_rescale}
\begin{equation}
\varTheta\,\equiv\,|\alpha|^{-1/2}\theta\,,\qquad\varPhi\,\equiv\,|\alpha|^{-1/2}\phi\,.
\label{DEF_rescale_spatial}
\end{equation}
The temporal scale of the wave period ($\lambda^{-1}=\mathrm{O}(|\alpha|^{-1/2})$) is similarly far from that of the migration of the wave train. The new shrunk time $T$, which is useful for measuring the latter, is given by
\begin{equation}
T\,\equiv\,|\alpha|\tau\,.
\label{DEF_rescale_temporal}
\end{equation}
\end{subequations}
The values of the exponents of $|\alpha|$ of the above variables are determined in Appendix \ref{SEC_derivations}.\par
%%%%%
We then introduce a locally defined wavenumber and angular frequency, which depend on the global coordinates $(\theta,\phi)$ and $T$, and subsequently derive a local dispersion relation and ray-tracing equations, which predict the movement of a wave packet. Their derivations are based on explanations in standard textbooks regarding wave dynamics \citep[e.g.][]{lighthill1978waves}, and we present an explanation of the corresponding details in Appendix \ref{SEC_derivations}. Here we summarise the results. The expression of the perturbations of the stream function postulated in Section \ref{SEC_mathematical} is rewritten here as $\psi_1\equiv\mathrm{Re}[M(\phi, \theta, T)\mathrm{e}^{\mathrm{i}\varphi_\mathrm{L}(\varPhi,\varTheta,\tau)}]$, where $M$ is the wave amplitude, and $\varphi_\mathrm{L}$ is the phase of the wave packet. With this ansatz, the local wavenumber and local nondimensional angular frequency are expressed as
\begin{equation}
k(\phi,\theta,T)\,\equiv\,\frac{1}{\sin\theta}\frac{\upartial\varphi_\mathrm{L}}{\upartial\varPhi}\,,\qquad l(\phi,\theta,T)\,\equiv\,\frac{\upartial\varphi_\mathrm{L}}{\upartial(-\varTheta)}\,,\qquad\lambda(\phi,\theta,T)\,\equiv\,-\frac{\upartial\varphi_\mathrm{L}}{\upartial\tau}\,.
\label{DEF_wavenumber_frequency}
\end{equation}
It should be noted that $\varphi_\mathrm{L}$ depends on the local coordinates $(\varPhi,\varTheta,\tau)$, while $M$, $k$, $l$, and $\lambda$ depend on the global ones $(\phi,\theta,T)$. The local dispersion relation for our present problem is obtained from the leading order terms in the governing equations \eqref{EQ_linear} in the form
\begin{equation}
\mathcal{D}(\phi,\theta,T,k,l,\lambda_\mathrm{s})\,\equiv\,\lambda_\mathrm{s}^2(k^2+l^2)\,+\,\lambda_\mathrm{s}k\sin\theta\,-\,k^2\mathcal{B}^2\sin^2\theta(k^2+l^2)\,=\,0\,,
\label{DEF_local_dispersion}
\end{equation}
where $\lambda_\mathrm{s}\equiv|\alpha|^{-1/2}\lambda$ is the scaled nondimensional angular frequency. On replacing $\sin\theta$ and $\mathcal{B}$ in \eqref{DEF_local_dispersion} with constants, we obtain an equivalent to the nondimensional dispersion relation on a middle latitude $\beta$ plane \citep{Zaqarashvili2007}. It should be noted that we denote $\lambda_\mathrm{s}=\lambda_\mathcal{H}(\phi, \theta, k, l, T)$ as the solution of \eqref{DEF_local_dispersion} for $\lambda_\mathrm{s}$. From \eqref{DEF_local_dispersion}, the components of the nondimensional local group velocity $\bm{c}_\mathrm{g}$ are given by
\begin{subequations}\label{DEF_group}
\begin{linenomath}\begin{align}
\frac{c_{\mathrm{g},\phi}}{|\alpha|}\,&\equiv\,\frac{\upartial\lambda_\mathcal{H}}{\upartial k}\,=\,-\frac{(\upartial\mathcal{D}/\upartial k)|_{\lambda_\mathrm{s}=\lambda_\mathcal{H}}}{(\upartial\mathcal{D}/\upartial \lambda_\mathrm{s})|_{\lambda_\mathrm{s}=\lambda_\mathcal{H}}}\,=\,\frac{2k\mathcal{B}^2\sin^2\theta(2k^2+l^2)-\lambda_\mathcal{H}(2k\lambda_\mathcal{H}+\sin\theta)}{2\lambda_\mathcal{H}(k^2+l^2)+k\sin\theta}\,,
\label{DEF_group_phi}\\
\frac{c_{\mathrm{g},-\theta}}{|\alpha|}\,&\equiv\,\frac{\upartial\lambda_\mathcal{H}}{\upartial l}\,=-\frac{(\upartial\mathcal{D}/\upartial l)|_{\lambda_\mathrm{s}=\lambda_\mathcal{H}}}{(\upartial\mathcal{D}/\upartial \lambda_\mathrm{s})|_{\lambda_\mathrm{s}=\lambda_\mathcal{H}}}\,=\,-\frac{2l(\lambda_\mathcal{H}^2-k^2\mathcal{B}^2\sin^2\theta)}{2\lambda_\mathcal{H}(k^2+l^2)+k\sin\theta}\,.
\label{DEF_group_theta}
\end{align}\end{linenomath}
\end{subequations}
We eventually derive the ray-tracing equations
\begin{subequations}\label{EQ_ray}
\begin{equation}
	\sin\theta\frac{\ud_\mathrm{g}\phi}{\ud T}\,=\,\frac{c_{\mathrm{g},\phi}}{|\alpha|}\,,\qquad\frac{\ud_\mathrm{g}(-\theta)}{\ud T}\,=\,\frac{c_{\mathrm{g},-\theta}}{|\alpha|}\,,
\label{EQ_ray_phi_theta}
\end{equation}
and 
\begin{linenomath}\begin{align}
\frac{\ud_\mathrm{g}(k\sin\theta)}{\ud T}\,&=\,-\left(\frac{\upartial \lambda_\mathcal{H}}{\upartial\phi}\right)_{k,l}\,=\,0\,,
\label{EQ_ray_k}\\
\frac{\ud_\mathrm{g}l}{\ud T}\,+\,\frac{c_{\mathrm{g},\phi}}{|\alpha|}k\cot\theta\,&=\,-\left[\frac{\upartial \lambda_\mathcal{H}}{\upartial(-\theta)}\right]_{k,l}\,=\,\frac{k^2[\ud(\mathcal{B}^2\sin^2\theta)/\ud\theta](k^2+l^2)-\lambda_\mathcal{H}k\cos\theta}{2\lambda_\mathcal{H}(k^2+l^2)+k\sin\theta}\,,
\label{EQ_ray_l}\\
\frac{\ud_\mathrm{g}\lambda_\mathrm{s}}{\ud T}\,&=\,\left(\frac{\upartial \lambda_\mathcal{H}}{\upartial T}\right)_{k,l}\,=\,0\,,
\label{EQ_ray_lambda}
\end{align}\end{linenomath}
where the material time derivative moving with the local group velocity is
\begin{equation}
\frac{\ud_\mathrm{g}}{\ud T}\,\equiv\,\frac{\upartial}{\upartial T}\,+\,\frac{c_{\mathrm{g},\phi}}{|\alpha|\sin\theta}\frac{\upartial}{\upartial\phi}\,+\,\frac{c_{\mathrm{g},-\theta}}{|\alpha|}\frac{\upartial}{\upartial(-\theta)}\,=\,\frac{\upartial}{\upartial T}\,+\,\frac{\bm{c}_\mathrm{g}}{|\alpha|}\bm{\cdot}\bm{\nabla}_\mathrm{G}\,,
\label{DEF_derivative}
\end{equation}
\end{subequations}
with $\bm{\nabla}_\mathrm{G}\equiv(\hat{\bm{e}}_\phi/\sin\theta)(\upartial/\upartial\phi)+\hat{\bm{e}}_{-\theta}[\upartial/\upartial(-\theta)]$. According to these equations, a wave train migrates with its group velocity depending on its latitudinal position and its dominant local wavenumber, which also varies with the colatitude $\theta$. Furthermore, \eqref{EQ_ray} shows that $k\sin\theta$ and $\lambda_\mathrm{s}$ (or $\lambda$) are invariant along a ray trajectory, but $l$ is not.\par
%%%%%
We conduct the numerical time integration of the ray-tracing equations \eqref{EQ_ray} with \eqref{DEF_group} for the movement of a wave packet originating at a given initial position $(\theta, \phi)$ with a given initial local wavenumber $(k, l)$ and a local dimensionless angular frequency $\lambda$ determined by the local dispersion relation \eqref{DEF_local_dispersion} \citep[e.g.][]{10.3389/fspas.2022.856912}. In our code, the initial local longitudinal wavenumber $k_\mathrm{init}$ is calculated from the relation \eqref{DEF_local_dispersion} with the \texttt{scipy.optimize.fsolve} function of the SciPy library after the initial values of $l$, $\lambda_\mathrm{s}$, $\phi$, and $\theta$ are specified. The succeeding time integration of \eqref{EQ_ray} is based on an explicit Runge--Kutta method of order eight (the \texttt{DOP853} algorithm in the {\tt \verb|scipy.integrate.solve_ivp|} function of the SciPy library). This integration is conducted without explicitly using \eqref{DEF_local_dispersion} on the way, and the numerical errors in our calculations are monitored based on the value of the function $\mathcal{D}$ of \eqref{DEF_local_dispersion}. Their results are also compared to those of Section \ref{SUBSEC_eigenfunctions} in terms of the evanescent property.\par
\begin{figure}
\begin{center}
	\resizebox*{150mm}{!}{\includegraphics{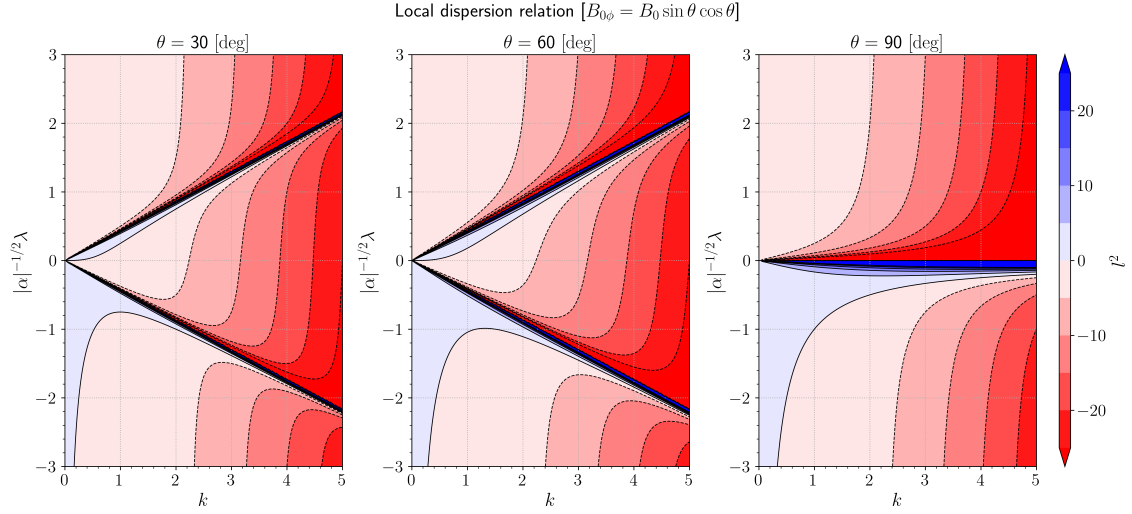}}%
	\caption{Local dispersion relation written as \eqref{EQ_l2} when the simplest equatorially antisymmetric non-Malkus field $\mathcal{B}=\cos\theta$ is imposed. In these panels, the square $l^2$ of the local meridional wavenumber is displayed as a function of the local longitudinal wavenumber $k$ and the scaled dimensionless angular frequency $\lambda_\mathrm{s}=|\alpha|^{-1/2}\lambda$ for the colatitudes $\theta=30^\circ$ (left panel), $60^\circ$ (middle), and $90^\circ$ (right).}
	\label{FIG_local_sincos_klambda}
\end{center}
\end{figure}

%%%%%
Before demonstrating the trajectories obtained numerically, we examine some properties of the local dispersion relation. In the following preliminary considerations, we assume that the physical variables satisfy the relation $\lambda_\mathrm{s}=\lambda_\mathcal{H}(\phi, \theta, k, l, T)$ at any time. Figures \ref{FIG_local_sincos_kl} and \ref{FIG_local_sin_kl} present contour plots of the scaled dimensionless angular frequency $\lambda_\mathrm{s}$ as a function of $k$ and $l$ for three different latitudes, when $\mathcal{B}=\cos\theta$ and $\mathcal{B}=1$, respectively. For ease of understanding Figure \ref{FIG_local_sincos_kl}, we first explain Figure \ref{FIG_local_sin_kl}. These diagrams are obtained based on the equation transformed from \eqref{DEF_local_dispersion} in the form
\begin{subequations}
\begin{equation}
\lambda_\mathrm{s}\,=\,\frac{-k\sin\theta\pm|k|\sin\theta\sqrt{1+4\mathcal{B}^2(k^2+l^2)^2}}{2(k^2+l^2)}\,,
\label{DEF_local_dispersion2}
\end{equation}
which corresponds to \eqref{EQ_dispersion_Malkus} for global modes when $\mathcal{B}=1$. The wavenumber $|k|\sin\theta$ corresponds to $m$, and $k^2+l^2$ corresponds to $n(n+1)$ (see also Appendix \ref{SEC_behaviors}). In Figures \ref{FIG_local_sincos_kl} and \ref{FIG_local_sin_kl}, we take the plus of the plus-minus sign in \eqref{DEF_local_dispersion2} such that $\lambda_\mathrm{s}$ is positive. This means that the sign of $k$ signifies the longitudinal direction of the phase velocity. The nearly vertical contour lines for large absolute values of $l$ in Figure \ref{FIG_local_sin_kl} correspond to the relations describing the propagation properties of wave packets that belong to prograde ($\lambda_\mathrm{s}/k>0$) and retrograde ($\lambda_\mathrm{s}/k<0$) Alfv\'en waves $\lambda_\mathrm{s}\simeq\pm|k|\sin\theta$. Since this dispersion relation is independent of $l$, the contours become vertical. Furthermore, it can be concluded that, in Figure \ref{FIG_local_sin_kl}, the circular contour lines that are tangent to the line $k=0$ represent the dispersion relation for fast MR waves $\lambda_\mathrm{s}\simeq-k\sin\theta/(k^2+l^2)$, and that the slightly curved part of the nearly vertical contour lines near the line $l=0$ on the half plane $\lambda_\mathrm{s}/k>0$ explains how slow MR waves $\lambda_\mathrm{s}\simeq|k|\sin\theta(k^2+l^2)$ propagate. The similarities between Figure \ref{FIG_local_sin_kl} and the left and middle panels of Figure \ref{FIG_local_sincos_kl} suggest that the same is true for the case where $\mathcal{B}=\cos\theta$. However, for $\mathcal{B}=\cos\theta$, no branches of Alfv\'en and slow MR waves exist at the equator ($\theta=90^\circ$), as shown in the right panel of Figure \ref{FIG_local_sincos_kl}, since the main field vanishes there. It should be noted that the direction of the gradient $(\upartial\lambda_\mathcal{H}/\upartial k, \upartial\lambda_\mathcal{H}/\upartial l)$ at a point $(k, l)$ in these plots is the same as that of the group velocity of a wave packet having a dominant local wavenumber $(k, l)$ at the colatitude $\theta$. For a wave packet belonging to either Alfv\'en or slow MR waves, the sign of the azimuthal component $c_{\mathrm{g},\phi}$ of its group velocity is identical to that of the azimuthal component $|\alpha|(\lambda_\mathrm{s}/k)$ of its nondimensional local phase velocity, while those of the meridional components ($c_{\mathrm{g},-\theta}$ and $|\alpha|(\lambda_\mathrm{s}/l)$) are the opposite for the retrograde Alfv\'en packet.\par
%%%%%
Figure \ref{FIG_local_sincos_kl} illustrates the remarkable feature that the north--south component $c_{\mathrm{g},-\theta}$ of the group velocity vanishes at $l=0$ and $l=\pm\infty$, as can also be observed from \eqref{DEF_group_theta}. The wave train can then be refracted at or absorbed into the latitude, moving only in the $\phi$ direction there \citep[e.g.][]{acheson_1972, mckenzie_1973, doi:10.1098/rsta.1977.0106, eltayeb_mckenzie_1977, doi:10.1080/03091927908244549}. In particular, from \eqref{DEF_local_dispersion}, the latter situation $l^2\to\infty$ with a reasonable condition $\lambda_\mathrm{s}k\sin\theta\neq0$ results in the limit $\lambda_\mathrm{s}^2-k^2\mathcal{B}^2\sin^2\theta\to0$, which indicates that the latitude is a critical one ($\varLambda=0$). It follows that the nearly vertical lines for the Alfv\'en waves in Figure \ref{FIG_local_sincos_kl} should be linked to the rays corresponding to the Alfv\'en continuous modes observed in the results in Section \ref{SEC_numerical}. It should be noted that, although the nearly vertical lines in Figure \ref{FIG_local_sin_kl} are similar to those in Figure \ref{FIG_local_sincos_kl}, the Malkus field $\mathcal{B}=1$ does not yield any continuous modes, since $\lambda_\mathrm{s}^2-k^2\mathcal{B}^2\sin^2\theta$ is constant and the critical latitude does not exist.\par
\begin{figure}
\begin{center}
	\resizebox*{150mm}{!}{\includegraphics{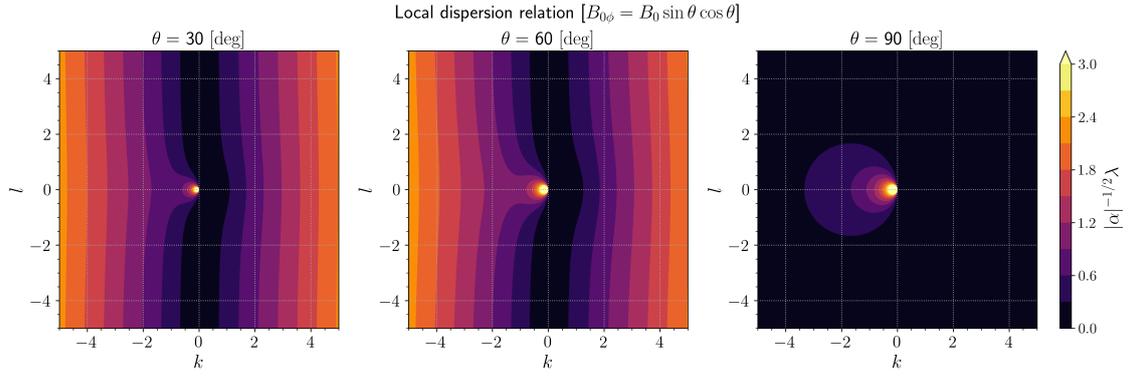}}%
	\caption{Local dispersion relation given by \eqref{DEF_local_dispersion2} when the background field is the simplest equatorially antisymmetric non-Malkus one ($\mathcal{B}=\cos\theta$). In these panels, the scaled nondimensional angular frequency $\lambda_\mathrm{s}=|\alpha|^{-1/2}\lambda$ is presented as a function of the local wavenumber $(k,l)$ for the colatitudes $\theta=30^\circ$ (left panel), $60^\circ$ (middle), and $90^\circ$ (right).}
	\label{FIG_local_sincos_kl}
\end{center}
\end{figure}
\begin{figure}
\begin{center}
	\resizebox*{150mm}{!}{\includegraphics{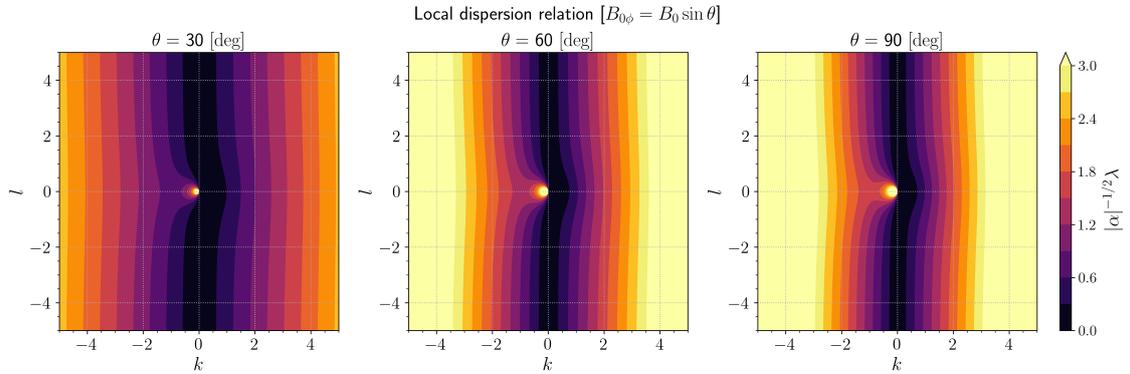}}%
	\caption{Same as Figure \ref{FIG_local_sincos_kl}, but for $\mathcal{B}=1$.}
	\label{FIG_local_sin_kl}
\end{center}
\end{figure}
%%%%%
The local dispersion relation \eqref{DEF_local_dispersion} also allows us to understand the evanescent property of waves from the sign of the squared local meridional wavenumber $l^2$. This value can be calculated from
\begin{equation}
l^2\,=\,-k^2\,-\,\frac{\lambda_\mathrm{s}k\sin\theta}{\lambda_\mathrm{s}^2-k^2\mathcal{B}^2\sin^2\theta}\,.
\label{EQ_l2}
\end{equation}
\end{subequations}
Figure \ref{FIG_local_sincos_klambda} presents contour plots of $l^2$ as a function of $k$ and $\lambda_\mathrm{s}$ for three different latitudes in the case where $\mathcal{B}=\cos\theta$. The waves can propagate only when their wavenumbers fall within the parameter domains where $\l^2>0$ in these panels, and these regions are classified into three groups. The two thin regions near the lines $\lambda_\mathrm{s}=\pm k|\mathcal{B}|\sin\theta$ in the left and middle panels correspond to the relations for prograde Alfv\'en and slow MR waves ($\lambda_\mathrm{s}/k>0$) and retrograde Alfv\'en waves ($\lambda_\mathrm{s}/k<0$). Again, as shown in the right panel, no areas for Alfv\'en and slow MR waves are found at the equator ($\theta=90^\circ$), since the background field vanishes there. The propagation properties of the fast MR waves are indicated by the domain near the line $k=0$ on the half plane $\lambda_\mathrm{s}/k<0$. It should be noted that the wave packets that approach the curved lines $l=0$ and the lines $\lambda_\mathrm{s}=\pm k|\mathcal{B}|\sin\theta$, where $l^2\to\infty$, are refracted at or absorbed into the corresponding latitude.\par
\begin{figure}
\begin{center}
	\resizebox*{150mm}{!}{\includegraphics{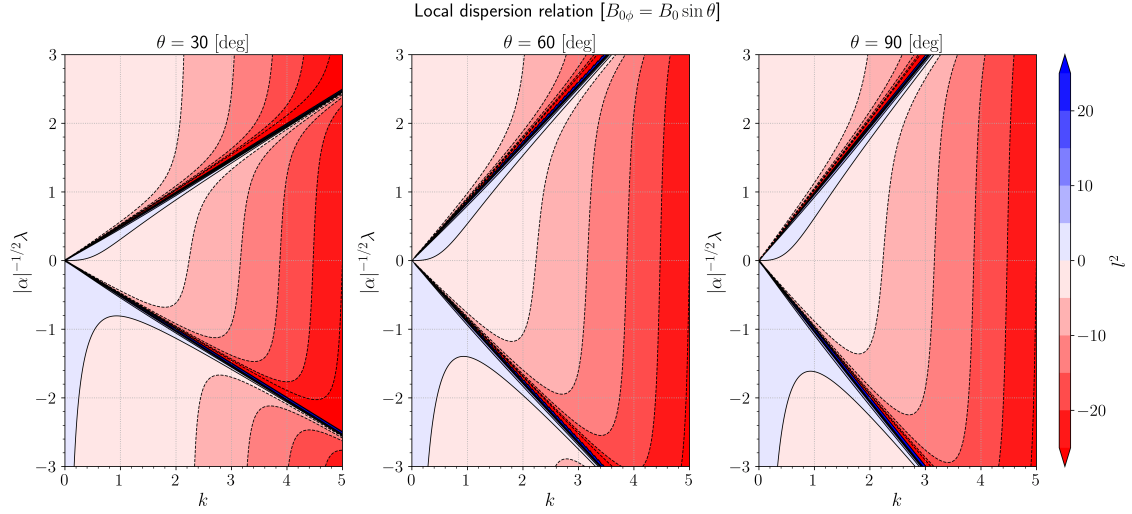}}%
	\caption{Same as Figure \ref{FIG_local_sincos_klambda}, but for $\mathcal{B}=1$.}
	\label{FIG_local_sin_klambda}
\end{center}
\end{figure}

%%%%%
Figure \ref{FIG_local_sincos_klambda} aids in the short-term prediction of the migration of a wave packet. We now consider the two cases: (i) when the wave packet moves toward its corresponding critical latitude, and (ii) when the packet proceeds in the direction away from the critical latitude. Since $k\sin\theta$ and $\lambda_\mathrm{s}$ remain constant during the movement of a wave packet, its migration in the north--south direction can be converted into the movement of the point $(k, \lambda_\mathrm{s})$ in the horizontal direction of the panels in Figure \ref{FIG_local_sincos_klambda} (unless the outlines of their contour plots change significantly depending on the latitude).
\begin{enumerate}
\item When a wave train moves toward the equator with $\lambda_\mathrm{s}/k$ and $l^2$ positive (then $\lambda_\mathrm{s}/l<0$ in the northern hemisphere from \eqref{DEF_group_theta} or Figure \ref{FIG_local_sincos_kl}), $k$ decreases and the point $(k, \lambda_\mathrm{s})$ approaches the line $\lambda_\mathrm{s}=k|\mathcal{B}|\sin\theta$ from the right on the plots of Figure \ref{FIG_local_sincos_klambda}. This means that the train belonging to either prograde Alfv\'en or slow MR waves approaches its corresponding critical latitude from the polar side and is refracted or absorbed there. If the train went beyond the latitude, $l^2$ would become negative, which results in evanescent waves. It should be noted that, although plots for the Malkus field $\mathcal{B}=1$, which are illustrated in Figure \ref{FIG_local_sin_klambda}, are similar to those in Figure \ref{FIG_local_sincos_klambda}, except for its right panel ($\theta=90^\circ$), the point $(k, \lambda_\mathrm{s})$ never reaches near the line $\lambda_\mathrm{s}=\pm k|\mathcal{B}|\sin\theta$ owing to the constancy of $k\sin\theta$ and $\lambda_\mathrm{s}$ (unless the initial condition has already approximately satisfied this equality). When $\lambda_\mathrm{s}/k$ is negative (under $\mathcal{B}=\cos\theta$), a wave train that travels poleward ($\lambda_\mathrm{s}/l<0$ in the northern hemisphere) and that does not belong to fast MR waves approaches its corresponding critical latitude, because $k$ increases while it migrates and the point $(k, \lambda_\mathrm{s})$ approaches the line $\lambda_\mathrm{s}=-k|\mathcal{B}|\sin\theta$ from the left. It follows that the train belonging to retrograde Alfv\'en waves approaches the critical latitude from the equatorial side.
\item A wave packet moving in the opposite direction from its corresponding critical latitude is realised by a local meridional phase velocity $|\alpha|(\lambda_\mathrm{s}/l)$ which has a sign opposite to that in case (i). In other words, we focus on a packet that travels poleward ($\lambda_\mathrm{s}/l>0$ in the northern hemisphere) with $\lambda_\mathrm{s}/k$ being positive and one that moves equatorward ($\lambda_\mathrm{s}/l>0$ in the northern hemisphere) with $\lambda_\mathrm{s}/k$ being negative. The point $(k, \lambda_\mathrm{s})$ then approaches the curved line $l=0$, resulting in its refraction or absorption. It can be concluded that the packet that belongs to either prograde Alfv\'en or slow MR waves approaches the latitude where $l=0$ from the equatorial side and that the packet belonging to retrograde Alfv\'en waves approaches from the polar side.
\end{enumerate}\par
%%%%%
Whether a wave packet is to be refracted at or absorbed into a latitude where $l^2\to\infty$, or $\lambda_\mathrm{s}^2=k^2\mathcal{B}^2\sin^2\theta$, is considered now. For a general profile of $\mathcal{B}$, near such a colatitude $\theta=\theta_\mathrm{c}$, \eqref{EQ_l2} becomes
\begin{equation}
l^2\,\simeq\,\frac{\lambda_\mathrm{s}/(k\sin\theta)}{(\ud\mathcal{B}^2/\ud\theta)|_{\theta=\theta_\mathrm{c}}(\theta-\theta_\mathrm{c})}\,,
\label{EQ_local_dispersion_critical}
\end{equation}
if $(\ud\mathcal{B}^2/\ud\theta)|_{\theta=\theta_\mathrm{c}}\neq0$. In particular, when $\mathcal{B}=\cos\theta$, the oscillatory condition $l^2>0$ requires that $(\theta-\theta_\mathrm{c})\cos\theta_\mathrm{c}$ has the sign opposite to that of $\lambda_\mathrm{s}/k$; the packet that belongs to either prograde Alfv\'en or slow MR waves ($\lambda_\mathrm{s}/k>0$) approaches the critical colatitude $\theta_\mathrm{c}$ from the polar side ($(\theta-\theta_\mathrm{c})\cos\theta_\mathrm{c}<0$), while the retrograde Alfv\'en one ($\lambda_\mathrm{s}/k<0$) does from the equatorial side ($(\theta-\theta_\mathrm{c})\cos\theta_\mathrm{c}>0$). This is just a mathematical paraphrasing of the consideration of the manner in which the packets approach the critical latitudes mentioned in the previous paragraph. Since the term $\lambda_\mathrm{s}k\sin\theta$ in \eqref{DEF_local_dispersion} -- which has the same sign as the numerator of \eqref{EQ_local_dispersion_critical} -- represents the planetary $\beta$ effect, the aforementioned distinction between the prograde and retrograde waves is caused by the $\beta$ effect. Moreover, on using \eqref{DEF_group} and \eqref{EQ_local_dispersion_critical}, we can obtain an asymptotic expression of the group velocity
\begin{subequations}
\begin{linenomath}\begin{align}
\frac{c_{\mathrm{g},\phi}}{|\alpha|}\,&\simeq\,\frac{\lambda_\mathrm{s}}{k}\,=\,\mathrm{O}(|\theta-\theta_\mathrm{c}|^{0})\,,\label{EQ_asymptotic_phi}\\
\frac{c_{\mathrm{g},-\theta}}{|\alpha|}\,&\simeq\,\frac{(\ud\mathcal{B}^2/\ud\theta)|_{\theta=\theta_\mathrm{c}}(k\sin\theta)^2(\theta-\theta_\mathrm{c})}{\lambda_\mathrm{s}l}\,=\,\mathrm{O}(|\theta-\theta_\mathrm{c}|^{3/2})\,.
\end{align}\end{linenomath}
\end{subequations}
This expression provides the travel time of the packet from a given latitude $\theta$ in the vicinity of the critical latitude to the latter in the form
\begin{equation}
\int_\theta^{\theta_\mathrm{c}}\frac{|\alpha|}{-c_{\mathrm{g},-\theta}(\theta_*)}\ud\theta_*\,\simeq\,-\frac{\lambda_\mathrm{s}}{(\ud\mathcal{B}^2/\ud\theta)|_{\theta=\theta_\mathrm{c}}(k\sin\theta)^2}\int_\theta^{\theta_\mathrm{c}}\frac{l(\theta_*)}{\theta_*-\theta_\mathrm{c}}\ud\theta_*\,=\,\mathrm{O}(|\theta-\theta_\mathrm{c}|^{-1/2})\,.
\label{EQ_time_critical}
\end{equation}
Any packets therefore never reach their corresponding critical latitudes in a finite time and are absorbed there. In contrast, wave packets are refracted at the latitudes where $l$ vanishes, as explained in a similar fashion in Appendix \ref{SEC_behaviors}. The latitudes are often referred to as ``turning latitudes''.\par
\begin{figure}
\begin{center}
\begin{minipage}{75mm}
	\subfigure[Initial local meridional wavenumber $l_\mathrm{init}=2$.]{
	\resizebox*{70mm}{!}{\includegraphics{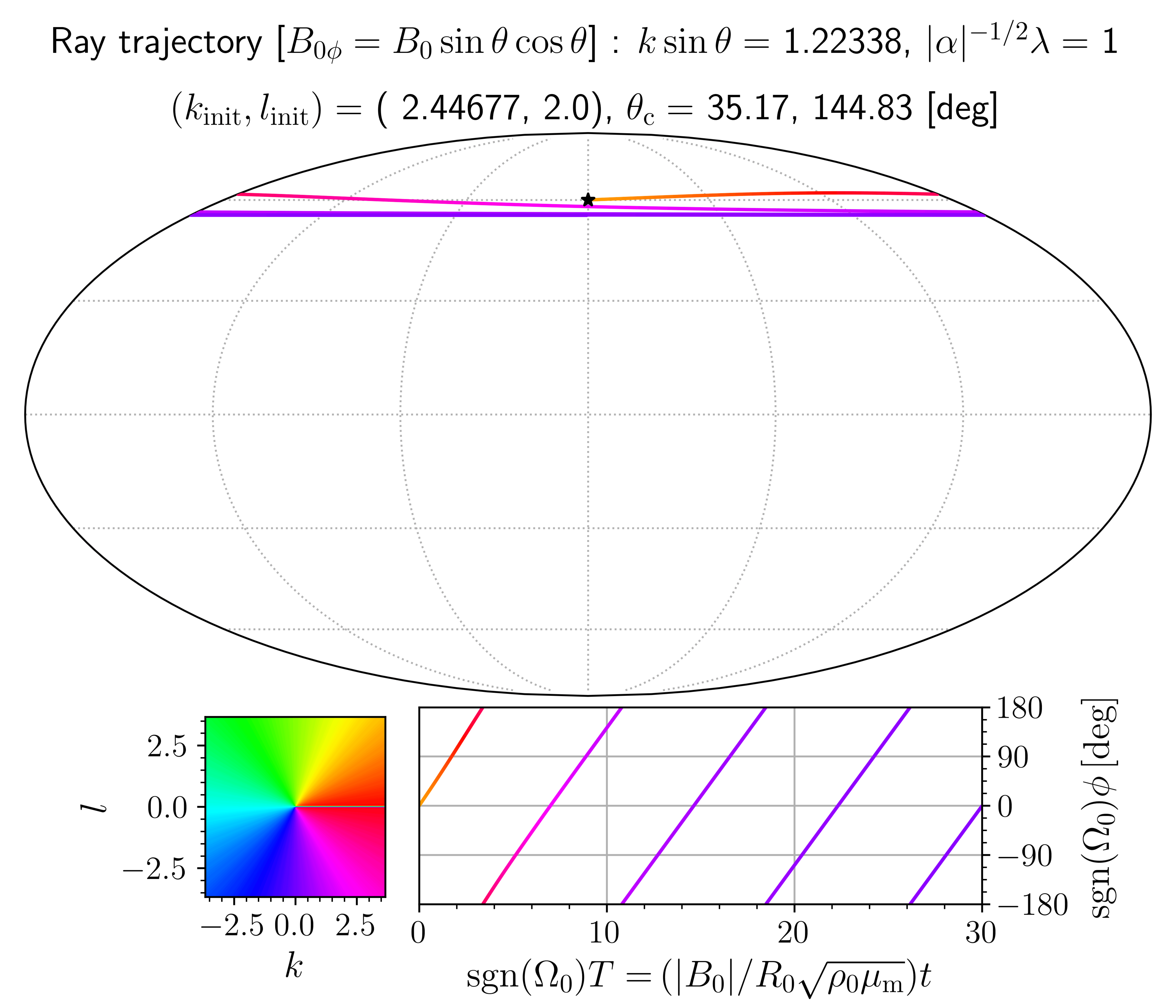}}}%
\end{minipage}
\begin{minipage}{75mm}
	\subfigure[Initial local meridional wavenumber $l_\mathrm{init}=-2$.]{
	\resizebox*{70mm}{!}{\includegraphics{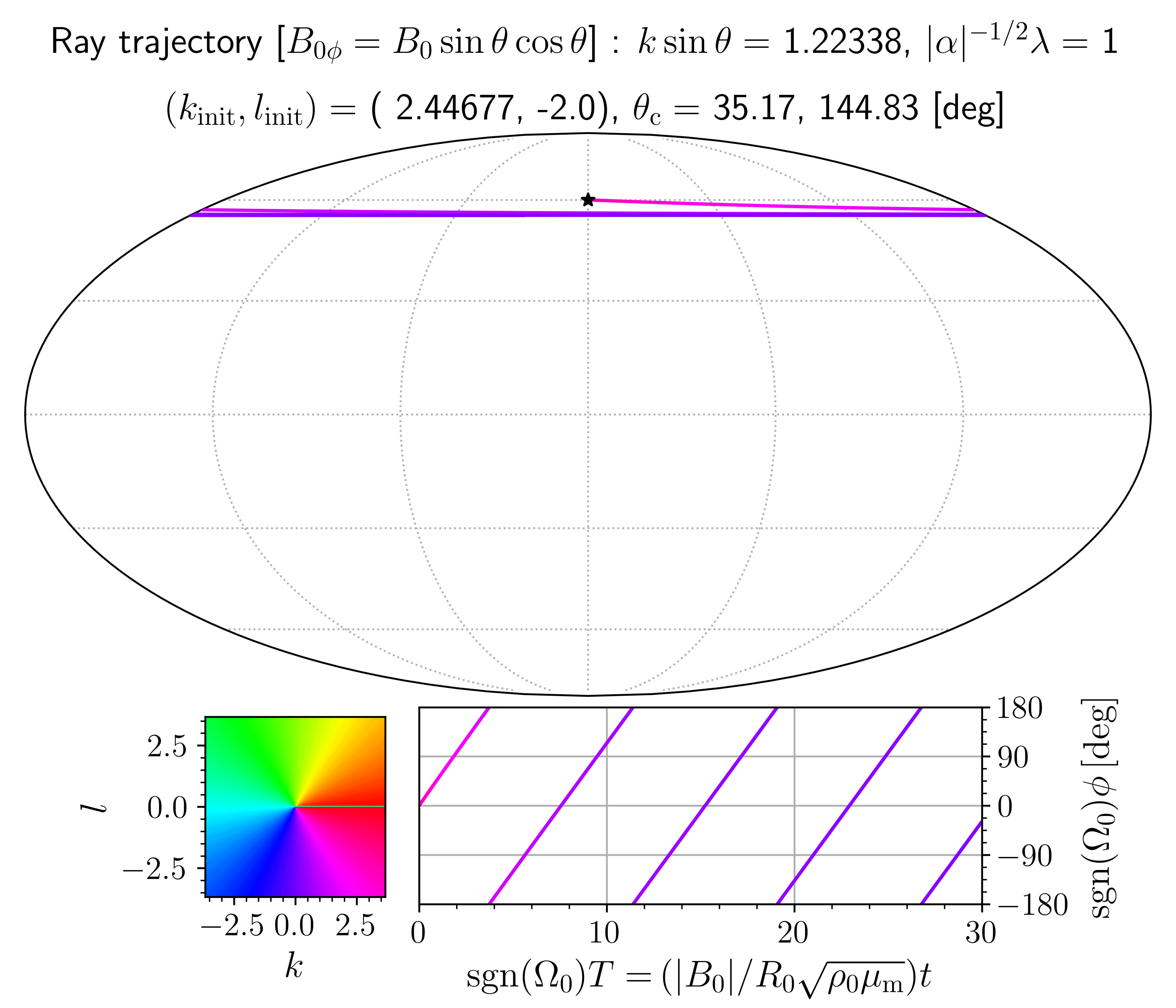}}}%
\end{minipage}
	\caption{Ray trajectories in the Mollweide projection for wave packets belonging to prograde Alfv\'en waves (the scaled zonal wavenumber $k\sin\theta\approx1.22338$, and the critical colatitude $\theta_\mathrm{c}\approx35.17^\circ$ in the northern hemisphere) when the simplest equatorially antisymmetric non-Malkus field $\mathcal{B}=\cos\theta$ pervades the system (upper panels). The scaled dimensionless angular frequency $\lambda_\mathrm{s}=|\alpha|^{-1/2}\lambda=1$. The black asterisks correspond to their initial position $(\theta_\mathrm{init}, \phi_\mathrm{init})=(30^\circ, 0^\circ)$. The directions $(k, l)$ of their local phase velocities are indicated by hue, the colour scales of which are presented in the left lower panels. The lower right panels present the longitudes $\phi$ of their positions against the scaled nondimensional time $T=|\alpha|\tau$ with the colour denoting the directions of their local phase velocities. It should be noted that both $\phi$ and $T$ are multiplied by $\operatorname{sgn}(\varOmega_0)$ such that the ``time'' $\operatorname{sgn}(\varOmega_0)T$ is always positive even when the rotation rate $\varOmega_0$ of the sphere is negative.}
	\label{FIG_ray_sincos_12}
\end{center}
\begin{center}
\begin{minipage}{75mm}
	\subfigure[Initial local meridional wavenumber $l_\mathrm{init}=2$.]{
	\resizebox*{70mm}{!}{\includegraphics{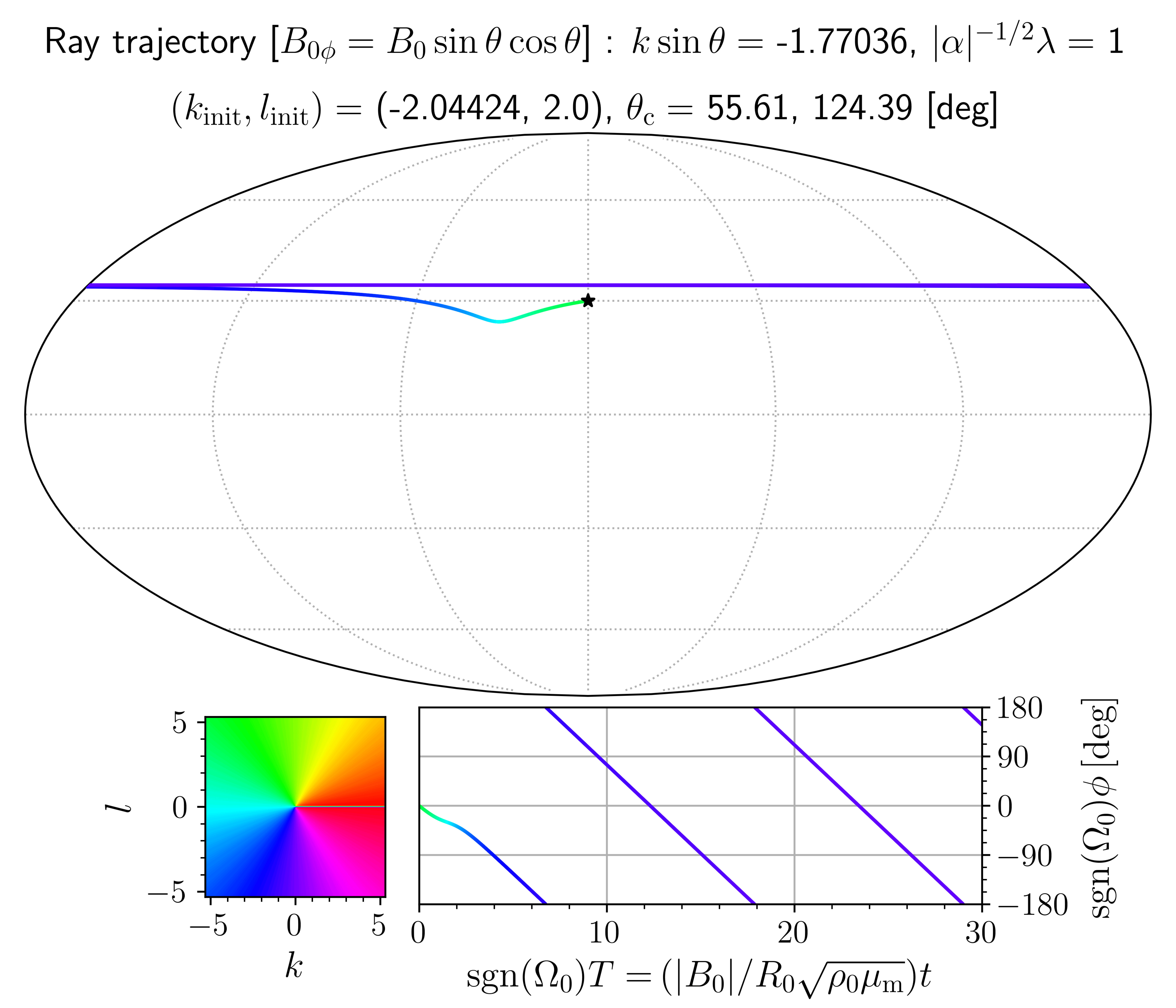}}}%
\end{minipage}
\begin{minipage}{75mm}
	\subfigure[Initial local meridional wavenumber $l_\mathrm{init}=-2$.]{
	\resizebox*{70mm}{!}{\includegraphics{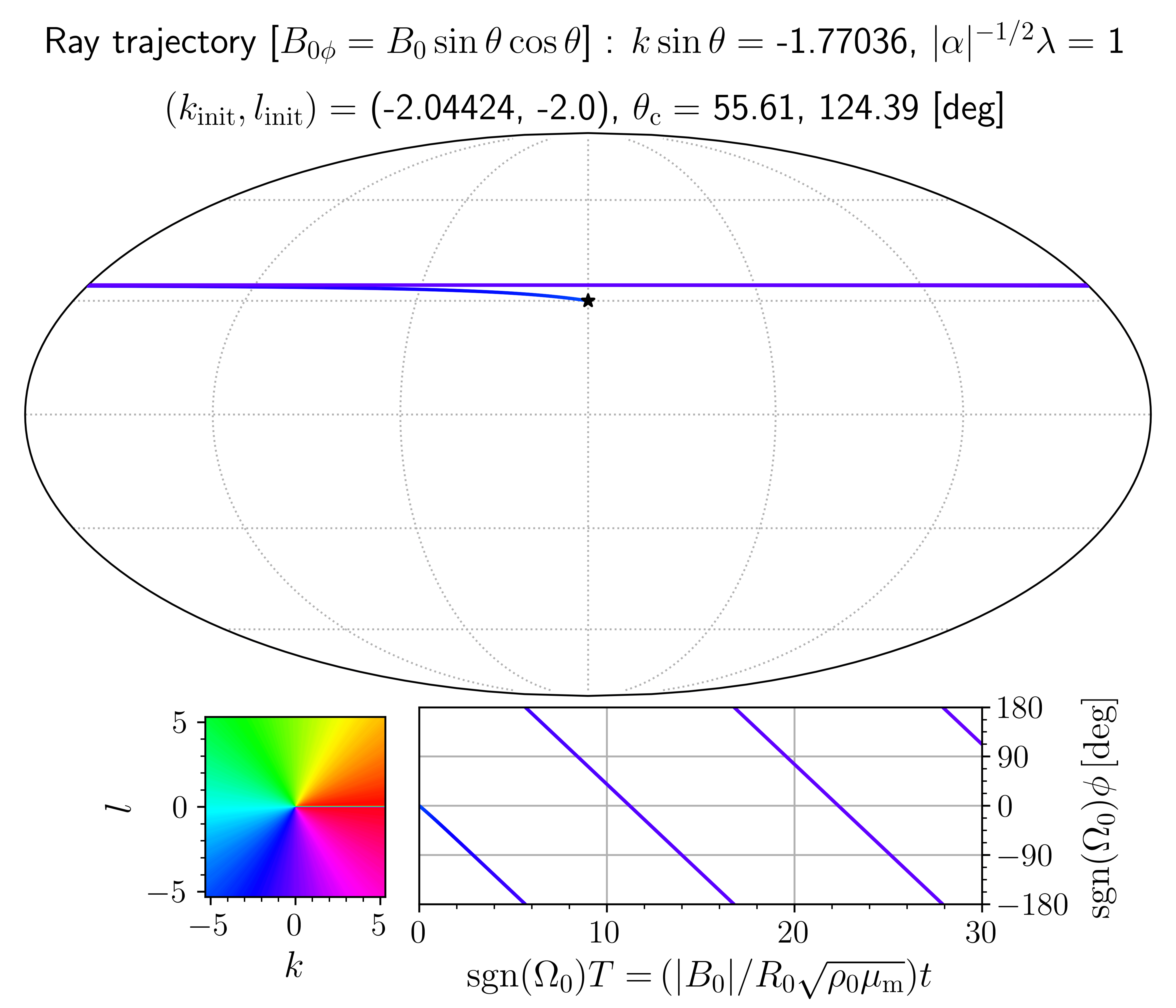}}}%
\end{minipage}
	\caption{Same as Figure \ref{FIG_ray_sincos_12}, but for retrograde Alfv\'en waves (the scaled zonal wavenumber $k\sin\theta\approx-1.77036$, and the critical colatitude $\theta_\mathrm{c}\approx55.61^\circ$ in the northern hemisphere). The initial colatitude $\theta_\mathrm{init}=60^\circ$.}
	\label{FIG_ray_sincos_34}
\end{center}
\end{figure}

%%%%%
We finally demonstrate the ray trajectories obtained from the numerical time integration of the ray-tracing equations, although rough trajectories can be sketched from the above examinations. Figures \ref{FIG_ray_sincos_12} and \ref{FIG_ray_sincos_34} present two of the trajectories for wave trains belonging to prograde and retrograde Alfv\'en waves, respectively. Each of the trains is injected at the black asterisk in each of their upper panels, and its position evolves in accordance with \eqref{EQ_ray_phi_theta}. The colours in the trajectories represent their local wavenumbers, or the directions of their local phase velocities by hue (see their lower left panels for the colour scale). In their lower right panels, the longitudes $\phi$ of their positions at time $T$ are recorded using the same colour scheme as their upper panels. The numerical errors for the results shown in Figures \ref{FIG_ray_sincos_12} and \ref{FIG_ray_sincos_34} are diagnosed through $\mathcal{D}$ in \eqref{DEF_local_dispersion} as $\mathcal{D}\lesssim10^{-3}$ throughout their numerical integrations. The trajectories agree with the above predictions in terms of the refraction at the turning latitudes, the absorption into the critical ones, and the incident directions to those. From the asymptotic expression \eqref{EQ_asymptotic_phi} of the group velocity near a critical latitude, we obtain the period for a wave packet to circle along the latitude line as $\operatorname{sgn}(\varOmega_0)T=|2\upi k\sin\theta/\lambda_\mathrm{s}|$, which approximates the periods read from the lower right panels of Figures \ref{FIG_ray_sincos_12} and \ref{FIG_ray_sincos_34}.
\par
\begin{figure}
\begin{center}
\begin{minipage}{75mm}
	\subfigure[Initial local meridional wavenumber $l_\mathrm{init}=2$.]{
	\resizebox*{70mm}{!}{\includegraphics{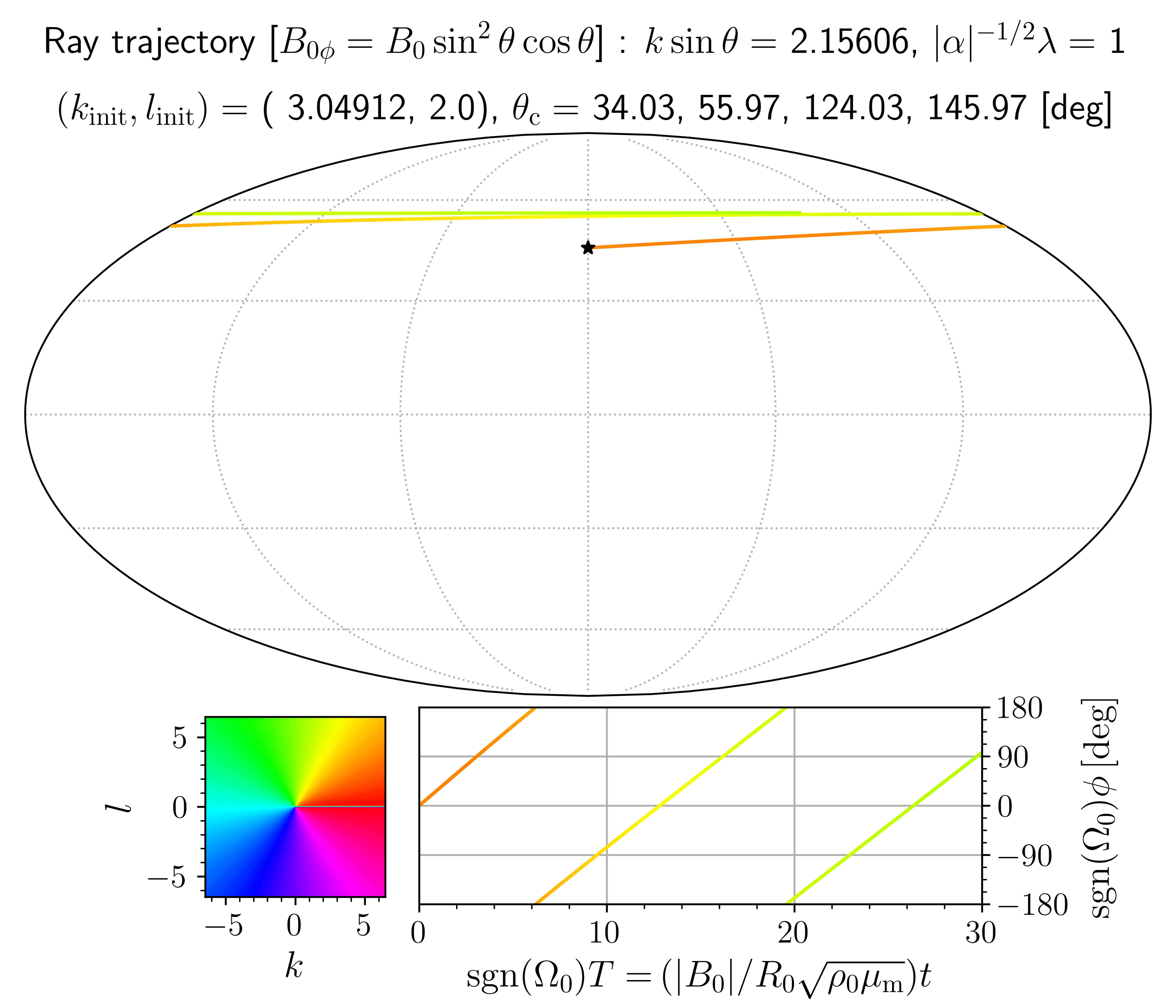}}}%
\end{minipage}
\begin{minipage}{75mm}
	\subfigure[Initial local meridional wavenumber $l_\mathrm{init}=-2$.]{
	\resizebox*{70mm}{!}{\includegraphics{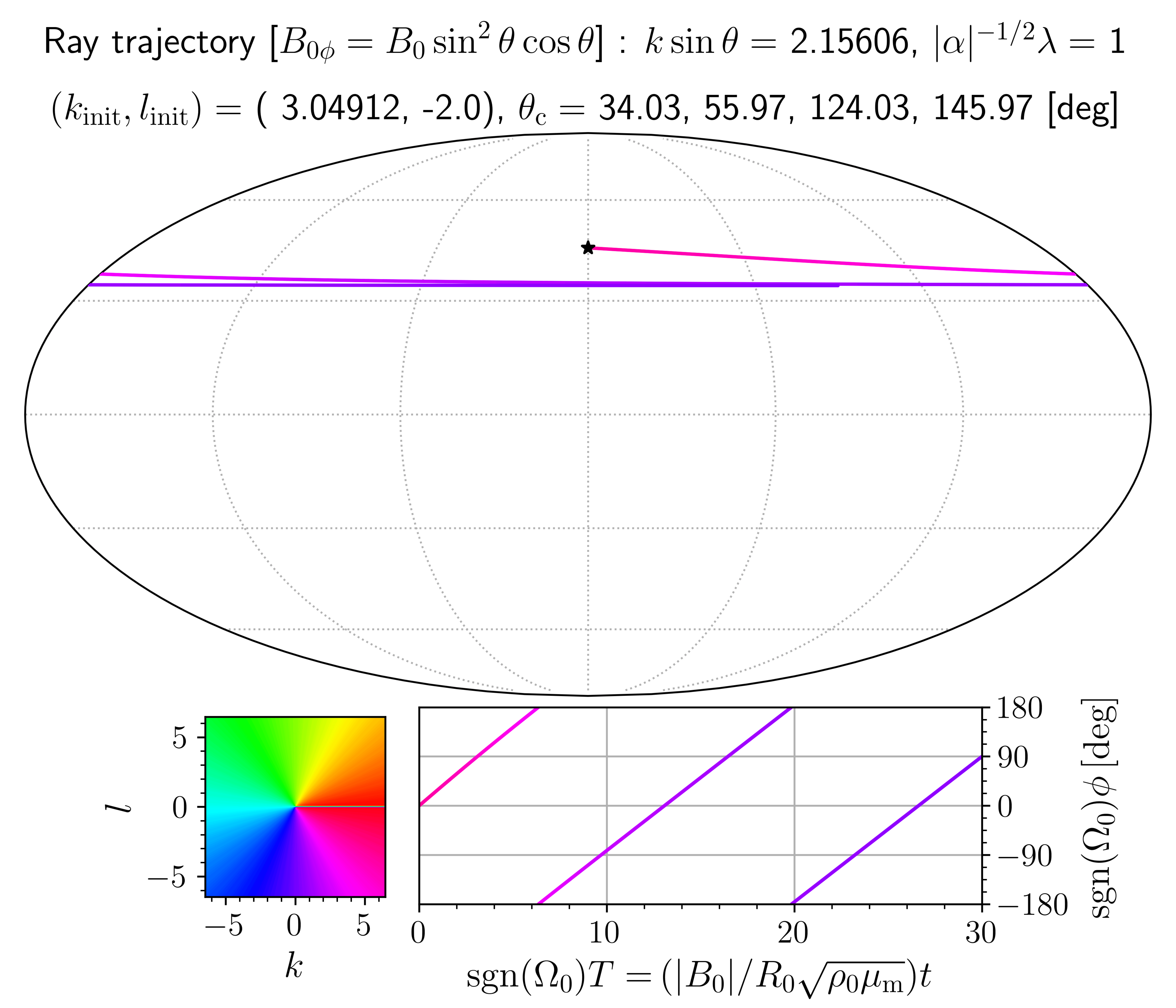}}}%
\end{minipage}
	\caption{Same as Figure \ref{FIG_ray_sincos_12}, but for $\mathcal{B}=\sin\theta\cos\theta$. Ray trajectories for wave packets that belong to prograde Alfv\'en waves (the scaled zonal wavenumber $k\sin\theta\approx2.15606$, and the critical colatitudes $\theta_\mathrm{c}\approx34.03^\circ, 55.97^\circ$ in the northern hemisphere) are presented. The initial colatitude $\theta_\mathrm{init}=45^\circ$.}
	\label{FIG_ray_sin2cos_12}
\end{center}
\begin{center}
\begin{minipage}{75mm}
	\subfigure[Scaled zonal wavenumber $k\sin\theta\approx-2.19789$, and the critical colatitudes $\theta_\mathrm{c}\approx32.75^\circ, 57.25^\circ$ in the northern hemisphere. The initial colatitude $\theta_\mathrm{init}=30^\circ$.]{
	\resizebox*{70mm}{!}{\includegraphics{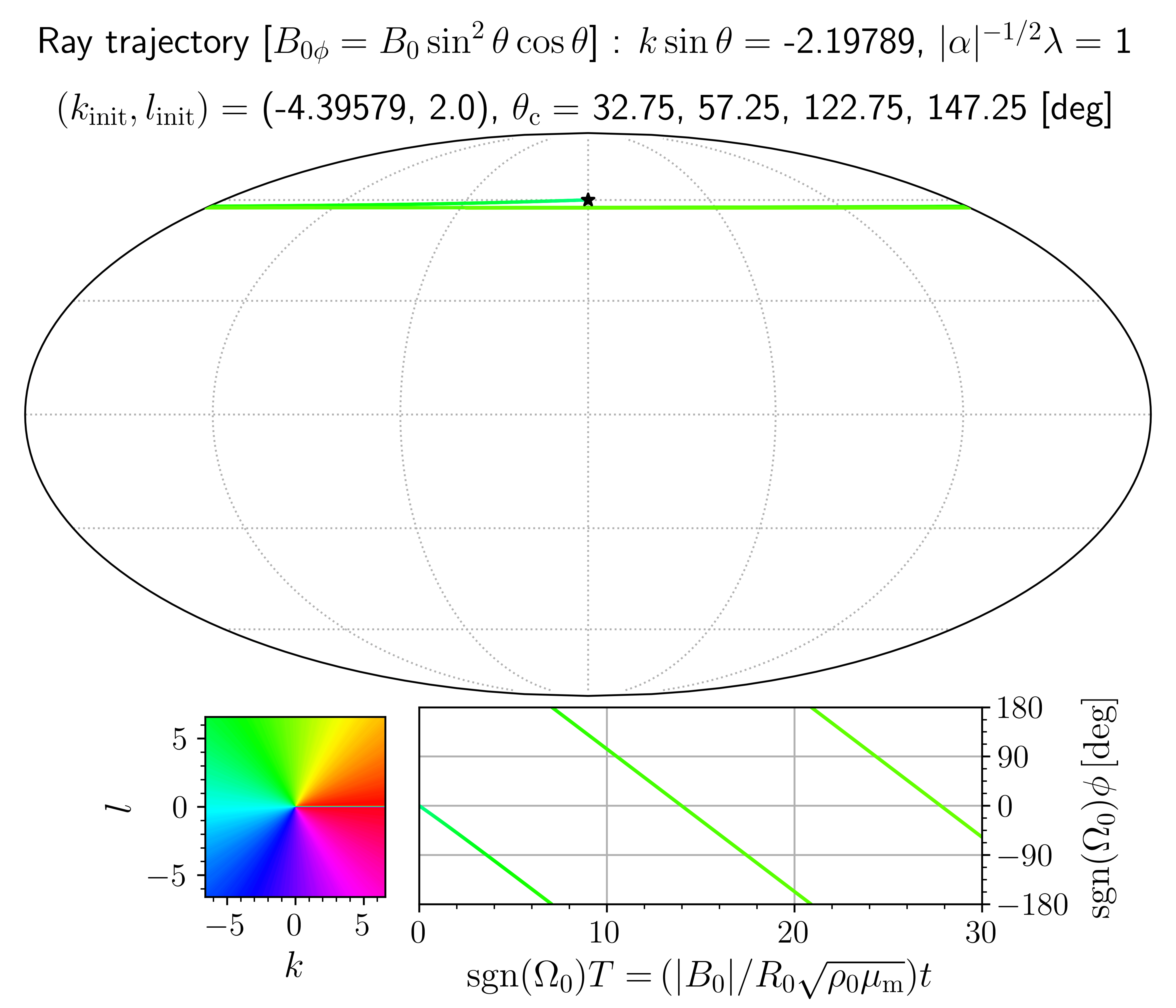}}}%
\end{minipage}
\begin{minipage}{75mm}
	\subfigure[Scaled zonal wavenumber $k\sin\theta\approx-2.04809$, and the critical colatitudes $\theta_\mathrm{c}\approx38.78^\circ, 51.22^\circ$ in the northern hemisphere. The initial colatitude $\theta_\mathrm{init}=60^\circ$.]{
	\resizebox*{70mm}{!}{\includegraphics{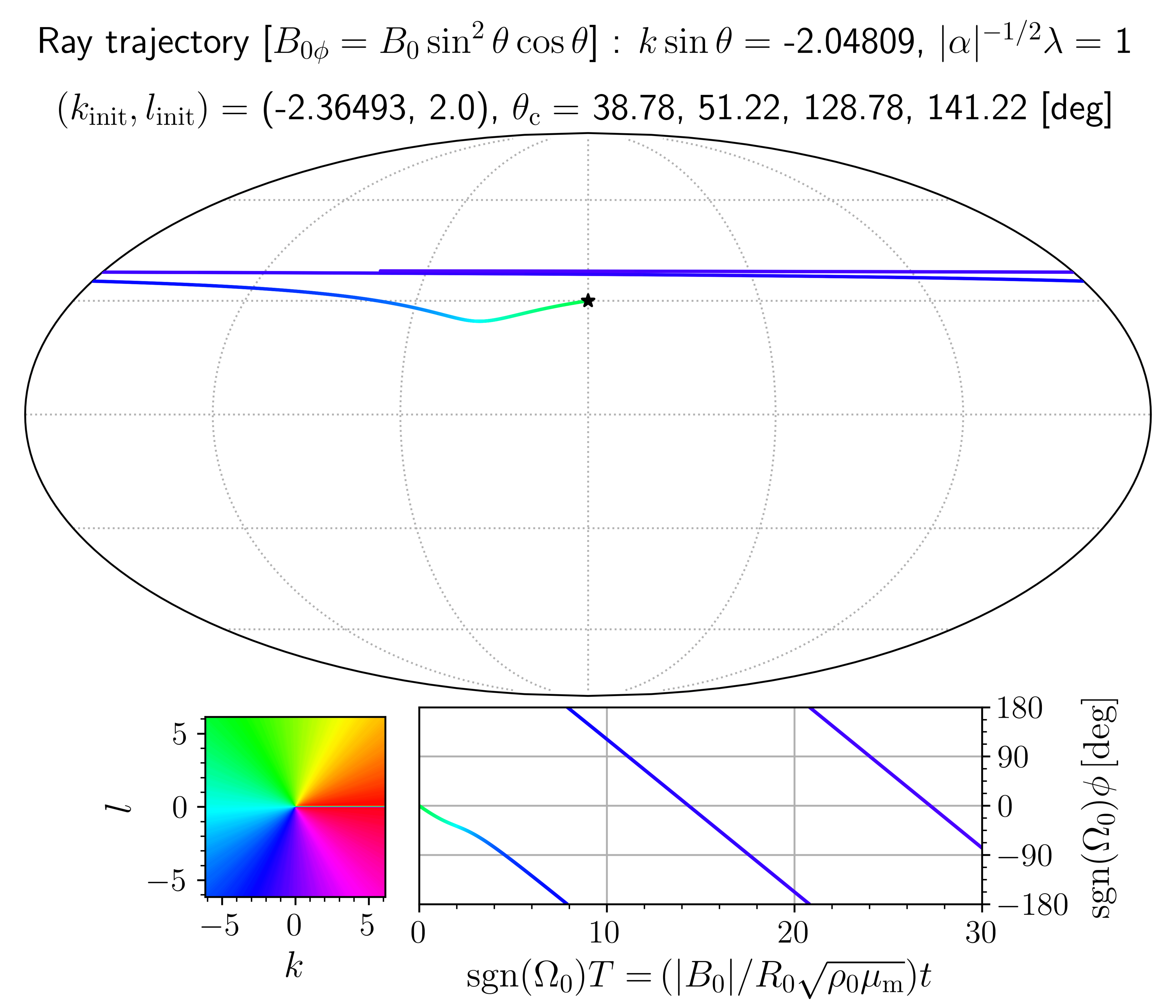}}}%
\end{minipage}
	\caption{Same as Figure \ref{FIG_ray_sin2cos_12}, but for retrograde Alfv\'en waves. The initial local meridional wavenumber $l_\mathrm{init}=2$.}
	\label{FIG_ray_sin2cos_34}
\end{center}
\end{figure}

%%%%%
The facts that the wave packets cannot cross their corresponding critical latitudes and that whether the packets approach there from the polar or equatorial sides depends on the sign of $\lambda_\mathrm{s}/k$ are consistent with the features observed from the numerical results for the continuous modes (Figures \ref{FIG_alleigenfunction_sincos_m1a01} and \ref{FIG_alleigenfunction_sincos_m1a001}) in the eigenvalue problem for global modes when $|\alpha|$ is small. Although the spatial scale of the waves focused on in this section is smaller than that of global modes, which were thematised in Section \ref{SEC_numerical}, this implies that, in the case where $|\alpha|\ll1$, the ray trajectories for Alfv\'en waves facilitate the rough prediction of the behaviours of the continuous modes without actually solving the eigenvalue problem. This approach is applicable even if the imposed field possesses more than two singular latitudes unlike the case where $\mathcal{B}=\cos\theta$. For instance, four of the trajectories for another equatorially antisymmetric field $\mathcal{B}=\sin\theta\cos\theta$ are displayed in Figures \ref{FIG_ray_sin2cos_12} and \ref{FIG_ray_sin2cos_34}. For this profile, two critical latitudes exist in each hemisphere when $0=\min(\mathcal{B}^2)<\lambda_\mathrm{s}^2/k^2\sin^2\theta<\max(\mathcal{B}^2)=1/4$. It should be noted that this field profile does not satisfy the condition for a smooth field at the poles, as mentioned in Section \ref{SEC_mathematical}. However, we have selected this profile for simplicity only for now. The starting points (the black asterisks) of the rays in these figures are selected such that $l^2 > 0$, and the rays must stay in regions where $l^2 > 0$. As discussed in the previous paragraphs, the regions where $l^2 > 0$ are bounded by critical latitudes (and turning latitudes). For prograde waves (Figure \ref{FIG_ray_sin2cos_12}; $\lambda_\mathrm{s}/k > 0$), the starting points should lie between the two critical latitudes in each hemisphere. For retrograde waves (Figure \ref{FIG_ray_sin2cos_34}; $\lambda_\mathrm{s}/k < 0$), the starting points should lie either on the polar side or the equatorial side in each hemisphere. This corresponds to the prograde and retrograde continuous modes being evanescent in the polar and equatorial regions and the mid-latitudes, respectively. The value of the function $\mathcal{L}^2$ of \eqref{DEF_alternative_differential_potential} also provides information regarding the evanescent property for global modes, as with the sign of $l^2$. This function explicitly includes the effects of the gradient of the background field, parts of which have been indirectly ignored in the derivation of the ray-tracing equations (see Appendix \ref{SEC_derivations}). However, its approximate formula \eqref{DEF_alternative_differential_potential_approximate} for a small $|\alpha|$ agrees with \eqref{EQ_local_dispersion_critical}. Figure \ref{FIG_waveeq}(b) depicts the values of $\mathcal{L}^2$ when $\mathcal{B}=\mu\sqrt{1-\mu^2}$ ($=\sin\theta\cos\theta$), and our prediction using the ray theory is consistent with this plot. Based on these facts, we can consider that wave packets belonging to Alfv\'en waves pertain to the continuous modes in Section \ref{SEC_numerical} as expected.\par
\begin{figure}
\begin{center}
	\resizebox*{75mm}{!}{\includegraphics{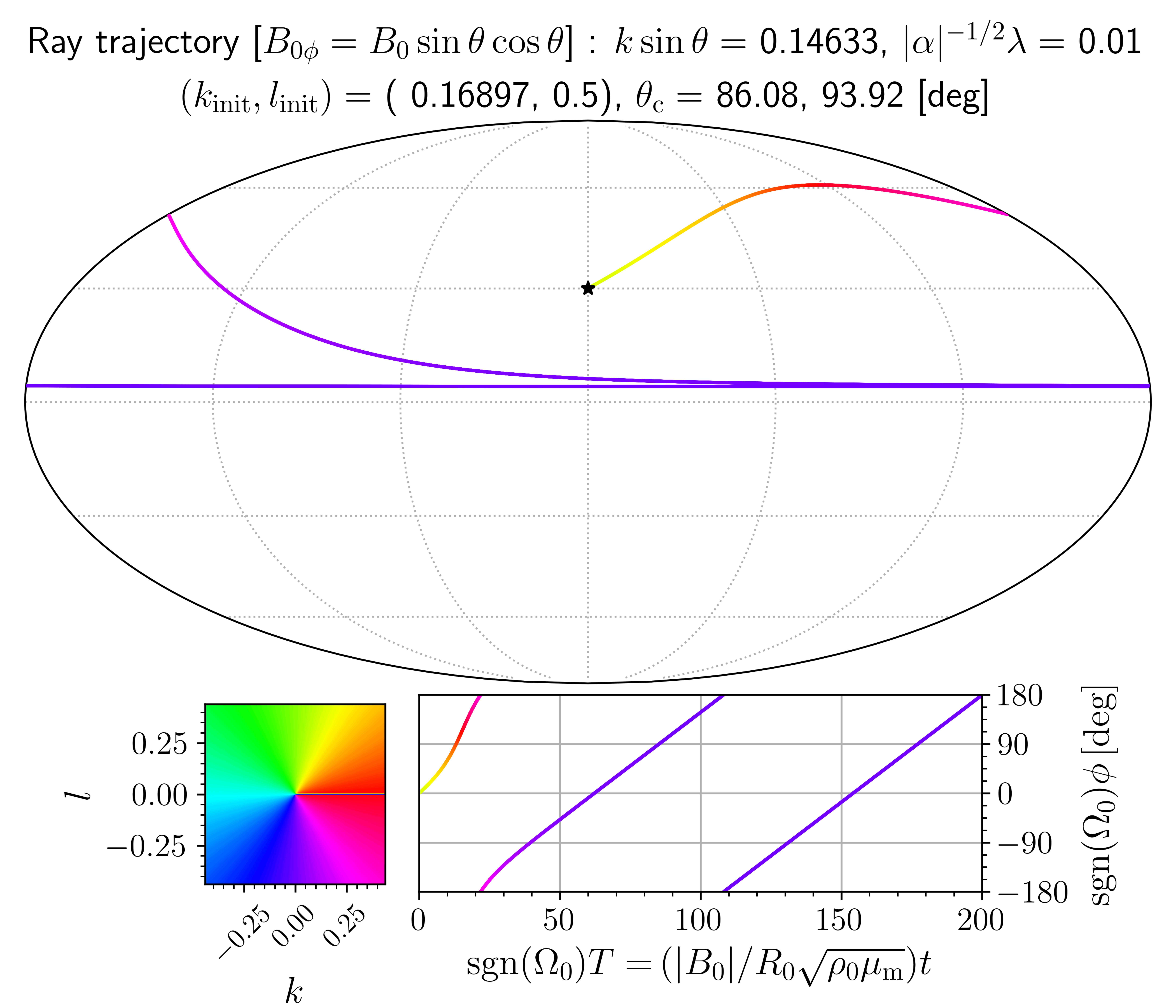}}%
	\caption{Same as Figure \ref{FIG_ray_sincos_12}, but for slow MR waves (the scaled zonal wavenumber $k\sin\theta\approx0.14633$, and the critical colatitude $\theta_\mathrm{c}\approx86.08^\circ$ in the northern hemisphere). The scaled nondimensional angular frequency $\lambda_\mathrm{s}=|\alpha|^{-1/2}\lambda=0.01$, the initial local meridional wavenumber $l_\mathrm{init}=0.5$, and the initial colatitude $\theta_\mathrm{init}=60^\circ$.}
	\label{FIG_ray_sincos_5}
\end{center}
\end{figure}

%%%%%
Moreover, we learn why discrete branches of slow MR waves disappear under non-Malkus fields, as shown in the numerical results in Section \ref{SEC_numerical}, from the ray-tracing approach. Figure \ref{FIG_ray_sincos_5} presents a trajectory for a wave packet that (at least initially) belongs to the slow MR wave, because the condition
\begin{equation}
\lambda_\mathrm{s}^2(k^2+l^2)\,\ll\,\lambda_\mathrm{s}k\sin\theta
\label{EQ_condition_slow_MR}
\end{equation}
has been satisfied by its initial condition. This inequality is obtained from the comparison between the first two terms in the local dispersion relation \eqref{DEF_local_dispersion} by analogy with slow MR waves in the case of the Malkus field. The illustrated trajectory is similar to those of prograde Alfv\'en waves (see Figure \ref{FIG_ray_sincos_12}). Since the propagation properties for slow MR waves and those of prograde Alfv\'en waves are continuous, as shown in Figures \ref{FIG_local_sincos_kl} and \ref{FIG_local_sincos_klambda}, the wave packets for slow MR waves transform into prograde Alfv\'en waves as they migrate in an inhomogeneous background field, changing their dominant local meridional wavenumber. The time evolution of $l$ can reverse the inequality sign of the condition \eqref{EQ_condition_slow_MR} as $l^2\to\infty$, thus the Alfv\'en balance $\lambda_\mathrm{s}^2\approx k^2\mathcal{B}^2\sin^2\theta$ is reached. From \eqref{EQ_l2}, the latitudinal variation of $l^2$ during the migration of a wave packet is written as
\begin{equation}
\left(\frac{\upartial l^2}{\upartial \theta}\right)_{k\sin\theta, \lambda_\mathrm{s}}\,=\,2k^2\cot\theta\,-\,\frac{\lambda_\mathrm{s}k^3\sin^3\theta}{(\lambda_\mathrm{s}^2-k^2\mathcal{B}^2\sin^2\theta)^2}\frac{\ud\mathcal{B}^2}{\ud\theta}\,,
\label{EQ_variation_l2}
\end{equation}
and we find that $(\upartial l^2/\upartial \theta)_{k\sin\theta, \lambda_\mathrm{s}}$ is always positive (negative) in the northern (southern) hemisphere when $\mathcal{B}=\cos\theta$ and $\lambda_\mathrm{s}/k>0$. Therefore, the mode conversion from slow MR into prograde Alfv\'en waves should have occurred between the initial colatitude $\theta_\mathrm{init}$ and the critical colatitude $\theta_\mathrm{c}$ in the example of Figure \ref{FIG_ray_sincos_5}, since $0\leq l^2\leq l_\mathrm{init}^2$ within the interval between $\theta_\mathrm{init}$ and the turning latitude. Although mode conversions can cause the valve effect \citep[e.g.][]{acheson_1972, mckenzie_1973, doi:10.1098/rsta.1977.0106, doi:10.1080/03091927908244549}, the effect does not occur in our system because there exists only one type of critical latitude: the Alfv\'en resonance $\lambda_\mathrm{s}^2=k^2\mathcal{B}^2\sin^2\theta$. On the other hand, wave trains belonging to fast MR waves, which have discrete branches when $\lambda<-m|\alpha|$ for global modes, move back and forth between two turning latitudes without transforming into Alfv\'en waves even though its dominant local wavenumber evolves (see Appendix \ref{SEC_fast}).\par
%%%%%
The invariants including the products of perturbations are useful for understanding waves and their associated phenomena. As derived in Appendix \ref{SEC_derivations}, our approximation leads to a conservation law in the form
\begin{equation}
\frac{\upartial}{\upartial T}\left(\frac{\upartial\mathcal{D}}{\upartial \lambda_\mathrm{s}}|M|^2\right)\,+\,\bm{\nabla}_\mathrm{G}\bm{\cdot}\left(\frac{\bm{c}_\mathrm{g}}{|\alpha|}\frac{\upartial\mathcal{D}}{\upartial \lambda_\mathrm{s}}|M|^2\right)\,=\,0\,.
\label{EQ_invariant}
\end{equation}
It follows from the constancy of $k\sin\theta$ and $\lambda_\mathrm{s}$ that the equation in which $(\upartial\mathcal{D}/\upartial \lambda_\mathrm{s})|M|^2$ is replaced by $(k\sin\theta/\lambda_\mathrm{s}^2)(\upartial\mathcal{D}/\upartial \lambda_\mathrm{s})|M|^2$ is also correct. In our subsequent paper, we intend to prove that the latter quantity is equivalent to the pseudomomentum density (up to constant factor) for the 2D ideal incompressible MHD system. If weak dissipations are introduced, a packet that is related to the Alfv\'en continuous modes would attenuate near its corresponding critical latitude, or within its corresponding thin inner boundary layer (see Section \ref{SEC_introduction}), owing to its long travel time \eqref{EQ_time_critical}. The fact that the packet carries the pseudomomentum in line with \eqref{EQ_invariant} implies that the mean flow would be accelerated there because the damping of waves can cause angular momentum exchange between the waves and a mean flow in accordance with the wave--mean flow interaction theory \citep[e.g.][]{buhler_2009}. This could possibly induce nonlinear oscillations such as the quasi-biennial oscillation (QBO) in the Earth's equatorial stratosphere \citep[e.g.][]{baldwin2001quasi}.\par
%%%%%
At the end of this section, we confirm the validity of the ray theory when $l^2\to\infty$. The approximation requires that the spatial scales at which the dominant local wavenumber $(k,l)$ and the amplitude $M$ of a wave packet vary are sufficiently larger than its wavelength. This condition can be written as
\begin{equation}
\min\left(\left|\frac{1}{l}\frac{\upartial l}{\upartial \theta}\right|^{-1}, \left|\frac{1}{M}\frac{\upartial M}{\upartial \theta}\right|^{-1}\right)\,\gg\,\frac{2\upi}{|\alpha|^{-1/2}|l|}\,.\qquad
\label{EQ_wkbj_validity}
\end{equation}
We now examine whether the condition \eqref{EQ_wkbj_validity} is valid near its corresponding critical latitudes, where $l$ diverges and $|M|$ may also diverge. To obtain an asymptotic expression of $|M|$ near the critical latitude $\theta_\mathrm{c}$, we take advantage of the conservation law \eqref{EQ_invariant} in the form
\begin{equation}
\iint_{S_\mathrm{g}}\left(\frac{\upartial\mathcal{D}}{\upartial \lambda_\mathrm{s}}|M|^2\right)\sin\theta\ud\theta\ud\phi\,=\,\text{const.}\,,
\end{equation}
where $S_\mathrm{g}$ is a region that moves and deforms with the local group velocity. Let $\theta_1(T)$ and $\theta_2(T)$ be the latitudinal positions, at an arbitrary time $T$, of the rear and front of an isolated wave packet, respectively. Since the local group velocity $\bm{c}_\mathrm{g}$ of the packet depends only on the latitudinal position $\theta$, we obtain
\begin{subequations}
\begin{equation}
T\,=\,\int_{\theta_1(0)}^{\theta_1(T)}\frac{|\alpha|}{-c_{\mathrm{g},-\theta}}\ud\theta\,=\,\int_{\theta_2(0)}^{\theta_2(T)}\frac{|\alpha|}{-c_{\mathrm{g},-\theta}}\ud\theta\,.
\end{equation}
Thus, the latitudinal length $|\theta_2-\theta_1|$ of the packet satisfies
\begin{equation}
\int_{\theta_1(T)}^{\theta_2(T)}\frac{|\alpha|}{-c_{\mathrm{g},-\theta}}\ud\theta\,=\,\text{const.}\,.
\end{equation}
\end{subequations}
Using these relations, we can estimate that $(\upartial\mathcal{D}/\upartial \lambda_\mathrm{s})|M|^2(\Delta\phi\sin\theta)\propto c_{\mathrm{g},-\theta}^{-1}$ with $\Delta\phi$ being the longitudinal length of the packet and obtain $M=\mathrm{O}(|\theta-\theta_\mathrm{c}|^{-1/4})$. It should be noted that an identical asymptotic expression is also obtained from the steady problem in the last two paragraphs of Appendix \ref{SEC_derivations}. Since $l=\mathrm{O}(|\theta-\theta_\mathrm{c}|^{-1/2})$ (from \eqref{EQ_local_dispersion_critical}) and $M=\mathrm{O}(|\theta-\theta_\mathrm{c}|^{-1/4})$, we cannot appropriately discuss the behaviour of a wave packet if it approaches within the distance $|\theta-\theta_\mathrm{c}|=\mathrm{O}(|\alpha|)$ from its corresponding critical latitudes. The case where $l=0$ is addressed in Appendix \ref{SEC_behaviors}.

% = = = = = = = = = = = = = = = = = = = = = = = = = = = = = = %
%                                                             %
%                         Section 5                           %
%                                                             %
% = = = = = = = = = = = = = = = = = = = = = = = = = = = = = = %

\section{Conclusion}\label{SEC_conclusion}
%%%%%
In the present paper, we numerically scrutinised 2D ideal incompressible MHD linear waves within a thin layer on a rotating sphere with latitudinally varying toroidal magnetic fields $B_{0\phi}=B_0\mathcal{B}(\cos\theta)\sin\theta$. In Section 3, we solved the eigenvalue problem for the simplest equatorially antisymmetric non-Malkus field $\mathcal{B}=\cos\theta$. The most important finding is that the Alfv\'en and slow MR discrete branches disappear and a continuous spectrum appears instead. The eigenfunctions of the continuous modes and their concomitant critical latitudes were investigated in unprecedented detail and compared with the Frobenius series solutions in Appendix \ref{SUBSEC_comparison}. In Section \ref{SEC_discussion}, we conducted a ray-theoretical study in an inhomogeneous magnetic field to clarify the nature of the continuous modes from a perspective different from the mode theory. Generally, the ray theory is the theory of wave packets with slowly varying frequencies, wavenumbers, and amplitudes. In our problem, this theory has the advantage that it can be used for small absolute values of the Lehnert number $\alpha$, where the mode theory encounters numerical difficulty. Accordingly, our results could act as a stepping stone to a deeper understanding of the dynamics of the outermost Earth's core and the solar tachocline.\par
%%%%%
The advent of the continuous mode can be interpreted in terms of the ray theory as wave packets that move toward a critical latitude and that are ultimately absorbed there. In particular, the wave packet belonging to the slow MR wave turns into the packet that belongs to the Alfv\'en wave before it is absorbed at its corresponding critical latitude. This indicates that the slow MR wave cannot be distinguished from the Alfv\'en wave in the case of a non-Malkus field and corroborates the idea that slow MR waves transform into Alfv\'en continuous modes.\par
%%%%%
We used the function $\mathcal{L}^2$ to show that the observed difference in the evanescent property between the prograde and retrograde continuous modes results from the planetary $\beta$ effect. This function corresponds with the square of the local meridional wavenumber $l^2$ in the ray theory, which provides the same explanation for the evanescent property. Whether the packet approaches the critical latitude from the polar or equatorial sides depends on the sign of the azimuthal component $|\alpha| (\lambda_s/k)$ of its nondimensional local phase velocity, and this packet-approaching side is the oscillatory one ($l^2>0$) of the corresponding mode.\par
%%%%%
We note some interesting topics below that should be investigated in the future studies.
\begin{enumerate}
\item It is a pivotal issue whether or not slow MR waves can remain discrete modes even under a non-Malkus field by recovering additional effects which we have omitted herein, since slow MR waves are considered to be possible causes of geomagnetic fluctuations.
\item An alternative explanation for the disappearance of the discrete slow MR modes could be their transformation into quasi-modes \citep[e.g.][]{doi:10.1063/1.872497, doi:10.1063/1.1289505, balmforth_llewellyn_smith_young_2001, turner_gilbert_2007, wang_gilbert_mason_2022}, or the Landau damping in a broad sense.  Although quasi-modes are not true eigenmodes, they play a crucial role in the time evolution of the system as non-diffusive decaying oscillations, the frequencies of which are close to the original real eigenvalues. To confirm this, we intend to conduct a study on quasi-modes \citep[e.g.][]{doi:10.1063/1.872497} with the method of spectral deformation \citep[e.g.][]{CRAWFORD1989265}.
\item The novel conservation law (39) derived from the ray approximation will provide insights into the interaction between waves and the mean flow and magnetic fields and thereby into the weakly nonlinear evolution of the background fields, although the spatial scales $O(|\alpha|^{1/2})$ of the waves in the theory may be too small to observe as geomagnetic variations.
\end{enumerate}

% = = = = = = = = = = = = = = = = = = = = = = = = = = = = = = %
%                                                             %
%                        References                           %
%                                                             %
% = = = = = = = = = = = = = = = = = = = = = = = = = = = = = = %

\section*{Acknowledgement}
We would like to thank the two anonymous referees for their valuable comments and suggestions, which helped improve our manuscript. We would like also to thank Editage (www.editage.jp) for English language editing. This work was supported in part by JSPS KAKENHI Grant Number JP24K07177 and JP24K00694, and performed with the support and under the auspices of the NIFS Collaboration Research program (NIFS24KIIC001).

\bibliographystyle{plain}
\bibliography{./ms.bib}

\newcommand{\noopsort}[1]{} \newcommand{\printfirst}[2]{#1} \newcommand{\singleletter}[1]{#1} \newcommand{\switchargs}[2]{#2#1}
\begin{thebibliography}{137}
\providecommand{\natexlab}[1]{#1}

\bibitem[\protect\citeauthoryear{Abramowitz and Stegun}{1964}]{abramowitz1964handbook}
Abramowitz, M. and Stegun, I.A., {\itshape Handbook of Mathematical Functions with Formulas, Graphs, and Mathematical Tables},  Vol. 55,  1964 (US Government printing office).

\bibitem[\protect\citeauthoryear{Acheson}{1972}]{acheson_1972}
Acheson, D.J., The critical level for hydromagnetic waves in a rotating fluid. {\itshape J. Fluid Mech.}, 1972, \textbf{53}, 401--415.

\bibitem[\protect\citeauthoryear{Adam}{1986}]{ADAM1986263}
Adam, J.A., Critical layer singularities and complex eigenvalues in some differential equations of mathematical physics. {\itshape Phys. Rep.}, 1986, \textbf{142}, 263--356.

\bibitem[\protect\citeauthoryear{Arlt {\itshape{et~al.}}}{2007{\natexlab{a}}}]{https://doi.org/10.1002/asna.200710882}
Arlt, R., Sule, A. and Filter, R., Stability of the solar tachocline with magnetic fields. {\itshape Astron. Nachr.}, 2007{\natexlab{a}}, \textbf{328}, 1142--1145.

\bibitem[\protect\citeauthoryear{Arlt {\itshape{et~al.}}}{2007{\natexlab{b}}}]{Arlt_2007_Stability}
Arlt, R., Sule, A. and R\"udiger, G., Stability of toroidal magnetic fields in the solar tachocline. {\itshape Astron. Astrophys.}, 2007{\natexlab{b}}, \textbf{461}, 295--301.

\bibitem[\protect\citeauthoryear{Baldwin {\itshape{et~al.}}}{2001}]{baldwin2001quasi}
Baldwin, M.P., Gray, L.J., Dunkerton, T.J., Hamilton, K., Haynes, P.H., Randel, W.J., Holton, J.R., Alexander, M.J., Hirota, I., Horinouchi, T., Jones, D.B.A., Kinnersley, J.S., Marquardt, C., Sato, K. and Takahashi, M., The quasi-biennial oscillation. {\itshape Rev. Geophys.}, 2001, \textbf{39}, 179--229.

\bibitem[\protect\citeauthoryear{Balmforth {\itshape{et~al.}}}{2001}]{balmforth_llewellyn_smith_young_2001}
Balmforth, N.J., Llewellyn~Smith, S.G. and Young, W.R., Disturbing vortices. {\itshape J. Fluid Mech.}, 2001, \textbf{426}, 95--133.

\bibitem[\protect\citeauthoryear{Balmforth and Morrison}{1995{\natexlab{a}}}]{balmforth1995normal}
Balmforth, N.J. and Morrison, P.J., Normal modes and continuous spectra. {\itshape Ann. N. Y. Acad. Sci.}, 1995{\natexlab{a}}, \textbf{773}, 80--94.

\bibitem[\protect\citeauthoryear{Balmforth and Morrison}{1995{\natexlab{b}}}]{Balmforth1995}
Balmforth, N.J. and Morrison, P.J., Singular eigenfunctions for shearing fluids I.  1995{\natexlab{b}}, Technical report, University of Texas at Austin. Institue for Fusion Studies.

\bibitem[\protect\citeauthoryear{Bardsley and Davidson}{2017}]{10.1093/gji/ggx143}
Bardsley, O.P. and Davidson, P.A., {The dispersion of magnetic-Coriolis waves in planetary cores}. {\itshape Geophys. J. Int.}, 2017, \textbf{210}, 18--26.

\bibitem[\protect\citeauthoryear{Barr\'e {\itshape{et~al.}}}{2015}]{BARRE2015723}
Barr\'e, J., Olivetti, A. and Yamaguchi, Y.Y., Landau damping and inhomogeneous reference states. {\itshape C. R. Phys.}, 2015, \textbf{16}, 723--728.

\bibitem[\protect\citeauthoryear{Barston}{1964}]{BARSTON1964282}
Barston, E.M., Electrostatic oscillations in inhomogeneous cold plasmas. {\itshape Ann. Phys.}, 1964, \textbf{29}, 282--303.

\bibitem[\protect\citeauthoryear{Bouffard {\itshape{et~al.}}}{2022}]{10.1093/gji/ggab343}
Bouffard, M., Favier, B., Lecoanet, D. and Le~Bars, M., {Internal gravity waves in a stratified layer atop a convecting liquid core in a non-rotating spherical shell}. {\itshape Geophys. J. Int.}, 2022, \textbf{228}, 337--354.

\bibitem[\protect\citeauthoryear{Boyd}{1981}]{doi:10.1063/1.525100}
Boyd, J.P., Sturm-Liouville eigenproblems with an interior pole. {\itshape J. Math. Phys.}, 1981, \textbf{22}, 1575--1590.

\bibitem[\protect\citeauthoryear{Braginsky}{1970}]{braginsky1970torsional}
Braginsky, S.I., Torsional magnetohydrodynamic vibrations in the Earth's core and variations in day length. {\itshape Geomagn. Aeron.}, 1970, \textbf{10}, 3--12.

\bibitem[\protect\citeauthoryear{Braginsky}{1984}]{doi:10.1080/03091928408210077}
Braginsky, S.I., Short-period geomagnetic secular variation. {\itshape Geophys. Astrophys. Fluid Dyn.}, 1984, \textbf{30}, 1--78.

\bibitem[\protect\citeauthoryear{Braginsky}{1993}]{19931517}
Braginsky, S.I., MAC-oscillations of the hidden ocean of the core. {\itshape J. Geomag. Geoelec.}, 1993, \textbf{45}, 1517--1538.

\bibitem[\protect\citeauthoryear{Braginsky}{1998}]{Braginsky1998}
Braginsky, S.I., Magnetic Rossby waves in the stratified ocean of the core, and topographic core-mantle coupling. {\itshape Earth Planets Space}, 1998, \textbf{50}, 641--649.

\bibitem[\protect\citeauthoryear{Braginsky}{1999}]{BRAGINSKY199921}
Braginsky, S.I., Dynamics of the stably stratified ocean at the top of the core. {\itshape Phys. Earth Planet. Inter.}, 1999, \textbf{111}, 21--34.

\bibitem[\protect\citeauthoryear{Braun}{1975}]{braun1975differential}
Braun, M., {\itshape Differential Equations and Their Applications},  1975 (Springer-Verlag, New York).

\bibitem[\protect\citeauthoryear{Bretherton}{1966}]{https://doi.org/10.1002/qj.49709239403}
Bretherton, F.P., The propagation of groups of internal gravity waves in a shear flow. {\itshape Q. J. R. Meteorol. Soc.}, 1966, \textbf{92}, 466--480.

\bibitem[\protect\citeauthoryear{Briggs {\itshape{et~al.}}}{1970}]{doi:10.1063/1.1692936}
Briggs, R.J., Daugherty, J.D. and Levy, R.H., Role of Landau damping in crossed-field electron beams and inviscid shear flow. {\itshape Phys. Fluids}, 1970, \textbf{13}, 421--432.

\bibitem[\protect\citeauthoryear{Brodholt and Badro}{2017}]{brodholt2017composition}
Brodholt, J. and Badro, J., Composition of the low seismic velocity E' layer at the top of Earth's core. {\itshape Geophys. Res. Lett.}, 2017, \textbf{44}, 8303--8310.

\bibitem[\protect\citeauthoryear{Buffett}{2014}]{Buffett2014}
Buffett, B., Geomagnetic fluctuations reveal stable stratification at the top of the Earth's core. {\itshape Nature}, 2014, \textbf{507}, 484--487.

\bibitem[\protect\citeauthoryear{Buffett and Knezek}{2018}]{10.1093/gji/ggx492}
Buffett, B. and Knezek, N., {Stochastic generation of MAC waves and implications for convection in Earth's core}. {\itshape Geophys. J. Int.}, 2018, \textbf{212}, 1523--1535.

\bibitem[\protect\citeauthoryear{Buffett {\itshape{et~al.}}}{2016}]{10.1093/gji/ggv552}
Buffett, B., Knezek, N. and Holme, R., {Evidence for MAC waves at the top of Earth's core and implications for variations in length of day}. {\itshape Geophys. J. Int.}, 2016, \textbf{204}, 1789--1800.

\bibitem[\protect\citeauthoryear{Buffett and Matsui}{2019}]{10.1093/gji/ggz233}
Buffett, B. and Matsui, H., {Equatorially trapped waves in Earth's core}. {\itshape Geophys. J. Int.}, 2019, \textbf{218}, 1210--1225.

\bibitem[\protect\citeauthoryear{Buffett and Seagle}{2010}]{buffett2010stratification}
Buffett, B.A. and Seagle, C.T., Stratification of the top of the core due to chemical interactions with the mantle. {\itshape J. Geophys. Res. Solid Earth}, 2010, \textbf{115}, B04407.

\bibitem[\protect\citeauthoryear{B\"uhler}{2009}]{buhler_2009}
B\"uhler, O., {\itshape Waves and Mean Flows}, Cambridge Monographs on Mechanics 2009 (Cambridge University Press).

\bibitem[\protect\citeauthoryear{Cally}{2001}]{Cally2001}
Cally, P.S., Nonlinear evolution of 2{D} tachocline instabilities. {\itshape Sol. Phys.}, 2001, \textbf{199}, 231--249.

\bibitem[\protect\citeauthoryear{Cally}{2003}]{10.1046/j.1365-8711.2003.06236.x}
Cally, P.S., {Three-dimensional magneto-shear instabilities in the solar tachocline}. {\itshape Mon. Notices Royal Astron. Soc.}, 2003, \textbf{339}, 957--972.

\bibitem[\protect\citeauthoryear{Cally {\itshape{et~al.}}}{2004}]{cally_dikpati_gilman_2004}
Cally, P.S., Dikpati, M. and Gilman, P.A., The solar tachocline: Limiting magneto-tipping instabilities. {\itshape Symposium - International Astronomical Union}, 2004, \textbf{219}, 541--545.

\bibitem[\protect\citeauthoryear{Cally {\itshape{et~al.}}}{2003}]{Cally_2003}
Cally, P.S., Dikpati, M. and Gilman, P.A., Clamshell and tipping instabilities in a two-dimensional magnetohydrodynamic tachocline. {\itshape Astrophys. J.}, 2003, \textbf{582}, 1190--1205.

\bibitem[\protect\citeauthoryear{Cally {\itshape{et~al.}}}{2008}]{10.1111/j.1365-2966.2008.13934.x}
Cally, P.S., Dikpati, M. and Gilman, P.A., {Three-dimensional magneto-shear instabilities in the solar tachocline - II. Axisymmetric case}. {\itshape Mon. Notices Royal Astron. Soc.}, 2008, \textbf{391}, 891--900.

\bibitem[\protect\citeauthoryear{Carpenter and Guha}{2019}]{doi:10.1063/1.5116633}
Carpenter, J.R. and Guha, A., Instability of a smooth shear layer through wave interactions. {\itshape Phys. Fluids}, 2019, \textbf{31}, 081701.

\bibitem[\protect\citeauthoryear{Case}{1959}]{CASE1959349}
Case, K.M., Plasma oscillations. {\itshape Ann. Phys.}, 1959, \textbf{7}, 349--364.

\bibitem[\protect\citeauthoryear{Case}{1960}]{doi:10.1063/1.1706010}
Case, K.M., Stability of inviscid plane Couette flow. {\itshape Phys. Fluids}, 1960, \textbf{3}, 143--148.

\bibitem[\protect\citeauthoryear{Chi-Dur\'an {\itshape{et~al.}}}{2021}]{https://doi.org/10.1029/2021GL094692}
Chi-Dur\'an, R., Avery, M.S. and Buffett, B.A., Signatures of high-latitude waves in observations of geomagnetic acceleration. {\itshape Geophys. Res. Lett.}, 2021, \textbf{48}, e2021GL094692.

\bibitem[\protect\citeauthoryear{Chulliat {\itshape{et~al.}}}{2015}]{https://doi.org/10.1002/2015GL064067}
Chulliat, A., Alken, P. and Maus, S., Fast equatorial waves propagating at the top of the Earth's core. {\itshape Geophys. Res. Lett.}, 2015, \textbf{42}, 3321--3329.

\bibitem[\protect\citeauthoryear{Couston {\itshape{et~al.}}}{2017}]{PhysRevFluids.2.094804}
Couston, L.A., Lecoanet, D., Favier, B. and Le~Bars, M., Dynamics of mixed convective--stably-stratified fluids. {\itshape Phys. Rev. Fluids}, 2017, \textbf{2}, 094804.

\bibitem[\protect\citeauthoryear{Crawford and Hislop}{1989}]{CRAWFORD1989265}
Crawford, J.D. and Hislop, P.D., Application of the method of spectral deformation to the Vlasov-poisson system. {\itshape Ann. Phys.}, 1989, \textbf{189}, 265--317.

\bibitem[\protect\citeauthoryear{Davies {\itshape{et~al.}}}{2018}]{davies2018partitioning}
Davies, C.J., Pozzo, M., Gubbins, D. and Alf{\`e}, D., Partitioning of oxygen between ferropericlase and Earth's liquid core. {\itshape Geophys. Res. Lett.}, 2018, \textbf{45}, 6042--6050.

\bibitem[\protect\citeauthoryear{Dikpati {\itshape{et~al.}}}{2018{\natexlab{a}}}]{Dikpati_2018_Phase}
Dikpati, M., Belucz, B., Gilman, P.A. and McIntosh, S.W., Phase speed of magnetized Rossby waves that cause solar seasons. {\itshape Astrophys. J.}, 2018{\natexlab{a}}, \textbf{862}, 159.

\bibitem[\protect\citeauthoryear{Dikpati {\itshape{et~al.}}}{2004}]{Dikpati_2004}
Dikpati, M., Cally, P.S. and Gilman, P.A., Linear analysis and nonlinear evolution of two-dimensional global magnetohydrodynamic instabilities in a diffusive tachocline. {\itshape Astrophys. J.}, 2004, \textbf{610}, 597--615.

\bibitem[\protect\citeauthoryear{Dikpati {\itshape{et~al.}}}{2017}]{Dikpati2017}
Dikpati, M., Cally, P.S., McIntosh, S.W. and Heifetz, E., The origin of the ``seasons'' in space weather. {\itshape Sci. Rep.}, 2017, \textbf{7}, 14750.

\bibitem[\protect\citeauthoryear{Dikpati and Gilman}{1999}]{Dikpati_1999}
Dikpati, M. and Gilman, P.A., Joint instability of latitudinal differential rotation and concentrated toroidal fields below the solar convection zone. {\itshape Astrophys. J.}, 1999, \textbf{512}, 417--441.

\bibitem[\protect\citeauthoryear{Dikpati {\itshape{et~al.}}}{2009}]{Dikpati_2009}
Dikpati, M., Gilman, P.A., Cally, P.S. and Miesch, M.S., Axisymmetric MHD instabilities in solar/stellar tachoclines. {\itshape Astrophys. J.}, 2009, \textbf{692}, 1421--1431.

\bibitem[\protect\citeauthoryear{Dikpati {\itshape{et~al.}}}{2020}]{Dikpati_2020}
Dikpati, M., Gilman, P.A., Chatterjee, S., McIntosh, S.W. and Zaqarashvili, T.V., Physics of Magnetohydrodynamic Rossby Waves in the Sun. {\itshape Astrophys. J.}, 2020, \textbf{896}, 141.

\bibitem[\protect\citeauthoryear{Dikpati {\itshape{et~al.}}}{2003}]{Dikpati_2003}
Dikpati, M., Gilman, P.A. and Rempel, M., Stability analysis of tachocline latitudinal differential rotation and coexisting toroidal band using a shallow-water model. {\itshape Astrophys. J.}, 2003, \textbf{596}, 680--697.

\bibitem[\protect\citeauthoryear{Dikpati {\itshape{et~al.}}}{2018{\natexlab{b}}}]{Dikpati_2018}
Dikpati, M., McIntosh, S.W., Bothun, G., Cally, P.S., Ghosh, S.S., Gilman, P.A. and Umurhan, O.M., Role of interaction between magnetic Rossby waves and tachocline differential rotation in producing solar seasons. {\itshape Astrophys. J.}, 2018{\natexlab{b}}, \textbf{853}, 144.

\bibitem[\protect\citeauthoryear{Drazin and Reid}{1981}]{drazin_reid_1981}
Drazin, P.G. and Reid, W.H., {\itshape Hydrodynamic Stability}, Cambridge Mathematical Library 1981 (Cambridge University Press).

\bibitem[\protect\citeauthoryear{Eltayeb}{1977}]{doi:10.1098/rsta.1977.0106}
Eltayeb, I.A., On linear wave motions in magnetic-velocity shears. {\itshape Philos. Trans. R. Soc. Lond. A Math. Phys. Sci.}, 1977, \textbf{285}, 607--636.

\bibitem[\protect\citeauthoryear{Eltayeb and Mckenzie}{1977}]{eltayeb_mckenzie_1977}
Eltayeb, I.A. and Mckenzie, J.F., Propagation of hydromagnetic planetary waves on a beta-plane through magnetic and velocity shear. {\itshape J. Fluid Mech.}, 1977, \textbf{81}, 1--23.

\bibitem[\protect\citeauthoryear{Farrell}{1982}]{TheInitialGrowthofDisturbancesinaBaroclinicFlow}
Farrell, B.F., The initial growth of disturbances in a baroclinic flow. {\itshape J. Atmos. Sci.}, 1982, \textbf{39}, 1663 -- 1686.

\bibitem[\protect\citeauthoryear{Finlay}{2008}]{FINLAY2008403}
Finlay, C.C., Course 8 Waves in the presence of magnetic fields, rotation and convection. In {\itshape Dynamos}, Les Houches, edited by P.~Cardin and L.~Cugliandolo, Vol. ~88 of {\itshape Les Houches}, pp. 403--450, 2008 (Elsevier: Amsterdam).

\bibitem[\protect\citeauthoryear{Gachechiladze {\itshape{et~al.}}}{2019}]{Gachechiladze_2019}
Gachechiladze, T., Zaqarashvili, T.V., Gurgenashvili, E., Ramishvili, G., Carbonell, M., Oliver, R. and Ballester, J.L., Magneto-Rossby waves in the solar tachocline and the annual variations in solar activity. {\itshape Astrophys. J.}, 2019, \textbf{874}, 162.

\bibitem[\protect\citeauthoryear{Gastine {\itshape{et~al.}}}{2020}]{10.1093/gji/ggaa250}
Gastine, T., Aubert, J. and Fournier, A., {Dynamo-based limit to the extent of a stable layer atop Earth's core}. {\itshape Geophys. J. Int.}, 2020, \textbf{222}, 1433--1448.

\bibitem[\protect\citeauthoryear{Gerick {\itshape{et~al.}}}{2021}]{https://doi.org/10.1029/2020GL090803}
Gerick, F., Jault, D. and Noir, J., Fast quasi-geostrophic Magneto-Coriolis modes in the Earth's core. {\itshape Geophys. Res. Lett.}, 2021, \textbf{48}, e2020GL090803.

\bibitem[\protect\citeauthoryear{Gillet {\itshape{et~al.}}}{2022}]{Gillet2022}
Gillet, N., Gerick, F., Angappan, R. and Jault, D., A dynamical prospective on interannual geomagnetic field changes. {\itshape Surv. Geophys.}, 2022, \textbf{43}, 71--105.

\bibitem[\protect\citeauthoryear{Gillet {\itshape{et~al.}}}{2010}]{Gillet2010}
Gillet, N., Jault, D., Canet, E. and Fournier, A., Fast torsional waves and strong magnetic field within the Earth's core. {\itshape Nature}, 2010, \textbf{465}, 74--77.

\bibitem[\protect\citeauthoryear{Gilman}{2000}]{Gilman_2000_Magnetohydrodynamic}
Gilman, P.A., Magnetohydrodynamic {\textquotedblleft}shallow water{\textquotedblright} equations for the solar tachocline. {\itshape Astrophys. J.}, 2000, \textbf{544}, L79--L82.

\bibitem[\protect\citeauthoryear{Gilman and Dikpati}{2000}]{Gilman_2000}
Gilman, P.A. and Dikpati, M., Joint instability of latitudinal differential rotation and concentrated toroidal fields below the solar convection zone. {II}. Instability of narrow bands at all latitudes. {\itshape Astrophys. J.}, 2000, \textbf{528}, 552--572.

\bibitem[\protect\citeauthoryear{Gilman and Dikpati}{2002}]{Gilman_2002}
Gilman, P.A. and Dikpati, M., Analysis of instability of latitudinal differential rotation and toroidal field in the solar tachocline using a magnetohydrodynamic shallow-water model. I. Instability for broad toroidal field profile. {\itshape Astrophys. J.}, 2002, \textbf{576}, 1031--1047.

\bibitem[\protect\citeauthoryear{Gilman {\itshape{et~al.}}}{2007}]{Gilman_2007}
Gilman, P.A., Dikpati, M. and Miesch, M.S., Global {MHD} instabilities in a three-dimensional thin-shell model of solar tachocline. {\itshape Astrophys. J. Suppl. Ser.}, 2007, \textbf{170}, 203--227.

\bibitem[\protect\citeauthoryear{Gilman and Fox}{1997}]{Gilman_1997}
Gilman, P.A. and Fox, P.A., Joint instability of latitudinal differential rotation and toroidal magnetic fields below the solar convection zone. {\itshape Astrophys. J.}, 1997, \textbf{484}, 439--454.

\bibitem[\protect\citeauthoryear{Gilman and Fox}{1999{\natexlab{a}}}]{Fox_1999}
Gilman, P.A. and Fox, P.A., Joint instability of latitudinal differential rotation and toroidal magnetic fields below the solar convection zone. {II}. Instability for toroidal fields that have a node between the equator and pole. {\itshape Astrophys. J.}, 1999{\natexlab{a}}, \textbf{510}, 1018--1044.

\bibitem[\protect\citeauthoryear{Gilman and Fox}{1999{\natexlab{b}}}]{Gilman_1999}
Gilman, P.A. and Fox, P.A., Joint instability of latitudinal differential rotation and toroidal magnetic fields below the solar convection zone. {III}. Unstable disturbance phenomenology and the solar cycle. {\itshape Astrophys. J.}, 1999{\natexlab{b}}, \textbf{522}, 1167--1189.

\bibitem[\protect\citeauthoryear{Gizon {\itshape{et~al.}}}{2020}]{Gizon_2020}
Gizon, L., Fournier, D. and Albekioni, M., Effect of latitudinal differential rotation on solar Rossby waves: Critical layers, eigenfunctions, and momentum fluxes in the equatorial $\beta$ plane. {\itshape Astron. Astrophys.}, 2020, \textbf{642}, A178.

\bibitem[\protect\citeauthoryear{Goedbloed and Poedts}{2004}]{goedbloed_poedts_2004}
Goedbloed, J.P.H. and Poedts, S., {\itshape Principles of Magnetohydrodynamics: With Applications to Laboratory and Astrophysical Plasmas},  2004 (Cambridge University Press).

\bibitem[\protect\citeauthoryear{Grimshaw}{1979}]{doi:10.1080/03091927908244549}
Grimshaw, R., A general theory of critical level absorption and valve effects for linear wave propagation. {\itshape Geophys. Astrophys. Fluid Dyn.}, 1979, \textbf{14}, 303--326.

\bibitem[\protect\citeauthoryear{Gubbins and Davies}{2013}]{GUBBINS201321}
Gubbins, D. and Davies, C.J., The stratified layer at the core-mantle boundary caused by barodiffusion of oxygen, sulphur and silicon. {\itshape Phys. Earth Planet. Inter.}, 2013, \textbf{215}, 21--28.

\bibitem[\protect\citeauthoryear{Gubbins {\itshape{et~al.}}}{1982}]{10.1111/j.1365-246X.1982.tb06972.x}
Gubbins, D., Thomson, C.J. and Whaler, K.A., {Stable regions in the Earth's liquid core}. {\itshape Geophys. J. Int.}, 1982, \textbf{68}, 241--251.

\bibitem[\protect\citeauthoryear{Hardy {\itshape{et~al.}}}{2020}]{10.1093/gji/ggaa260}
Hardy, C.M., Livermore, P.W. and Niesen, J., {Enhanced magnetic fields within a stratified layer}. {\itshape Geophys. J. Int.}, 2020, \textbf{222}, 1686--1703.

\bibitem[\protect\citeauthoryear{Heifetz {\itshape{et~al.}}}{2020}]{doi:10.1063/5.0011351}
Heifetz, E., Guha, A. and Carpenter, J.R., Wave interactions in neutrally stable shear layers: Regular and singular modes, and non-modal growth. {\itshape Phys. Fluids}, 2020, \textbf{32}, 074106.

\bibitem[\protect\citeauthoryear{Helffrich and Kaneshima}{2010}]{Helffrich2010}
Helffrich, G. and Kaneshima, S., Outer-core compositional stratification from observed core wave speed profiles. {\itshape Nature}, 2010, \textbf{468}, 807--810.

\bibitem[\protect\citeauthoryear{Heng and Spitkovsky}{2009}]{Heng_2009}
Heng, K. and Spitkovsky, A., Magnetohydrodynamic shallow water waves: Linear analysis. {\itshape Astrophys. J.}, 2009, \textbf{703}, 1819--1831.

\bibitem[\protect\citeauthoryear{Hide}{1966}]{doi:10.1098/rsta.1966.0026}
Hide, R., Free hydromagnetic oscillations of the Earth's core and the theory of the geomagnetic secular variation. {\itshape Philos. Trans. R. Soc. Lond. A Math. Phys. Sci.}, 1966, \textbf{259}, 615--647.

\bibitem[\protect\citeauthoryear{Hollerbach and Cally}{2009}]{Hollerbach2009}
Hollerbach, R. and Cally, P.S., Nonlinear evolution of axisymmetric twisted flux tubes in the solar tachocline. {\itshape Sol. Phys.}, 2009, \textbf{260}, 251--260.

\bibitem[\protect\citeauthoryear{Hori {\itshape{et~al.}}}{2023}]{Hori2022}
Hori, K., Nilsson, A. and Tobias, S.M., Waves in planetary dynamos. {\itshape Rev. Mod. Plasma Phys.}, 2023, \textbf{7}, 5.

\bibitem[\protect\citeauthoryear{Hughes and Tobias}{2001}]{doi:10.1098/rspa.2000.0725}
Hughes, D.W. and Tobias, S.M., On the instability of magnetohydrodynamic shear flows. {\itshape Proc. R. Soc. Lond. A: Math. Phys. Eng. Sci.}, 2001, \textbf{457}, 1365--1384.

\bibitem[\protect\citeauthoryear{Iga}{1999}]{Iga_1999}
Iga, K., Critical layer instability as a resonance between a non-singular mode and continuous modes. {\itshape Fluid Dyn. Res.}, 1999, \textbf{25}, 63--86.

\bibitem[\protect\citeauthoryear{Iga}{2013}]{iga_2013}
Iga, K., Shear instability as a resonance between neutral waves hidden in a shear flow. {\itshape J. Fluid Mech.}, 2013, \textbf{715}, 452--476.

\bibitem[\protect\citeauthoryear{Irving {\itshape{et~al.}}}{2018}]{doi:10.1126/sciadv.aar2538}
Irving, J.C.E., Cottaar, S. and Leki\'c, V., Seismically determined elastic parameters for Earth's outer core. {\itshape Sci. Adv.}, 2018, \textbf{4}, eaar2538.

\bibitem[\protect\citeauthoryear{Jaupart and Buffett}{2017}]{10.1093/gji/ggx088}
Jaupart, E. and Buffett, B., {Generation of MAC waves by convection in Earth's core}. {\itshape Geophys. J. Int.}, 2017, \textbf{209}, 1326--1336.

\bibitem[\protect\citeauthoryear{Kaneshima}{2018}]{KANESHIMA2018234}
Kaneshima, S., Array analyses of SmKS waves and the stratification of Earth's outermost core. {\itshape Phys. Earth Planet. Inter.}, 2018, \textbf{276}, 234--246.

\bibitem[\protect\citeauthoryear{Kitchatinov and R\"udiger}{2008}]{Kitchatinov_2008}
Kitchatinov, L.L. and R\"udiger, G., Stability of toroidal magnetic fields in rotating stellar radiation zones. {\itshape Astron. Astrophys.}, 2008, \textbf{478}, 1--8.

\bibitem[\protect\citeauthoryear{Knezek and Buffett}{2018}]{KNEZEK20181}
Knezek, N. and Buffett, B., Influence of magnetic field configuration on magnetohydrodynamic waves in Earth's core. {\itshape Phys. Earth Planet. Inter.}, 2018, \textbf{277}, 1--9.

\bibitem[\protect\citeauthoryear{Landau}{1946}]{landau}
Landau, L., On the vibration of the electronic plasma. {\itshape J. Phys. U.S.S.R.}, 1946, \textbf{10}, 25.

\bibitem[\protect\citeauthoryear{Landeau {\itshape{et~al.}}}{2016}]{Landeau2016}
Landeau, M., Olson, P., Deguen, R. and Hirsh, B.H., Core merging and stratification following giant impact. {\itshape Nat. Geosci.}, 2016, \textbf{9}, 786--789.

\bibitem[\protect\citeauthoryear{Lighthill}{1978}]{lighthill1978waves}
Lighthill, J., {\itshape Waves in Fluids},  1978 (Cambridge University Press).

\bibitem[\protect\citeauthoryear{Longuet-Higgins}{1968}]{doi:10.1098/rsta.1968.0003}
Longuet-Higgins, M.S., The eigenfunctions of Laplace's tidal equation over a sphere. {\itshape Philos. Trans. R. Soc. Lond. A Math. Phys. Sci.}, 1968, \textbf{262}, 511--607.

\bibitem[\protect\citeauthoryear{Luo {\itshape{et~al.}}}{2022}]{doi:10.1098/rspa.2022.0108}
Luo, J., Marti, P. and Jackson, A., Waves in the Earth's core. II. Magneto-Coriolis modes. {\itshape Proc. R. Soc. A: Math. Phys. Eng. Sci.}, 2022, \textbf{478}, 20220108.

\bibitem[\protect\citeauthoryear{Mak {\itshape{et~al.}}}{2016}]{mak_griffiths_hughes_2016}
Mak, J., Griffiths, S.D. and Hughes, D.W., Shear flow instabilities in shallow-water magnetohydrodynamics. {\itshape J. Fluid Mech.}, 2016, \textbf{788}, 767--796.

\bibitem[\protect\citeauthoryear{Malkus}{1967}]{malkus_1967}
Malkus, W.V.R., Hydromagnetic planetary waves. {\itshape J. Fluid Mech.}, 1967, \textbf{28}, 793--802.

\bibitem[\protect\citeauthoryear{M\'arquez-Artavia {\itshape{et~al.}}}{2017}]{doi:10.1080/03091929.2017.1301937}
M\'arquez-Artavia, X., Jones, C.A. and Tobias, S.M., Rotating magnetic shallow water waves and instabilities in a sphere. {\itshape Geophys. Astrophys. Fluid Dyn.}, 2017, \textbf{111}, 282--322.

\bibitem[\protect\citeauthoryear{Maslowe}{1986}]{doi:10.1146/annurev.fl.18.010186.002201}
Maslowe, S.A., Critical layers in shear flows. {\itshape Annu. Rev. Fluid Mech.}, 1986, \textbf{18}, 405--432.

\bibitem[\protect\citeauthoryear{McKenzie}{1973}]{mckenzie_1973}
McKenzie, J.F., On the existence of critical levels, with applications to hydromagnetic waves. {\itshape J. Fluid Mech.}, 1973, \textbf{58}, 709--726.

\bibitem[\protect\citeauthoryear{Miesch and Gilman}{2004}]{Miesch2004}
Miesch, M.S. and Gilman, P.A., Thin-shell magnetohydrodynamic equations for the solar tachocline. {\itshape Sol. Phys.}, 2004, \textbf{220}, 287--305.

\bibitem[\protect\citeauthoryear{Miesch {\itshape{et~al.}}}{2007}]{Miesch_2007}
Miesch, M.S., Gilman, P.A. and Dikpati, M., Nonlinear evolution of global magnetoshear instabilities in a three-dimensional thin-shell model of the solar tachocline. {\itshape Astrophys. J. Suppl. Ser.}, 2007, \textbf{168}, 337--361.

\bibitem[\protect\citeauthoryear{Moffatt}{1978}]{moffatt1978magnetic}
Moffatt, H.K., {\itshape Magnetic field generation in electrically conducting fluids}, Cambridge Monographs on Mechanics and Applied Mathematics 1978 (Cambridge University Press).

\bibitem[\protect\citeauthoryear{Mound {\itshape{et~al.}}}{2019}]{Mound2019}
Mound, J., Davies, C., Rost, S. and Aurnou, J., Regional stratification at the top of Earth's core due to core--mantle boundary heat flux variations. {\itshape Nat. Geosci.}, 2019, \textbf{12}, 575--580.

\bibitem[\protect\citeauthoryear{Pozzo {\itshape{et~al.}}}{2012}]{Pozzo2012}
Pozzo, M., Davies, C., Gubbins, D. and Alf{\`e}, D., Thermal and electrical conductivity of iron at Earth's core conditions. {\itshape Nature}, 2012, \textbf{485}, 355--358.

\bibitem[\protect\citeauthoryear{Raphaldini and Raupp}{2020}]{doi:10.1098/rspa.2020.0174}
Raphaldini, B. and Raupp, C.F.M., Nonlinear MHD Rossby wave interactions and persistent geomagnetic field structures. {\itshape Proc. R. Soc. A: Math. Phys. Eng. Sci.}, 2020, \textbf{476}, 20200174.

\bibitem[\protect\citeauthoryear{Reese {\itshape{et~al.}}}{2004}]{Reese_2004_Oscillations}
Reese, D., Rincon, F. and Rieutord, M., Oscillations of magnetic stars - II. Axisymmetric toroidal and non-axisymmetric shear Alfv\'en modes in a spherical shell. {\itshape Astron. Astrophys.}, 2004, \textbf{427}, 279--292.

\bibitem[\protect\citeauthoryear{Rincon and Rieutord}{2003}]{Rincon_2003_Oscillations}
Rincon, F. and Rieutord, M., Oscillations of magnetic stars: I. Axisymmetric shear Alfv\'en modes of a spherical shell in a dipolar magnetic field. {\itshape Astron. Astrophys.}, 2003, \textbf{398}, 663--675.

\bibitem[\protect\citeauthoryear{Schecter {\itshape{et~al.}}}{2000}]{doi:10.1063/1.1289505}
Schecter, D.A., Dubin, D.H.E., Cass, A.C., Driscoll, C.F., Lansky, I.M. and O'Neil, T.M., Inviscid damping of asymmetries on a two-dimensional vortex. {\itshape Phys. Fluids}, 2000, \textbf{12}, 2397--2412.

\bibitem[\protect\citeauthoryear{Schmitt}{2010}]{doi:10.1080/03091920903439746}
Schmitt, D., Magneto-inertial waves in a rotating sphere. {\itshape Geophys. Astrophys. Fluid Dyn.}, 2010, \textbf{104}, 135--151.

\bibitem[\protect\citeauthoryear{Sedl\'a\v{c}ek}{1971}]{sedlacek_1971}
Sedl\'a\v{c}ek, Z., Electrostatic oscillations in cold inhomogeneous plasma I. Differential equation approach. {\itshape J. Plasma Phys.}, 1971, \textbf{5}, 239--263.

\bibitem[\protect\citeauthoryear{Sharif and Jones}{2005}]{doi:10.1080/03091920500372084}
Sharif, B.W. and Jones, C.A., Rotational and magnetic instability in the diffusive tachocline. {\itshape Geophys. Astrophys. Fluid Dyn.}, 2005, \textbf{99}, 493--511.

\bibitem[\protect\citeauthoryear{Shivamoggi}{1992}]{Shivamoggi1992}
Shivamoggi, B.K., Ideal and resistive magnetohydrodynamic modes. {\itshape Int. J. Theor. Phys.}, 1992, \textbf{31}, 2121--2141.

\bibitem[\protect\citeauthoryear{Spencer and Rasband}{1997}]{doi:10.1063/1.872497}
Spencer, R.L. and Rasband, S.N., Damped diocotron quasi-modes of non-neutral plasmas and inviscid fluids. {\itshape Phys. Plasmas}, 1997, \textbf{4}, 53--60.

\bibitem[\protect\citeauthoryear{Spiegel and Zahn}{1992}]{1992A&A...265..106S}
Spiegel, E.A. and Zahn, J.P., {The solar tachocline}. {\itshape Astron. Astrophys.}, 1992, \textbf{265}, 106--114.

\bibitem[\protect\citeauthoryear{Steinolfson}{1985}]{Steinolfson1985ResistiveWD}
Steinolfson, R.S., Resistive wave dissipation on magnetic inhomogeneities Normal modes and phase mixing. {\itshape Astrophys. J.}, 1985, \textbf{295}, 213--219.

\bibitem[\protect\citeauthoryear{Stewartson}{1967}]{doi:10.1098/rspa.1967.0129}
Stewartson, K., Slow oscillations of fluid in a rotating cavity in the presence of a toroidal magnetic field. {\itshape Proc. R. Soc. Lond. A Math. Phys. Sci.}, 1967, \textbf{299}, 173--187.

\bibitem[\protect\citeauthoryear{Strogatz}{2000}]{STROGATZ20001}
Strogatz, S.H., From Kuramoto to Crawford: exploring the onset of synchronization in populations of coupled oscillators. {\itshape Phys. D: Nonlinear Phenom.}, 2000, \textbf{143}, 1--20.

\bibitem[\protect\citeauthoryear{Takehiro and Lister}{2001}]{TAKEHIRO2001357}
Takehiro, S. and Lister, J.R., Penetration of columnar convection into an outer stably stratified layer in rapidly rotating spherical fluid shells. {\itshape Earth Planet. Sci. Lett.}, 2001, \textbf{187}, 357--366.

\bibitem[\protect\citeauthoryear{Taniguchi and Ishiwatari}{2006}]{taniguchi_ishiwatari_2006}
Taniguchi, H. and Ishiwatari, M., Physical interpretation of unstable modes of a linear shear flow in shallow water on an equatorial beta-plane. {\itshape J. Fluid Mech.}, 2006, \textbf{567}, 1--26.

\bibitem[\protect\citeauthoryear{Tataronis and Grossmann}{1973}]{Tataronis1973}
Tataronis, J. and Grossmann, W., Decay of MHD waves by phase mixing. {\itshape Z. Phys.}, 1973, \textbf{261}, 203--216.

\bibitem[\protect\citeauthoryear{Teruya {\itshape{et~al.}}}{2022}]{10.3389/fspas.2022.856912}
Teruya, A.S.W., Raphaldini, B. and Raupp, C.F.M., Ray tracing of MHD Rossby waves in the solar tachocline: Meridional propagation and implications for the solar magnetic activity. {\itshape Front. Astron. Space Sci.}, 2022, \textbf{9}, 856912.

\bibitem[\protect\citeauthoryear{Triana {\itshape{et~al.}}}{2022}]{Triana2021}
Triana, S.A., Dumberry, M., C{\'e}bron, D., Vidal, J., Trinh, A., Gerick, F. and Rekier, J., Core eigenmodes and their impact on the Earth's rotation. {\itshape Surv. Geophys.}, 2022, \textbf{43}, 107--148.

\bibitem[\protect\citeauthoryear{Tung}{1979}]{ATheoryofStationaryLongWavesPartIIIQuasiNormalModesinaSingularWaveguide}
Tung, K.K., A theory of stationary long waves. Part III: Quasi-normal modes in a singular waveguide. {\itshape Mon. Weather Rev.}, 1979, \textbf{107}, 751--774.

\bibitem[\protect\citeauthoryear{Turner and Gilbert}{2007}]{turner_gilbert_2007}
Turner, M.R. and Gilbert, A.D., Linear and nonlinear decay of cat's eyes in two-dimensional vortices, and the link to Landau poles. {\itshape J. Fluid Mech.}, 2007, \textbf{593}, 255--279.

\bibitem[\protect\citeauthoryear{Uberoi}{1972}]{doi:10.1063/1.1694148}
Uberoi, C., Alfv\'en waves in inhomogeneous magnetic fields. {\itshape Phys. Fluids}, 1972, \textbf{15}, 1673--1675.

\bibitem[\protect\citeauthoryear{{Van Kampen}}{1955}]{VANKAMPEN1955949}
{Van Kampen}, N.G., On the theory of stationary waves in plasmas. {\itshape Physica}, 1955, \textbf{21}, 949--963.

\bibitem[\protect\citeauthoryear{van Tent {\itshape{et~al.}}}{2020}]{10.1093/gji/ggaa368}
van Tent, R., Deuss, A., Kaneshima, S. and Thomas, C., {The signal of outermost-core stratification in body-wave and normal-mode data}. {\itshape Geophys. J. Int.}, 2020, \textbf{223}, 1338--1354.

\bibitem[\protect\citeauthoryear{Wang {\itshape{et~al.}}}{2022{\natexlab{a}}}]{wang_gilbert_mason_2022}
Wang, C., Gilbert, A. and Mason, J., Critical-layer instability of shallow-water magnetohydrodynamic shear flows. {\itshape J. Fluid Mech.}, 2022{\natexlab{a}}, \textbf{943}, A24.

\bibitem[\protect\citeauthoryear{Wang {\itshape{et~al.}}}{2022{\natexlab{b}}}]{wang_gilbert_mason_2022_analytical}
Wang, C., Gilbert, A.D. and Mason, J., An analytical study of the MHD clamshell instability on a sphere. {\itshape J. Fluid Mech.}, 2022{\natexlab{b}}, \textbf{953}, A38.

\bibitem[\protect\citeauthoryear{Zaqarashvili {\itshape{et~al.}}}{2021}]{Zaqarashvili2021}
Zaqarashvili, T.V., Albekioni, M., Ballester, J.L., Bekki, Y., Biancofiore, L., Birch, A.C., Dikpati, M., Gizon, L., Gurgenashvili, E., Heifetz, E., Lanza, A.F., McIntosh, S.W., Ofman, L., Oliver, R., Proxauf, B., Umurhan, O.M. and Yellin-Bergovoy, R., Rossby waves in astrophysics. {\itshape Space Sci. Rev.}, 2021, \textbf{217}, 15.

\bibitem[\protect\citeauthoryear{Zaqarashvili {\itshape{et~al.}}}{2009}]{Zaqarashvili_2009}
Zaqarashvili, T.V., Oliver, R. and Ballester, J.L., Global shallow water magnetohydrodynamic waves in the solar tachocline. {\itshape Astrophys. J.}, 2009, \textbf{691}, L41--L44.

\bibitem[\protect\citeauthoryear{Zaqarashvili {\itshape{et~al.}}}{2011}]{Zaqarashvili_2011}
Zaqarashvili, T.V., Oliver, R., Ballester, J.L., Carbonell, M., Khodachenko, M.L., Lammer, H., Leitzinger, M. and Odert, P., Rossby waves and polar spots in rapidly rotating stars: Implications for stellar wind evolution. {\itshape Astron. Astrophys.}, 2011, \textbf{532}, A139.

\bibitem[\protect\citeauthoryear{Zaqarashvili {\itshape{et~al.}}}{2007}]{Zaqarashvili2007}
Zaqarashvili, T.V., Oliver, R., Ballester, J.L. and Shergelashvili, B.M., Rossby waves in "shallow water" magnetohydrodynamics. {\itshape Astron. Astrophys.}, 2007, \textbf{470}, 815--820.

\bibitem[\protect\citeauthoryear{Zaqarashvili}{2018}]{Zaqarashvili_2018}
Zaqarashvili, T., Equatorial magnetohydrodynamic shallow water waves in the solar tachocline. {\itshape Astrophys. J.}, 2018, \textbf{856}, 32.

\bibitem[\protect\citeauthoryear{Zaqarashvili {\itshape{et~al.}}}{2010{\natexlab{a}}}]{Zaqarashvili_2010}
Zaqarashvili, T.V., Carbonell, M., Oliver, R. and Ballester, J.L., Magnetic Rossby waves in the solar tachocline and Rieger-type periodicities. {\itshape Astrophys. J.}, 2010{\natexlab{a}}, \textbf{709}, 749--758.

\bibitem[\protect\citeauthoryear{Zaqarashvili {\itshape{et~al.}}}{2010{\natexlab{b}}}]{Zaqarashvili_2010_Quasi-biennial}
Zaqarashvili, T.V., Carbonell, M., Oliver, R. and Ballester, J.L., Quasi-biennial oscillations in the solar tachocline caused by magnetic Rossby wave instabilities. {\itshape Astrophys. J.}, 2010{\natexlab{b}}, \textbf{724}, L95--L98.

\bibitem[\protect\citeauthoryear{Zaqarashvili {\itshape{et~al.}}}{2015}]{Zaqarashvili_2015}
Zaqarashvili, T.V., Oliver, R., Hanslmeier, A., Carbonell, M., Ballester, J.L., Gachechiladze, T. and Usoskin, I.G., Long-term variation in the Sun's activity caused by magnetic Rossby waves in the tachocline. {\itshape Astrophys. J.}, 2015, \textbf{805}, L14.

\bibitem[\protect\citeauthoryear{Zatman and Bloxham}{1997}]{Zatman1997}
Zatman, S. and Bloxham, J., Torsional oscillations and the magnetic field within the Earth's core. {\itshape Nature}, 1997, \textbf{388}, 760--763.

\bibitem[\protect\citeauthoryear{Zhang {\itshape{et~al.}}}{2022}]{Zhange2119001119}
Zhang, Y., Luo, K., Hou, M., Driscoll, P., Salke, N.P., Min{\'a}r, J., Prakapenka, V.B., Greenberg, E., Hemley, R.J., Cohen, R.E. and Lin, J.F., Thermal conductivity of Fe-Si alloys and thermal stratification in Earth{\textquoteright}s core. {\itshape Proc. Natl. Acad. Sci.}, 2022, \textbf{119}, e2119001119.

\end{thebibliography}

% = = = = = = = = = = = = = = = = = = = = = = = = = = = = = = %
%                                                             %
%                         Appendix                            %
%                                                             %
% = = = = = = = = = = = = = = = = = = = = = = = = = = = = = = %

\appendix

\section{Proofs of \eqref{EQ_semicircle} and \eqref{EQ_bound}}\label{proofs}
%%%%%
In this first appendix, we prove two theorems \eqref{EQ_semicircle} and \eqref{EQ_bound} giving eigenvalue bounds. Integrating \eqref{EQ_differential} with respect to $\mu$ from $-1$ to $1$ after it is multiplied by the complex conjugate of $\tilde{\psi}$, we find
\begin{equation}
\int_{-1}^1\varLambda|\tilde{Q}|^2\ud\mu\,+\,m\int_{-1}^1\left[\lambda+2m\alpha^2\mathcal{B}\frac{\ud(\mathcal{B}\mu)}{\ud\mu}\right]|\tilde{\psi}|^2\ud\mu\,=\,0\,,
\label{EQ_integral}
\end{equation}
where $|\tilde{Q}|^2\equiv(1-\mu^2)|\ud\tilde{\psi}/\ud\mu|^2+[m^2/(1-\mu^2)]|\tilde{\psi}|^2$. The real part of \eqref{EQ_integral} is given by
\begin{subequations}
\begin{equation}
\begin{aligned}
\left\{[\mathrm{Re}(\lambda)]^2-[\mathrm{Im}(\lambda)]^2\right\}&\int_{-1}^1|\tilde{Q}|^2\ud\mu\,-\,m^2\alpha^2\int_{-1}^{1}\mathcal{B}^2|\tilde{Q}|^2\ud\mu\\
\,&+\,m\mathrm{Re}(\lambda)\int_{-1}^1|\tilde{\psi}|^2\ud\mu\,+\,2m^2\alpha^2\int_{-1}^1\mathcal{B}\frac{\ud(\mathcal{B}\mu)}{\ud\mu}|\tilde{\psi}|^2\ud\mu\,=\,0\,.
\end{aligned}
\label{EQ_necessary_real}
\end{equation}
On the other hand, the imaginary part is
\begin{equation}
\mathrm{Im}(\lambda)\left(2\mathrm{Re}(\lambda)\int_{-1}^1|\tilde{Q}|^2\ud\mu\,+\,m\int_{-1}^1|\tilde{\psi}|^2\ud\mu\right)\,=\,0\,.
\label{EQ_necessary_imag}
\end{equation}
\end{subequations}
It should be noted that $\mathrm{Re}(\lambda)\leq0$ when $\mathrm{Im}(\lambda)\neq0$, as can be seen from \eqref{EQ_necessary_imag}. The combination of the expressions \eqref{EQ_necessary_real} and \eqref{EQ_necessary_imag} with $\mathrm{Im}(\lambda)\neq0$ yields
\begin{linenomath}\begin{align}
\left[\frac{\mathrm{Re}(\lambda)}{m}\right]^2\,+\,\left[\frac{\mathrm{Im}(\lambda)}{m}\right]^2\,&=\,-\alpha^2\frac{\int_{-1}^1\mathcal{B}^2|\tilde{Q}|^2\ud\mu}{\int_{-1}^1|\tilde{Q}|^2\ud\mu}\,+\,4\frac{-\mathrm{Re}(\lambda)}{m}\alpha^2\frac{\int_{-1}^1\mathcal{B}[\ud(\mathcal{B}\mu)/\ud\mu]|\tilde{\psi}|^2\ud\mu}{\int_{-1}^1|\tilde{\psi}|^2\ud\mu}\notag\\
\,&\leq\,-\alpha^2\min(\mathcal{B}^2)\,+\,2\frac{-\mathrm{Re}(\lambda)}{m}\alpha^2\max\left[2\mathcal{B}\frac{\ud(\mathcal{B}\mu)}{\ud\mu}\right]\,.
\end{align}\end{linenomath}
This inequality gives \eqref{EQ_semicircle}.\par
We can derive another bound \eqref{EQ_bound} by using \eqref{EQ_necessary_imag} when $\mathrm{Im}(\lambda)\neq0$. The property of the Rayleigh quotient of the associated Legendre operator in the form
\begin{equation}
m(m+1)\,=\,\min_{\tilde{\psi}(\mu)}\frac{\int_{-1}^1|\tilde{Q}|^2\ud\mu}{\int_{-1}^1|\tilde{\psi}|^2\ud\mu}
\end{equation}
then becomes
\begin{equation}
-\frac{m}{2\mathrm{Re}(\lambda)}\geq m(m+1)\,,
\end{equation} 
unless $\mathrm{Im}(\lambda)=0$. This inequality is equal to \eqref{EQ_bound} since $\mathrm{Re}(\lambda)\leq0$.

% = = = = = = = = = = = = = = = = = = = = = = = = = = = = = = %

\section{Fast MR waves when $\mathcal{B}=\mu$}\label{SEC_fast}
\begin{figure}
\begin{center}
	\resizebox*{150mm}{!}{\includegraphics{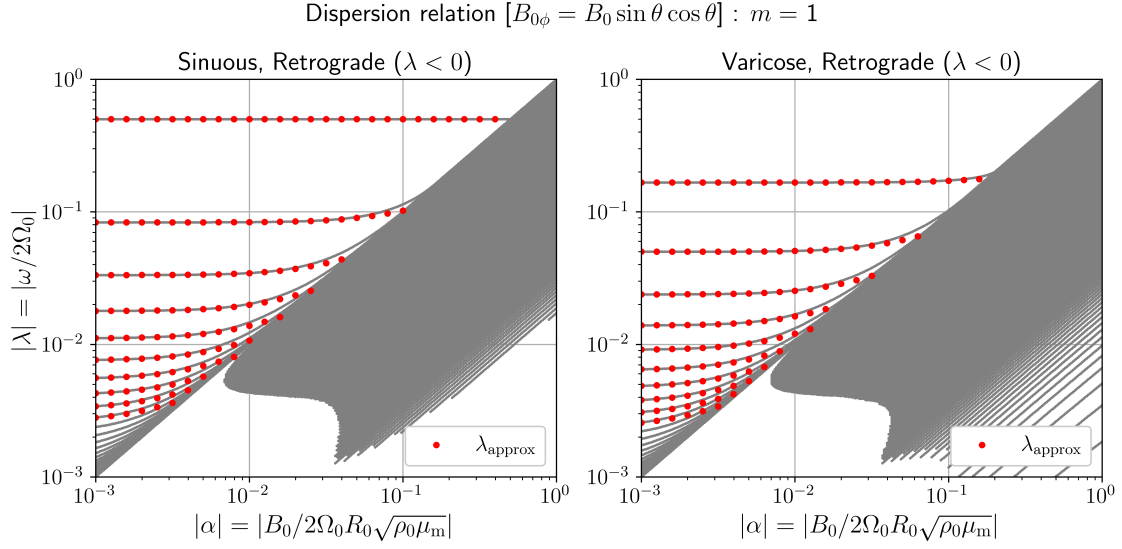}}%
	\caption{Same as the left panels of Figure \ref{FIG_dispersion_sincos_m1}, but for a different plotting range (grey markers). The red circles represent the approximate dimensionless angular frequencies $\lambda_\mathrm{approx}$ calculated by the approximate dispersion relation \eqref{EQ_dispersion_fast}.}
	\label{FIG_rossby}
\end{center}
\end{figure}

\begin{figure}
\begin{center}
\begin{minipage}{75mm}
	\subfigure[Varicose mode with $\lambda\approx-0.17272$ and the approximate eigenmode with $\lambda_\mathrm{approx}\approx-0.17151$ ($n=2$, $c^2\approx0.39758$).]{
	\resizebox*{70mm}{!}{\includegraphics{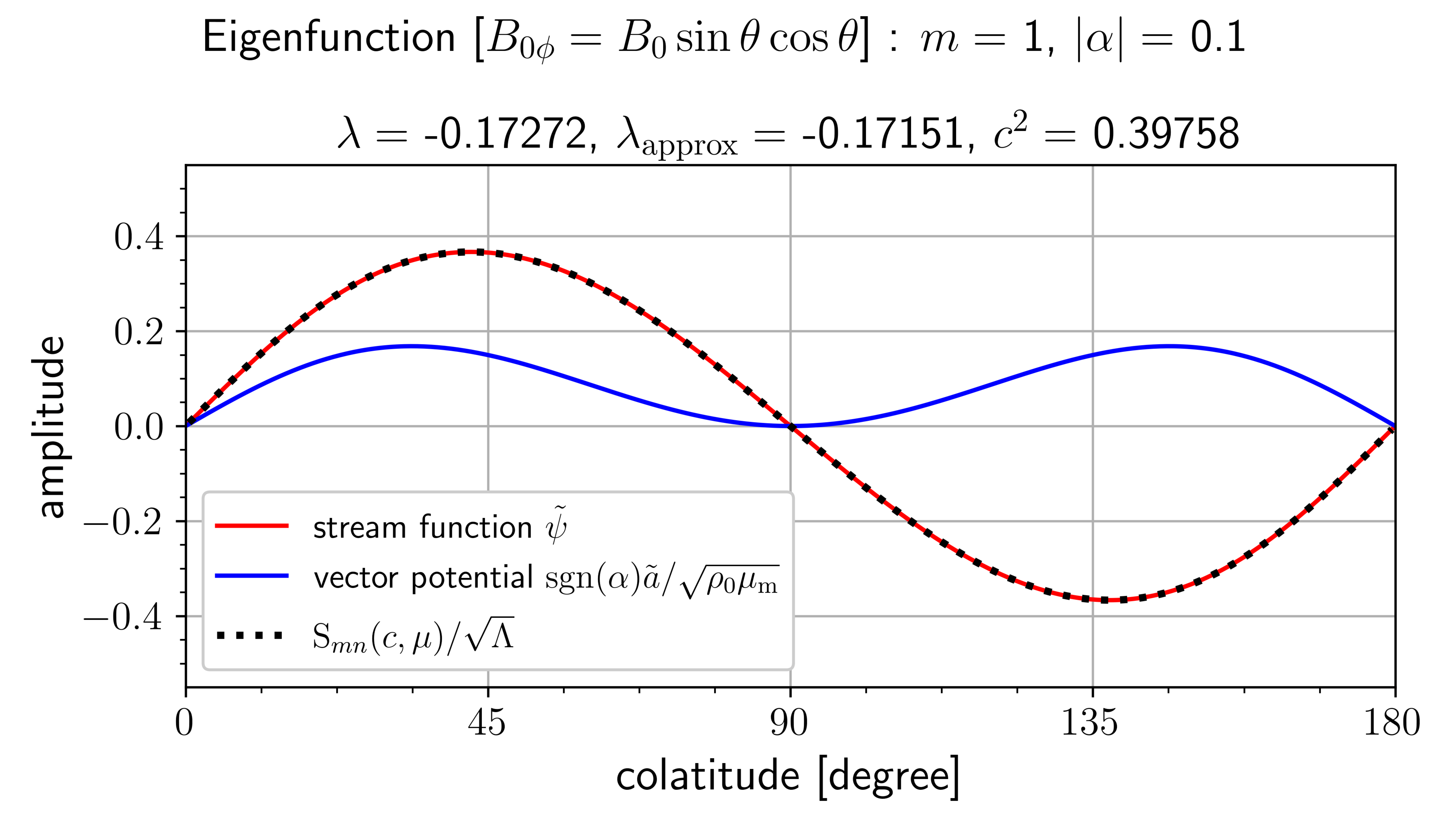}}}%
\end{minipage}
\begin{minipage}{75mm}
	\subfigure[Sinuous mode with $\lambda\approx-0.11407$ and the approximate eigenmode with $\lambda_\mathrm{approx}\approx-0.10212$ ($n=3$, $c^2\approx-2.67696$).]{
	\resizebox*{70mm}{!}{\includegraphics{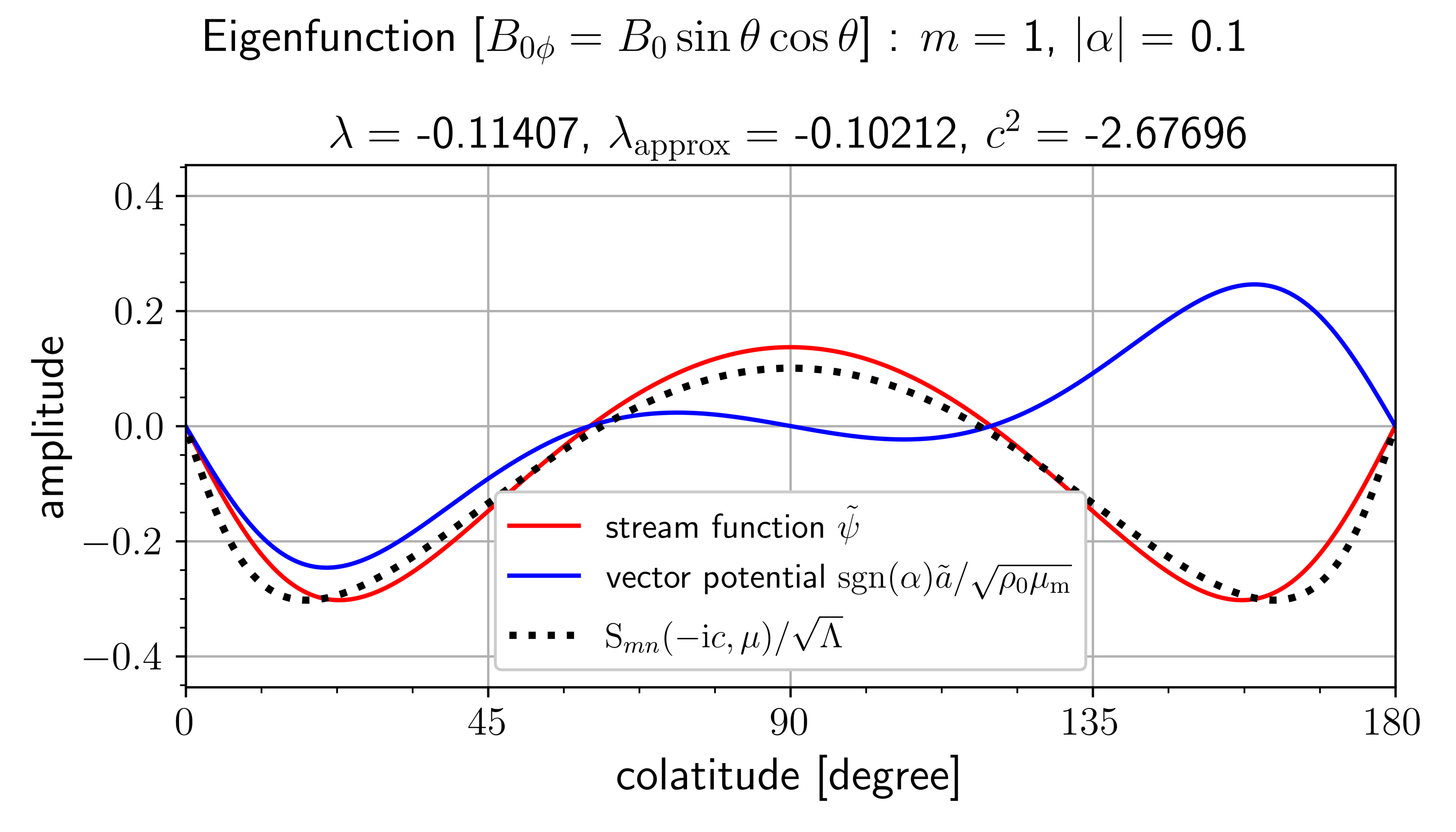}}}%
\end{minipage}
	\caption{Same as Figure \ref{FIG_eigenfunction_sincos_m1a01}(a), but for fast MR waves outside the continuous spectrum. The black dashed curves illustrate their approximate eigenfunctions $\mathrm{S}_{mn}/\sqrt{\varLambda}$. The eigenfunctions have been scaled such that their amplitudes become comparable to those of their corresponding eigenfunctions which are obtained numerically from the full eigenvalue problem.}
	\label{FIG_rossby_eigfunc}
\end{center}
\end{figure}

%%%%%
This appendix and Appendix \ref{SEC_discrete} cover the fast MR waves when the simplest equatorially antisymmetric non-Malkus field $\mathcal{B}=\mu$ is imposed. In particular, we now narrow our interests down to discrete branches $\lambda<-m|\alpha|$ outside the continuous spectrum. If we focus on waves with frequencies $|\lambda|\gg m|\alpha|$, sufficiently higher than the continuous modes, \eqref{DEF_alternative_differential_potential} is reduced to
\begin{equation}
\mathcal{L}^2\,=\,-m^2\,+\,(1-\mu^2)\left[\left(-\frac{m}{\lambda}+\frac{m^2\alpha^2}{\lambda^2}\right)-c^2\mu^2+\mathrm{O}\left(\frac{\alpha^4}{\lambda^4}\right)\right]\,,
\label{EQ_alternative_differential_potential_high-frequency}
\end{equation}
where $c^2\equiv(m^2\alpha^2/\lambda^2)(7+m/\lambda)$, and then \eqref{EQ_alternative_differential} approximately becomes the same form as the differential equation for the angular prolate spheroidal wave function $\mathrm{S}_{mn}(c,\mu)$ (e.g. \citealt{abramowitz1964handbook}; see also \citealt{Zaqarashvili_2009}). The power series expansion for the eigenvalues $\lambda_{mn}\equiv-m/\lambda+m^2\alpha^2/\lambda^2$ of this differential equation yields the approximate dispersion relation for the fast MR waves outside the continuous spectrum in the form
\begin{equation}
\lambda_{mn}\,=\,n(n+1)\,+\,\frac{1}{2}\left[1-\frac{(2m-1)(2m+1)}{(2n-1)(2n+3)}\right]c^2\,+\,\mathrm{O}\left(\frac{\alpha^4}{\lambda^4}\right)\,.
\label{EQ_dispersion_fast}
\end{equation}
Their approximate angular frequencies $\lambda_\mathrm{approx}$ were calculated from the above relation with the \texttt{scipy.optimize.fsolve} function of the SciPy library. Their results for $m=1$ are shown in Figure \ref{FIG_rossby}, overlaid on the same graphs as the left panels of Figure \ref{FIG_dispersion_sincos_m1}. When $|\lambda|\gg m|\alpha|$, their values surely approximate the angular frequencies obtained numerically from the full eigenvalue problem. Figure \ref{FIG_rossby_eigfunc} displays two of their approximate eigenfunctions $\mathrm{S}_{mn}/\sqrt{\lambda_\mathrm{approx}^2-m^2\alpha^2\mu^2}$ ($\simeq\tilde{\psi}$), where $\mathrm{S}_{mn}$ were given by the \texttt{scipy.special.pro\_ang1} function of the SciPy library, with their corresponding eigenfunctions calculated from the full problem. It should be noted that, if $c^2=(m^2\alpha^2/\lambda_\mathrm{approx}^2)(7+m/\lambda_\mathrm{approx})<0$, the parameter $c$ in $\mathrm{S}_{mn}(c,\mu)$ is replaced by $-\mathrm{i}c$, and then $\mathrm{S}_{mn}(-\mathrm{i}c,\mu)$ is regarded as the angular oblate spheroidal wave function. In this case, we used the \texttt{scipy.special.obl\_ang1} function of the SciPy library.\par
\begin{figure}
\begin{center}
\begin{minipage}{75mm}
	\subfigure[Scaled zonal wavenumber $k\sin\theta\approx-0.87264$, and the initial local meridional wavenumber $l_\mathrm{init}=0.25$.]{
	\resizebox*{70mm}{!}{\includegraphics{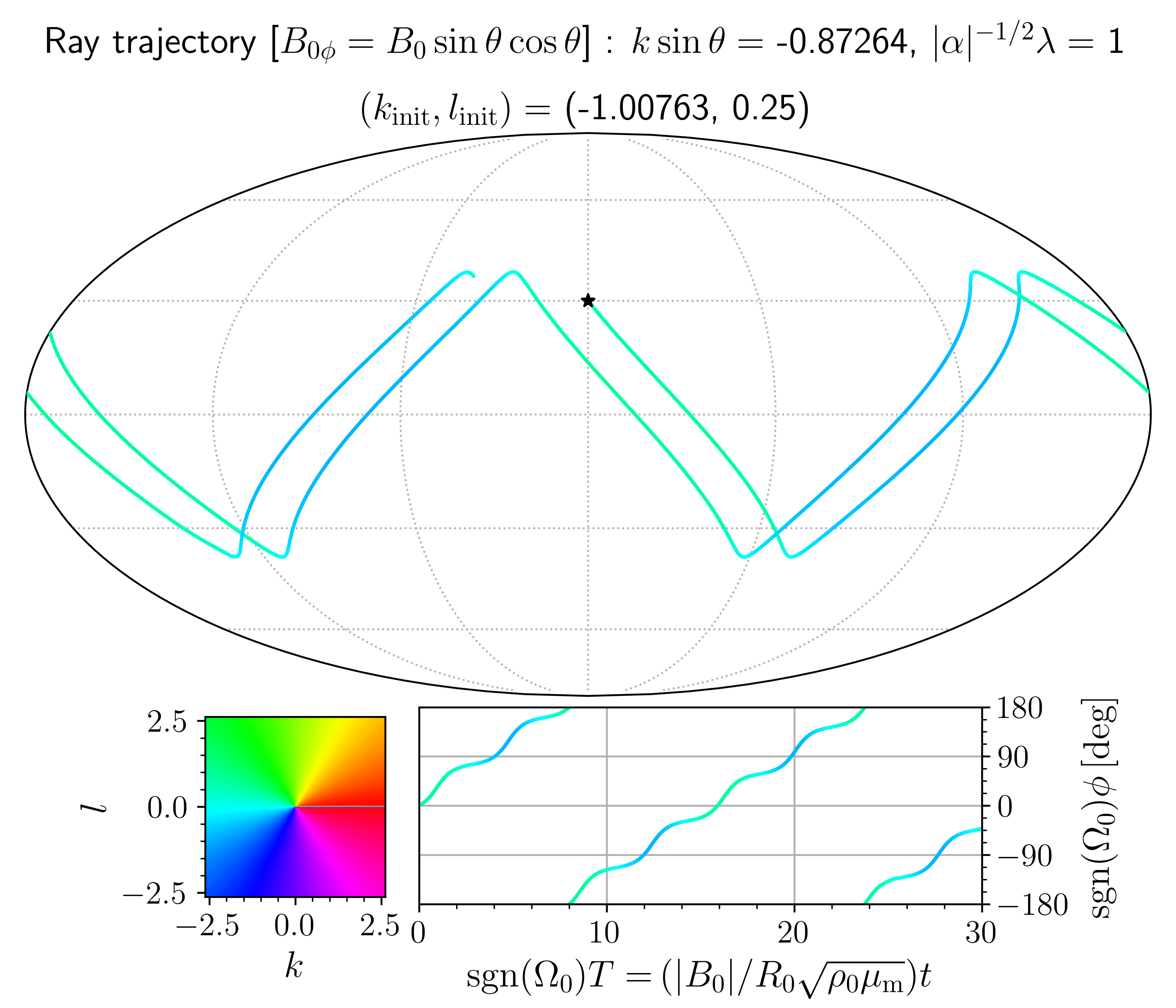}}}%
\end{minipage}
\begin{minipage}{75mm}
	\subfigure[Scaled zonal wavenumber $k\sin\theta\approx-0.43762$, and the initial local meridional wavenumber $l_\mathrm{init}=0.5$.]{
	\resizebox*{70mm}{!}{\includegraphics{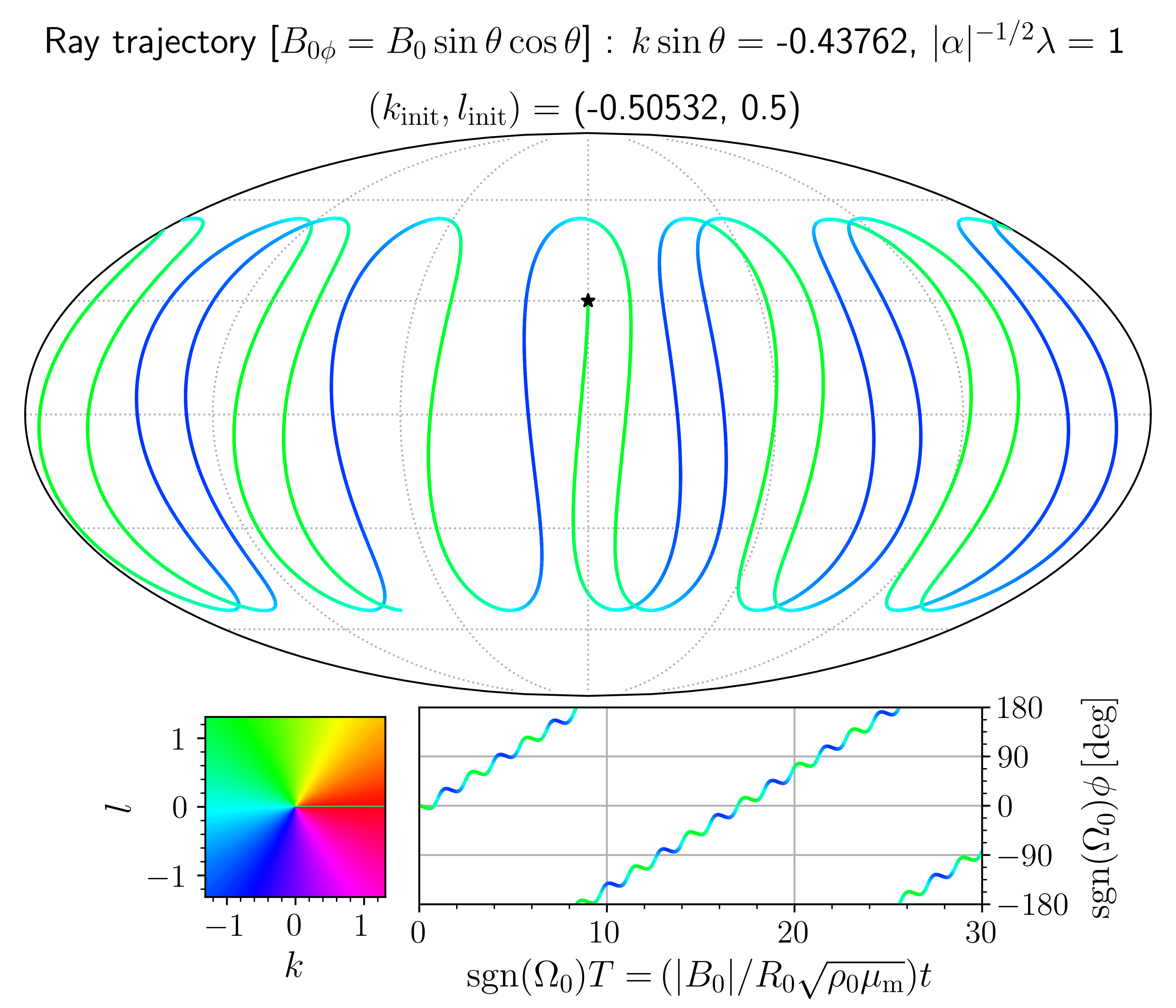}}}%
\end{minipage}
\begin{minipage}{75mm}
	\subfigure[Scaled zonal wavenumber $k\sin\theta\approx-0.06871$, and the initial local meridional wavenumber $l_\mathrm{init}=0.25$.]{
	\resizebox*{70mm}{!}{\includegraphics{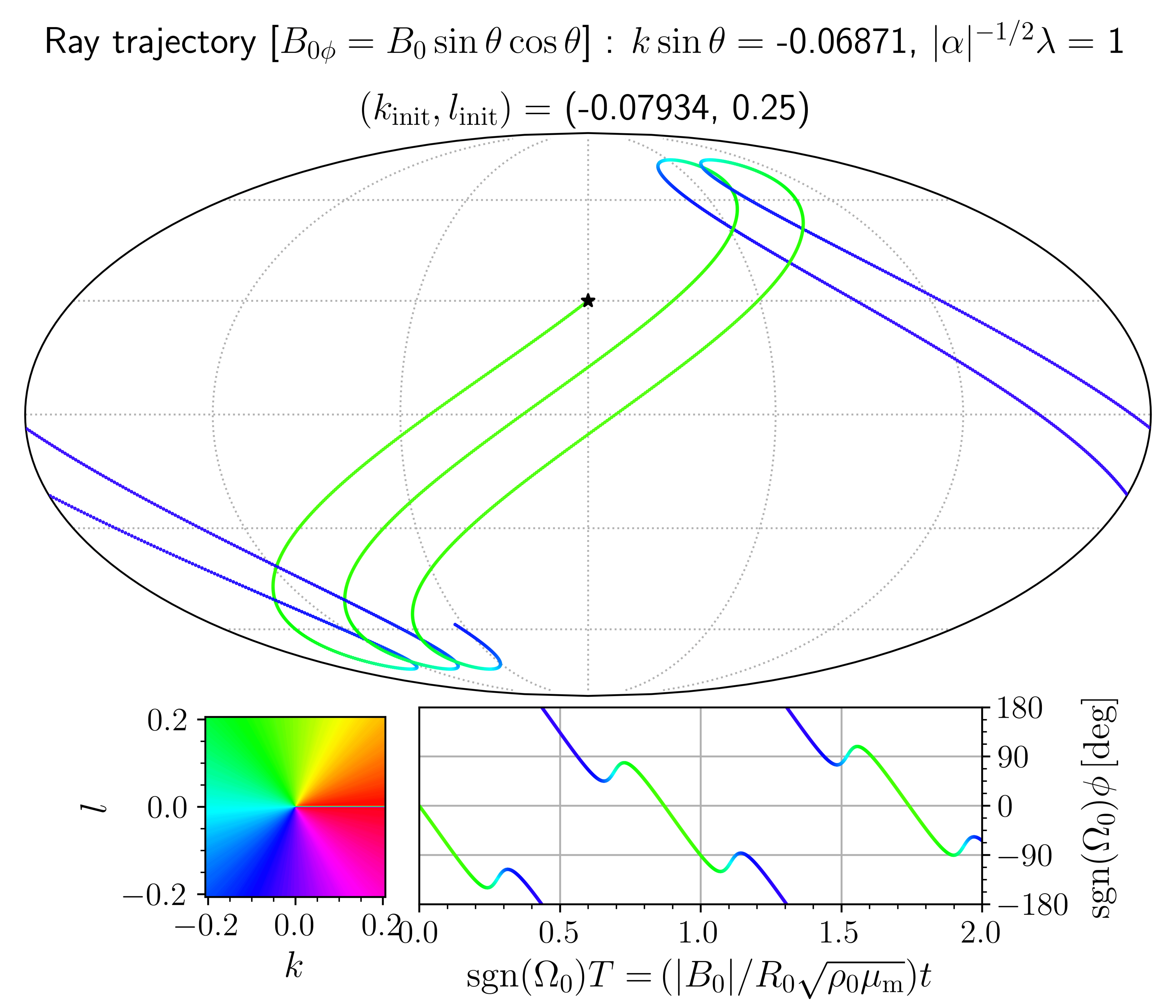}}}%
\end{minipage}
	\caption{Same as Figure \ref{FIG_ray_sincos_12}, but for fast MR waves. The initial colatitude $\theta_\mathrm{init}=60^\circ$.}
	\label{FIG_ray_sincos_rossby}
\end{center}
\end{figure}

%%%%%
We also integrated the ray-tracing equations (see Section \ref{SEC_discussion}) numerically for wave packets that belong to fast MR waves. Figure \ref{FIG_ray_sincos_rossby} shows three of their trajectories when $\mathcal{B}=\cos\theta$. They move back and forth between two turning latitudes, at which $l=0$, in the latitudinal direction, while their manners of migration in the longitudinal direction are complicated by the fact that the radii of the circular contour lines which represent the propagation properties of the packets belonging to the fast modes depend on the latitude (see Figure \ref{FIG_local_sincos_kl}); when the point $(k,l)$ corresponding to the dominant local wavenumber of a wave packet is located on the left half of one of the contour lines, the azimuthal component $c_{\mathrm{g},\phi}$ of its group velocity is positive (see the cyanish markers in Figures \ref{FIG_ray_sincos_rossby}(a), (b) and (c)), and conversely $c_{\mathrm{g},\phi}<0$ for the points on the right half (see the green and blue markers in Figures \ref{FIG_ray_sincos_rossby}(b) and (c)). Since no critical latitudes exist, their behaviours are decidedly different from those of Alfv\'en (see Figures \ref{FIG_ray_sincos_12} and \ref{FIG_ray_sincos_34}) and slow MR waves (see Figure \ref{FIG_ray_sincos_5}).

% = = = = = = = = = = = = = = = = = = = = = = = = = = = = = = %

\section{Discrete eigenvalue buried in the continuous spectrum}\label{SEC_discrete}
\begin{figure}
\begin{center}
\begin{minipage}{75mm}
	\subfigure[$|\alpha|=0.1$.]{
	\resizebox*{70mm}{!}{\includegraphics{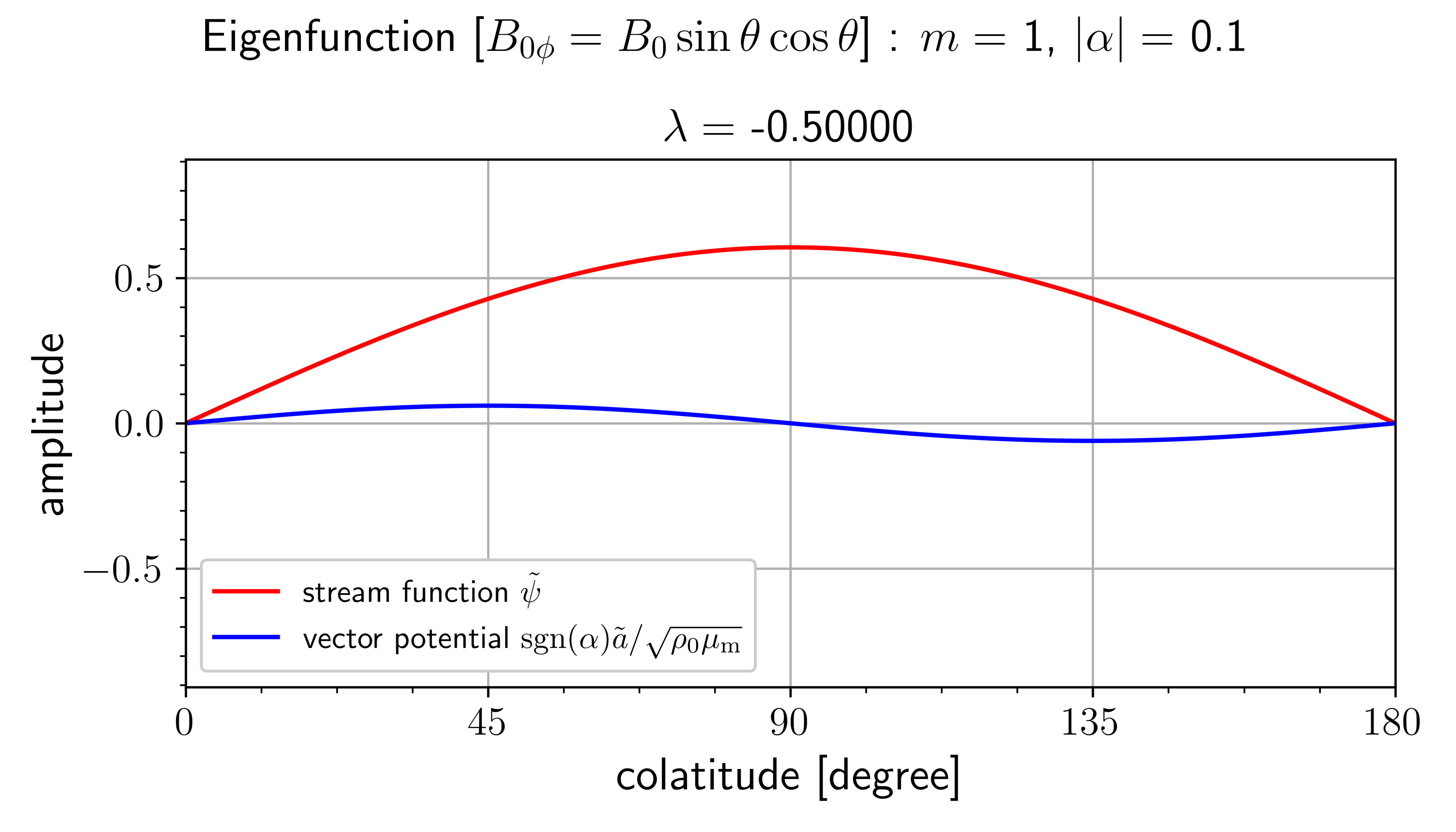}}}%
\end{minipage}
\begin{minipage}{75mm}
	\subfigure[$|\alpha|=1$.]{
	\resizebox*{70mm}{!}{\includegraphics{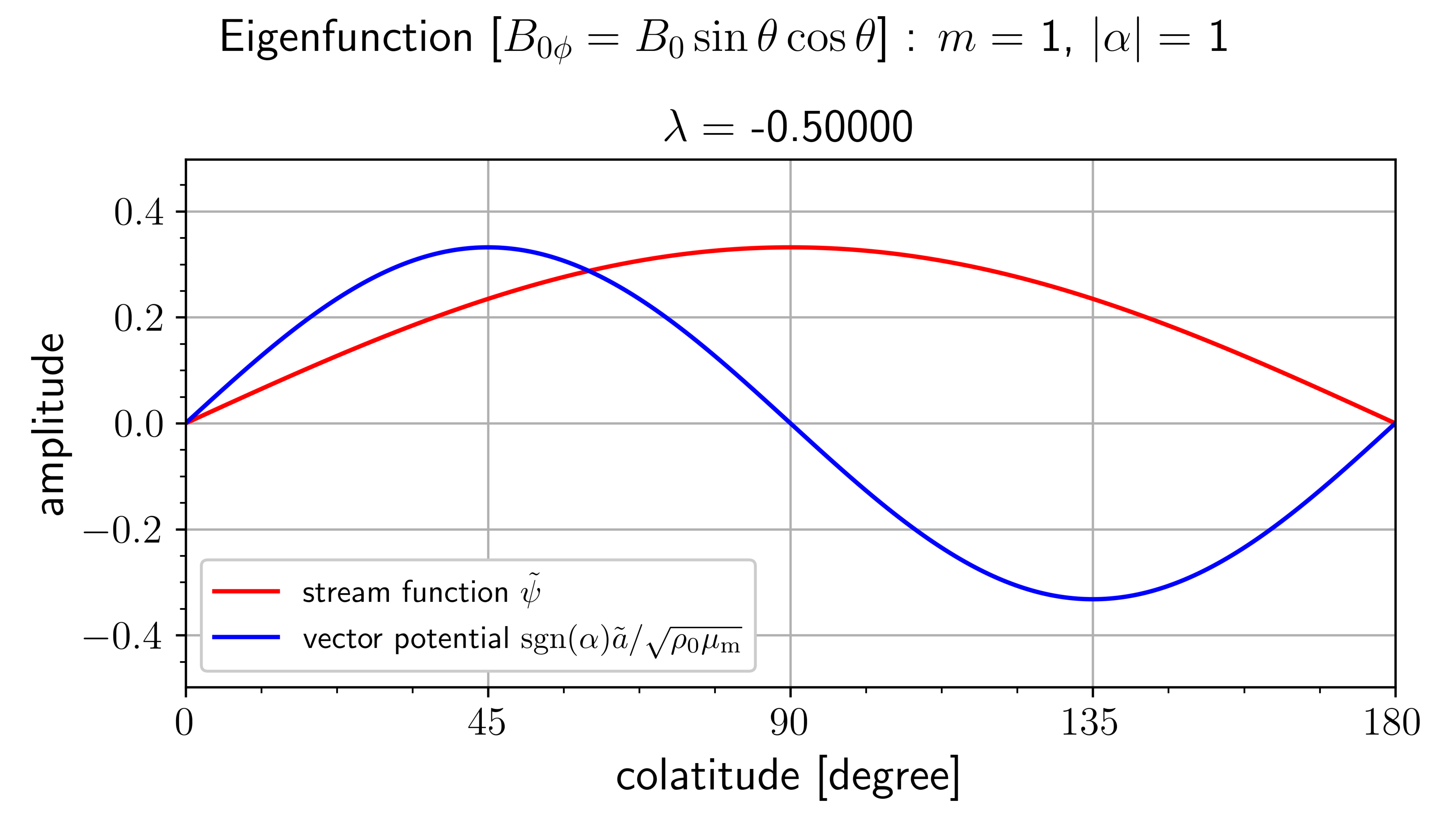}}}%
\end{minipage}
	\caption{Same as Figure \ref{FIG_eigenfunction_sincos_m1a01}(a), but for the fast MR wave buried in the Alfv\'en continuous spectrum (the zonal wavenumber $m=1$ and the nondimensional angular frequency $\lambda=-0.5$).}
	\label{FIG_buried_rossby_eigfunc}
\end{center}
\end{figure}

%%%%%
We here explain, for the simplest equatorially antisymmetric non-Malkus background field $\mathcal{B}=\mu$, why the lowermost wavenumber branch of the $m=1$ sinuous modes of the fast MR wave has no critical latitudes (see the upper panels of Figure \ref{FIG_alleigenfunction_sincos_m2a1}), and why its branch can penetrate the Alfv\'en continuous spectrum while being discrete (see the upper left panel of Figure \ref{FIG_dispersion_sincos_m1}). According to Section \ref{SEC_numerical}, its angular frequency $\lambda=-1/2$ is constant regardless of the value of $|\alpha|$. When $m=1$ and $\lambda=-1/2$, \eqref{EQ_nondim} are simplified as
\begin{subequations}\label{EQ_nondim_discrete}
\begin{linenomath}\begin{align}
\left(\frac{1}{2}\nabla_\mathrm{h}^2+1\right)\tilde{\psi}\,&=\,|\alpha|\mu(\nabla_\mathrm{h}^2+6)\left[\frac{\operatorname{sgn}(\alpha)\tilde{a}}{\sqrt{\rho_0\mu_\mathrm{m}}}\right]\,,
\label{EQ_vorticity_nondim_discrete}\\
\frac{1}{2}\left[\frac{\operatorname{sgn}(\alpha)\tilde{a}}{\sqrt{\rho_0\mu_\mathrm{m}}}\right]\,&=\,|\alpha|\mu\tilde{\psi}\,.
\label{EQ_potential_nondim_discrete}
\end{align}\end{linenomath}
\end{subequations}
If both sides of both \eqref{EQ_vorticity_nondim_discrete} and \eqref{EQ_potential_nondim_discrete} did not vanish, the above equations would be transformed into the same form as \eqref{EQ_differential} and inevitably possess critical latitudes $\mu_\mathrm{c}^2=1/4\alpha^2$. We, therefore, presume that a solution for which both sides of \eqref{EQ_vorticity_nondim_discrete} are identically equal to zero exists. Because $\nabla_\mathrm{h}^2\mathrm{P}_1^1=-2\mathrm{P}_1^1$, $\nabla_\mathrm{h}^2\mathrm{P}_2^1=-6\mathrm{P}_2^1$, and $\mu\mathrm{P}_1^1=\mathrm{P}_2^1/3$ are derived from \eqref{EQ_recurrence}, the presumption requires that its eigenfunction is written as $\tilde{\psi}=\tilde{\psi}^{[1]}\mathcal{N}_1^1\mathrm{P}_1^1$ and $\operatorname{sgn}(\alpha)\tilde{a}/\sqrt{\rho_0\mu_\mathrm{m}}=\tilde{a}^{[2]}\mathcal{N}_2^1\mathrm{P}_2^1$, where $2|\tilde{\psi}^{[1]}|^2+6|\tilde{a}^{[2]}|^2=1$ and $\tilde{a}^{[2]}/\tilde{\psi}^{[1]}=2|\alpha|/\sqrt{5}$ (see Figure \ref{FIG_buried_rossby_eigfunc}). This implies that this discrete mode buried in the continuous spectrum is exceptional. On its branch, the equipartition between $\mathrm{MKE}$ and $\mathrm{MME}$ occurs only when $|\alpha|=\sqrt{5/12}\approx0.6455$, and we found that the value agrees with that which one can read from the colours of markers in the upper left panel of Figure \ref{FIG_dispersion_sincos_m1}. Furthermore, we notice that its angular frequency $\lambda=-1/2$ satisfies even the approximate dispersion relation \eqref{EQ_dispersion_fast} for fast MR waves ``outside'' the continuous spectrum, since one then has $c^2=20\alpha^2$ and $\lambda_{11}=2+4\alpha^2$. Nevertheless, its (exact) eigenfunction $\tilde{\psi}^{[1]}\mathcal{N}_1^1\mathrm{P}_1^1$ is not identical with the approximate eigenfunctions $\mathrm{S}_{11}/\sqrt{\varLambda}$ for the fast modes outside the continuous spectrum (see Appendix \ref{SEC_fast}).\par
%%%%%
It should be added, at the end of this appendix, that \citet{wang_gilbert_mason_2022_analytical} recently studied two distinctive modes in the same system as our problem. One is the above buried discrete mode ($\lambda=-1/2$), and the other is the stationary mode ($\lambda=0$) whose eigenfunction is represented as $\psi_1=0$ and $a_1\propto\mathrm{P}_2^1\mathrm{e}^{\mathrm{i}\phi}$ in \eqref{EQ_nondim_discrete}. Their paper showed that the latter is closely linked to ``clamshell instability'' (see Section \ref{SEC_introduction}) and referred to this mode as the ``tilting mode''. On the other hand, we consider that the former is also interesting, because this mode remains stationary in the non-rotating frame although this is a rotating fluid problem, as shown in \citet{wang_gilbert_mason_2022_analytical}. Those two modes are similar in terms of the tilt of a toroidal magnetic field. For the buried mode of the fast MR wave, the perturbation $\psi_1$ in the stream function, which is proportional to $\mathrm{P}_1^1\mathrm{e}^{\mathrm{i}\phi}$, makes the imposed field $B_{0\phi}=B_0\sin\theta\cos\theta$ tilt, hence the perturbation $a_1\propto\mathrm{P}_2^1\mathrm{e}^{\mathrm{i}\phi}$ in the magnetic vector potential. Such a total magnetic field does not obviously induce the Lorentz force acting on the perturbation. It should be noted that the right-hand side of \eqref{EQ_vorticity_nondim_discrete} vanishes when $\operatorname{sgn}(\alpha)\tilde{a}/\sqrt{\rho_0\mu_\mathrm{m}}\propto\mathrm{P}_2^1$.

% = = = = = = = = = = = = = = = = = = = = = = = = = = = = = = %

\section{Comparison of the numerical results with the Frobenius series solutions}\label{SUBSEC_comparison}
\begin{figure}
\begin{center}
	\resizebox*{125mm}{!}{\includegraphics{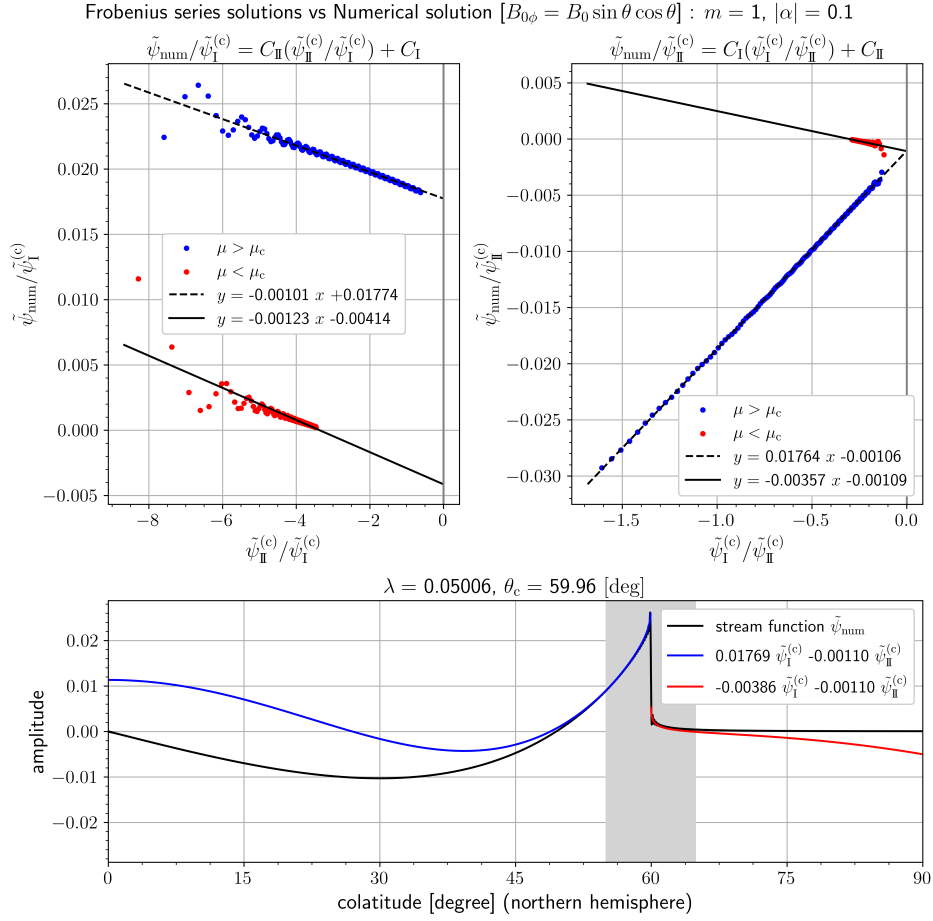}}%
	\caption{Comparison between the numerical eigenfunction $\tilde{\psi}_\mathrm{num}$ of the sinuous mode with the dimensionless angular frequency $\lambda\approx0.05006$ (the critical latitude $\theta_\mathrm{c}=59.96^\circ$ in the northern hemisphere) and linear combinations $C_\mathrm{I}\tilde{\psi}_\mathrm{I}^{(\mathrm{c})}+C_\mathrm{I\!I}\tilde{\psi}_\mathrm{I\!I}^{(\mathrm{c})}$ of the Frobenius series solutions \eqref{DEF_Frobenius} with \eqref{DEF_coefficient} when the zonal wavenumber $m=1$, the absolute value of the Lehnert number $|\alpha|=0.1$, and the simplest equatorially antisymmetric non-Malkus field $\mathcal{B}=\mu$ permeates the system. The undetermined coefficients $C_\mathrm{I}$ and $C_\mathrm{I\!I}$ are estimated from least squares fittings of $\tilde{\psi}_\mathrm{num}/\tilde{\psi}_\mathrm{I}^{(\mathrm{c})}=C_\mathrm{I\!I}(\tilde{\psi}_\mathrm{I\!I}^{(\mathrm{c})}/\tilde{\psi}_\mathrm{I}^{(\mathrm{c})})+C_\mathrm{I}$ (upper left panel) and $\tilde{\psi}_\mathrm{num}/\tilde{\psi}_\mathrm{I\!I}^{(\mathrm{c})}=C_\mathrm{I}(\tilde{\psi}_\mathrm{I}^{(\mathrm{c})}/\tilde{\psi}_\mathrm{I\!I}^{(\mathrm{c})})+C_\mathrm{I\!I}$ (upper right panel) with $N_\mathrm{data}=200$ points on the equatorial (red circles and solid lines) and the polar (blue circles and dashed lines) sides of the critical latitude for $N_\theta=7201$. The lower panel shows $\tilde{\psi}_\mathrm{num}$ (black curve) and $C_\mathrm{I}\tilde{\psi}_\mathrm{I}^{(\mathrm{c})}+C_\mathrm{I\!I}\tilde{\psi}_\mathrm{I\!I}^{(\mathrm{c})}$ with $C_\mathrm{I}$ and $C_\mathrm{I\!I}$ determined by the fittings (red and blue curves) as functions of the colatitude. The vertical grey shaded area of this panel contains the $2N_\mathrm{data}=400$ points used in the fittings.}
	\label{FIG_frobenius}
\end{center}
\end{figure}

%%%%%
Here, we confirm that the eigenfunctions obtained numerically can be approximated by linear combinations of the linearly independent Frobenius series solutions \eqref{DEF_Frobenius}. Let $\tilde{\psi}_\mathrm{num}$ be one of the numerical stream functions, and we fit it into the form $C_\mathrm{I}\tilde{\psi}_\mathrm{I}^{(\mathrm{c})}+C_\mathrm{I\!I}\tilde{\psi}_\mathrm{I\!I}^{(\mathrm{c})}$ by adjusting the coefficients $C_\mathrm{I}$ and $C_\mathrm{I\!I}$ on each side of the critical latitudes. The fitting procedure is as follows. The colatitude ($0\leq\theta\leq\upi$) is divided into $N_\theta$ points at even intervals. On each side of the nearest point of a singular latitude, the $N_\mathrm{data}$ points closest to the point are chosen among the $N_\theta$ points for the fitting. For these points $\theta_i$ ($1\leq i\leq N_\mathrm{data}$) either on the equatorial or polar sides, we may write
\begin{equation}
\frac{\tilde{\psi}_\mathrm{num}(\cos\theta_i)}{\tilde{\psi}_\mathrm{I}^{(\mathrm{c})}(\cos\theta_i)}\,\approx\,C_\mathrm{I\!I}\frac{\tilde{\psi}_\mathrm{I\!I}^{(\mathrm{c})}(\cos\theta_i)}{\tilde{\psi}_\mathrm{I}^{(\mathrm{c})}(\cos\theta_i)}\,+\,C_\mathrm{I}\quad\text{and}\quad\frac{\tilde{\psi}_\mathrm{num}(\cos\theta_i)}{\tilde{\psi}_\mathrm{I\!I}^{(\mathrm{c})}(\cos\theta_i)}\,\approx\,C_\mathrm{I}\frac{\tilde{\psi}_\mathrm{I}^{(\mathrm{c})}(\cos\theta_i)}{\tilde{\psi}_\mathrm{I\!I}^{(\mathrm{c})}(\cos\theta_i)}\,+\,C_\mathrm{I\!I}\,.
\label{EQ_fitting}
\end{equation}
Now, $\tilde{\psi}_\mathrm{I}^{(\mathrm{c})}$ and $\tilde{\psi}_\mathrm{I\!I}^{(\mathrm{c})}$ are approximated by the second-order Frobenius solutions with \eqref{DEF_coefficient}. We obtain a candidate value for each of $C_\mathrm{I}$ and $C_\mathrm{I\!I}$ from each equation, through least squares fittings of \eqref{EQ_fitting} with the \texttt{numpy.polyfit} function of the NumPy library. Thereby we have four candidate values for each of $C_\mathrm{I}$ and $C_\mathrm{I\!I}$ for one critical latitude, two from the equatorial side, and two from the polar side. The upper panels of Figure \ref{FIG_frobenius} show a result when we perform this procedure with $N_\theta=7201$, and $N_\mathrm{data}=200$ individually for the equatorial (red circles and solid lines) and the polar (blue circles and dashed lines) sides. These fittings demonstrate that $C_\mathrm{I}$ typically has different values ($C_\mathrm{I}^{(\mathrm{e})}$ and $C_\mathrm{I}^{(\mathrm{p})}$, say) between the two sides of a critical latitude, whilst $C_\mathrm{I\!I}$ has the same value on both sides. Since $C_\mathrm{I\!I\!I}$ in \eqref{EQ_solution_all} can also be written as $\operatorname{sgn}(\mu_\mathrm{c})[C_\mathrm{I}^{(\mathrm{p})}-C_\mathrm{I}^{(\mathrm{e})}]$, our numerical eigenmodes are consistent with the general results as described in Sections \ref{SUBSEC_existence} and \ref{SUBSEC_frobenius}. We accordingly adopt the mean values of the two candidate values for each of $C_\mathrm{I}^{(\mathrm{e})}$ and $C_\mathrm{I}^{(\mathrm{p})}$ and four candidate ones of $C_\mathrm{I\!I}$ ($C_\mathrm{I\!I}^{(\mathrm{e})}=C_\mathrm{I\!I}^{(\mathrm{p})}$) as their definite values, which are used in the graph comparing $\tilde{\psi}_\mathrm{num}$ with $C_\mathrm{I}\tilde{\psi}_\mathrm{I}^{(\mathrm{c})}+C_\mathrm{I\!I}\tilde{\psi}_\mathrm{I\!I}^{(\mathrm{c})}$ (the lower panel of Figure \ref{FIG_frobenius}).\par
\begin{figure}
\begin{center}
	\resizebox*{150mm}{!}{\includegraphics{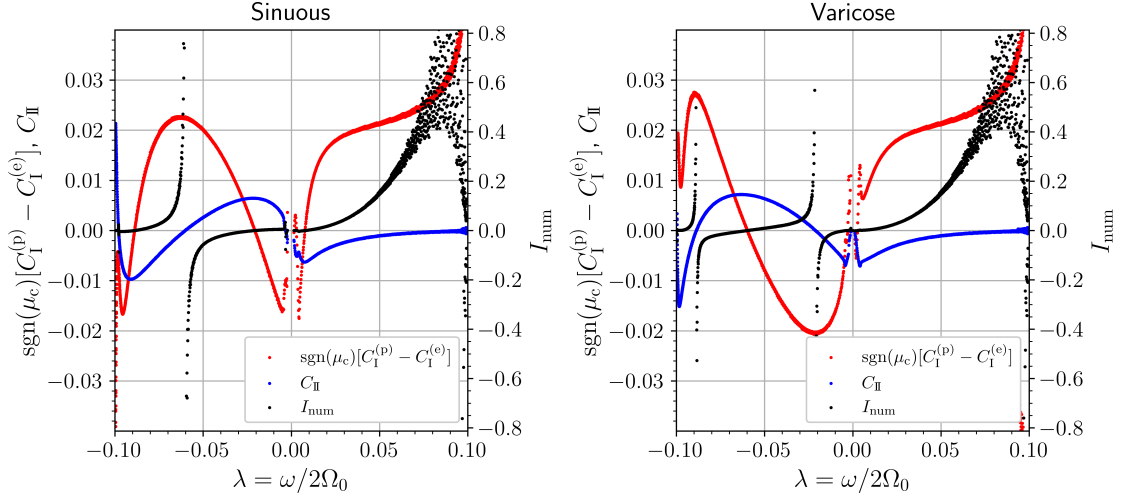}}%
	\caption{Values of $\operatorname{sgn}(\mu_\mathrm{c})[C_\mathrm{I}^{(\mathrm{p})}-C_\mathrm{I}^{(\mathrm{e})}]$ (red circles and the left vertical axes), $C_\mathrm{I\!I}$ (blue circles and the left vertical axes), and $I_\mathrm{num}$ defined as \eqref{DEF_integral} (black circles and the right vertical axes), relevant to the coefficients of the Frobenius series solutions against the nondimensional angular frequency $\lambda$. The case where the zonal wavenumber $m=1$, the absolute value of the Lehnert number $|\alpha|=0.1$, and the simplest equatorially antisymmetric non-Malkus field $\mathcal{B}=\mu$ is imposed is shown in the left and the right panels for sinuous and varicose modes, respectively. $N_\theta$ and $N_\mathrm{data}$ are set to $7201$ and $200$, respectively, in the same way as Figure \ref{FIG_frobenius}.}
	\label{FIG_allfrobenius}
\end{center}
\end{figure}

The above procedure is also applied to all the continuous modes obtained from our numerical calculations. Figure \ref{FIG_allfrobenius} depicts their values of $\operatorname{sgn}(\mu_\mathrm{c})[C_\mathrm{I}^{(\mathrm{p})}-C_\mathrm{I}^{(\mathrm{e})}]$ ($=C_\mathrm{I\!I\!I}$, red circles) and $C_\mathrm{I\!I}$ (blue circles) for $m=1$ and $|\alpha|=0.1$ in the left ordinates as functions of $\lambda$. Their results for the sinuous and varicose modes are shown in the left and right panels, respectively. Meanwhile, the right vertical axes of these panels represent the numerical counterpart of \eqref{DEF_detour}, which is written in the present instance as
\begin{equation}
I_\mathrm{num}\,\equiv\,-\operatorname{sgn}(\mu_\mathrm{c})\frac{C_\mathrm{I}^{(\mathrm{p})}-C_\mathrm{I}^{(\mathrm{e})}}{C_\mathrm{I\!I}}(1-\mu_{\mathrm{c}}^{2})m^2\alpha^2(\mathcal{B}_{\mathrm{c}}^2)'\left[\tilde{\psi}_\mathrm{I}^{(\mathrm{c})}(\mu_{\mathrm{c}})\right]^2\,.
\label{DEF_integral}
\end{equation}
This outcome demonstrates that the values of $I_\mathrm{num}$ appear to be compatible with our expectation stated in Section \ref{SUBSEC_existence}; the values are arbitrary numbers and adjusted to satisfy boundary conditions. In addition, we observe that the value of $\operatorname{sgn}(\mu_\mathrm{c})[C_\mathrm{I}^{(\mathrm{p})}-C_\mathrm{I}^{(\mathrm{e})}]$ vanishes at the extremum points of $C_\mathrm{I\!I}$, and that $\operatorname{sgn}(\mu_\mathrm{c})[C_\mathrm{I}^{(\mathrm{p})}-C_\mathrm{I}^{(\mathrm{e})}]$ has extremums at the zeros of $C_\mathrm{I\!I}$. If discrete eigenmodes without logarithmic and step function singularities are buried in the continuum, the two values must simultaneously vanish. In fact, we attempted to use these values as a means to reveal buried discrete eigenmodes, though no such eigenmodes have been found.\par
%

% = = = = = = = = = = = = = = = = = = = = = = = = = = = = = = %

\section{Derivations of the ray-tracing equations and their related formulae}\label{SEC_derivations}
%%%%%
This appendix explains detailed derivations of the ray-tracing equations and their related formulae, which were skipped in Section \ref{SEC_discussion}. We first introduce the parameter given by $s\equiv-\operatorname{sgn}\left(\ln|\alpha|\right)$. Then, $|\alpha|^s\leq1$ is always satisfied regardless of the magnitude of $|\alpha|$. If we only consider the situation where either $|\alpha|\gg1$ or $|\alpha|\ll1$, the parameter $|\alpha|^s$ becomes sufficiently smaller than unity. Utilising this small parameter, we can express the separations of the spatial and the temporal scales, as explained in Section \ref{SEC_discussion}. In other words, with the local coordinates $(\varPhi, \varTheta, \tau)$ and the global ones $(\phi, \theta, T)$, we can write
\begin{equation}
\varTheta\,\equiv\,|\alpha|^{-sp}\theta\,,\qquad\varPhi\,\equiv\,|\alpha|^{-sp}\phi\,,\qquad T\,\equiv\,|\alpha|^{sq}\tau\,,
\label{DEF_rescale_spatial_temporal_general}
\end{equation}
where $p$, $q>0$.\par
%%%%%
In such separations, the local coordinates are supposed to be suitable to measure the phase $\varphi_\mathrm{L}$ of waves. The local wavenumber and local nondimensional angular frequency, which may depend only on the global coordinates, are accordingly defined as
\begin{subequations}\label{DEF_wavenumber_frequency_general}
\begin{linenomath}\begin{align}
k(\phi,\theta,T)\,&\equiv\,\frac{1}{\sin\theta}\frac{\upartial\varphi_\mathrm{L}}{\upartial\varPhi}\,=\,\frac{1}{\sin\theta}\frac{\upartial(|\alpha|^{sp}\varphi_\mathrm{G})}{\upartial\phi}\,=\,|\alpha|^{sp}\frac{m}{\sin\theta}\,,
\label{DEF_k_general}\\
l(\phi,\theta,T)\,&\equiv\,\frac{\upartial\varphi_\mathrm{L}}{\upartial(-\varTheta)}\,=\,\frac{\upartial(|\alpha|^{sp}\varphi_\mathrm{G})}{\upartial(-\theta)}\,,
\label{DEF_l_general}\\
\lambda(\phi,\theta,T)\,&\equiv\,-\frac{\upartial\varphi_\mathrm{L}}{\upartial\tau}\,=\,-|\alpha|^{s(q-p)}\frac{\upartial(|\alpha|^{sp}\varphi_\mathrm{G})}{\upartial T}\,,
\label{DEF_frequency_general}
\end{align}\end{linenomath}
\end{subequations}
where $\varphi_\mathrm{G}(\phi, \theta, T)$ ($=\varphi_\mathrm{L}(\varPhi,\varTheta,\tau)$) denotes the phase function whose independent variables are the global coordinates. Now we recognise that $k$, $l=\mathrm{O}(|\alpha|^0)$ by construction, and that the zonal wavenumber $m=\mathrm{O}(|\alpha|^{-sp})$ measured in the global coordinates is large. This implies that $\lambda=\mathrm{O}(|\alpha|^{s(q-p)})$, and we introduce the scaled nondimensional angular frequency $\lambda_\mathrm{s}\equiv|\alpha|^{s(p-q)}\lambda$. In addition, \eqref{DEF_wavenumber_frequency_general} yield the curl-free conditions
\begin{equation}
\frac{\upartial(k\sin\theta)}{\upartial(-\theta)}\,=\,\frac{\upartial l}{\upartial\phi}\,,\qquad-\frac{\upartial (k\sin\theta)}{\upartial T}\,=\,\frac{\upartial \lambda_\mathrm{s}}{\upartial \phi}\,,\qquad-\frac{\upartial l}{\upartial T}\,=\,\frac{\upartial \lambda_\mathrm{s}}{\upartial(-\theta)}\,.
\label{DEF_wavenumber_frequency_relation_general}
\end{equation}
If we consider $(\phi,-\theta)$ to be generalised coordinates in analytical mechanics, $(k\sin\theta, l)$ and $|\alpha|^{sp}\varphi_\mathrm{G}$, from \eqref{DEF_k_general} and \eqref{DEF_l_general}, correspond to generalised momentums and a generating function, respectively. The first equation of \eqref{DEF_wavenumber_frequency_relation_general} is also equivalent to the property $\{k\sin\theta, l\}=0$ of the Poisson bracket. On the basis of these facts, we may raise the topic of the Hamilton--Jacobi equation 
\begin{equation}
\frac{\upartial(|\alpha|^{sp}\varphi_\mathrm{G})}{\upartial T}\,+\,\lambda_\mathcal{H}(\phi,\theta,k,l,T)\,=\,0\,,
\label{DEF_Hamilton-Jacobi}
\end{equation}
where $\lambda_\mathcal{H}$ is the ``Hamiltonian'' for our problem. We however will not pursue the solution $|\alpha|^{sp}\varphi_\mathrm{G}$ of this equation. Upon comparing \eqref{DEF_frequency_general} to \eqref{DEF_Hamilton-Jacobi} and treating $\lambda_\mathrm{s}$ as an independent variable, we obtain a dispersion relation in the form
\begin{equation}
\mathcal{D}_\mathcal{H}(\phi,\theta,T,k,l,\lambda_\mathrm{s})\,\equiv\,\lambda_\mathrm{s}-\lambda_\mathcal{H}(\phi,\theta,k,l,T)\,=\,0\,.
\label{DEF_dispersion_lambda_general}
\end{equation}
The nondimensional local group velocity with respect to the dimensionless time $\tau$ and the coordinates $(\phi, -\theta)$ is then written as
\begin{subequations}\label{DEF_group_general}
\begin{linenomath}\begin{align}
c_{\mathrm{g},\phi}\,&\equiv\,-\frac{[\upartial \mathcal{D}_\mathcal{H}/\upartial (|\alpha|^{-sp}k)]|_{\lambda_\mathrm{s}=\lambda_\mathcal{H}}}{[\upartial \mathcal{D}_\mathcal{H}/\upartial (|\alpha|^{s(q-p)}\lambda_\mathrm{s})]|_{\lambda_\mathrm{s}=\lambda_\mathcal{H}}}\,=\,|\alpha|^{sq}\frac{\upartial\lambda_\mathcal{H}}{\upartial k}\,,\\
c_{\mathrm{g},-\theta}\,&\equiv\,-\frac{[\upartial \mathcal{D}_\mathcal{H}/\upartial (|\alpha|^{-sp}l)]|_{\lambda_\mathrm{s}=\lambda_\mathcal{H}}}{[\upartial \mathcal{D}_\mathcal{H}/\upartial (|\alpha|^{s(q-p)}\lambda_\mathrm{s})]|_{\lambda_\mathrm{s}=\lambda_\mathcal{H}}}\,=\,|\alpha|^{sq}\frac{\upartial\lambda_\mathcal{H}}{\upartial l}\,.
\end{align}\end{linenomath}
\end{subequations}
Instead of the Hamilton--Jacobi equation \eqref{DEF_Hamilton-Jacobi}, we will solve the ray-tracing equations, which are the counterparts of Hamilton's equations in analytical mechanics, in the forms
\begin{subequations}\label{EQ_ray_general}
\begin{equation}
\sin\theta\frac{\ud_\mathrm{g}\phi}{\ud T}\,=\,\frac{\upartial\lambda_\mathcal{H}}{\upartial k}\,=\,|\alpha|^{-sq}c_{\mathrm{g},\phi}\,,\qquad\frac{\ud_\mathrm{g}(-\theta)}{\ud T}\,=\,\frac{\upartial\lambda_\mathcal{H}}{\upartial l}\,=\,|\alpha|^{-sq}c_{\mathrm{g},-\theta}\,,
\label{EQ_ray_phi_theta_general}
\end{equation}
\begin{equation}
\frac{\ud_\mathrm{g}(k\sin\theta)}{\ud T}\,=\,-\left(\frac{\upartial \lambda_\mathcal{H}}{\upartial\phi}\right)_{k,l}\,,\qquad\frac{\ud_\mathrm{g}l}{\ud T}\,+\,|\alpha|^{-sq}c_{\mathrm{g},\phi}k\cot\theta\,=\,-\left[\frac{\upartial \lambda_\mathcal{H}}{\upartial(-\theta)}\right]_{k,l}\,,
\label{EQ_ray_kl_general}
\end{equation}
\begin{equation}
\frac{\ud_\mathrm{g}\lambda_\mathrm{s}}{\ud T}\,=\,\left(\frac{\upartial \lambda_\mathcal{H}}{\upartial T}\right)_{k,l}\,,
\label{EQ_ray_lambda_general}
\end{equation}
\end{subequations}
where the material time derivative moving with the local group velocity is expressed as $(\ud_\mathrm{g}/\ud T)\equiv(\upartial/\upartial T)+|\alpha|^{-sq}(\bm{c}_\mathrm{g}\bm{\cdot}\bm{\nabla}_\mathrm{G})$. Using the chain rule, one can also get \eqref{EQ_ray_kl_general} and \eqref{EQ_ray_lambda_general} from \eqref{DEF_wavenumber_frequency_general}, \eqref{DEF_wavenumber_frequency_relation_general}, \eqref{DEF_Hamilton-Jacobi}, and \eqref{DEF_group_general}.\par
%%%%%
The difference between our ray-tracing approach and standard analytical mechanics is that we do not initially know an explicit expression of the Hamiltonian. To obtain it, we need to substitute the ansatz $\psi_1\equiv\mathrm{Re}[M(\phi,\theta,T)\mathrm{e}^{\mathrm{i}\varphi_\mathrm{L}(\varPhi,\varTheta,\tau)}]$ into the original partial differential equation
\begin{equation}
\left(\frac{\upartial^2}{\upartial \tau^2}\nabla_\mathrm{h}^2+\frac{\upartial^2}{\upartial\tau\upartial\phi}\right)\psi_1\,=\,\alpha^2\left[\mathcal{B}^2\frac{\upartial^2}{\upartial\phi^2}\nabla_\mathrm{h}^2+\frac{\ud\mathcal{B}^2}{\ud\theta}\frac{\upartial^3}{\upartial\phi^2\upartial\theta}-\frac{2\mathcal{B}}{\sin\theta}\frac{\ud(\mathcal{B}\cos\theta)}{\ud\theta}\frac{\upartial^2}{\upartial\phi^2}\right]\psi_1\,,
\label{EQ_differential_general}
\end{equation}
which is constructed from the perturbation equations \eqref{EQ_linear}. For an arbitrary function $f(\phi,\theta,T)$, the relation
\begin{equation}
\frac{1}{\sin\theta}\frac{\upartial(fM\mathrm{e}^{\mathrm{i}\varphi_\mathrm{L}})}{\upartial\phi}\,=\,\left[\frac{1}{\sin\theta}\frac{\upartial f}{\upartial\phi}+f\left(\frac{1}{\sin\theta}\frac{\upartial\ln M}{\upartial\phi}+\mathrm{i}|\alpha|^{-sp}k\right)\right]M\mathrm{e}^{\mathrm{i}\varphi_\mathrm{L}}\,,
\label{EQ_derivative_ansatz}
\end{equation}
and so forth, and nests of these are useful in the above tedious substitution. Then, we find that the leading order terms of each term in \eqref{EQ_differential_general} are $\mathrm{O}(|\alpha|^{2s(q-2p)})$, $\mathrm{O}(|\alpha|^{s(q-2p)})$, $\mathrm{O}(|\alpha|^{2-4sp})$, $\mathrm{O}(|\alpha|^{2-3sp})$, and $\mathrm{O}(|\alpha|^{2-2sp})$ from the first term on the left-hand side to the third term on the right-hand side. A plausible balance among them requires that 
\begin{equation}
2(q-2p)\,=\,q-2p\,=\,2s-4p\quad(\,<\,2s-3p\,<\,2s-2p\,).
\label{EQ_balance}
\end{equation}
We then have $s=1$ ($|\alpha|\ll1$), $p=1/2$, and $q=1$. Because possible balances for $s=-1$ do not exist, our approximation is probably inappropriate for the case where $|\alpha|\gg1$. It follows that the leading order terms $\mathrm{O}(|\alpha|^0)$ in \eqref{EQ_differential_general} yield the local dispersion relation \eqref{DEF_local_dispersion}, or $\mathcal{D}(\phi, \theta, T, k, l, \lambda_\mathrm{s})=0$. Solving this relation for $\lambda_\mathrm{s}$, we can also obtain \eqref{DEF_dispersion_lambda_general}. Additionally, the next-to-leading order terms $\mathrm{O}(|\alpha|^{1/2})$ are written as
\begin{linenomath}\begin{align}
\frac{\upartial \ln M}{\upartial T}\frac{\upartial\mathcal{D}}{\upartial \lambda_\mathrm{s}}\,&-\,\frac{1}{\sin\theta}\frac{\upartial \ln M}{\upartial\phi}\frac{\upartial\mathcal{D}}{\upartial k}\,-\,\frac{\upartial \ln M}{\upartial(-\theta)}\frac{\upartial\mathcal{D}}{\upartial l}\,=\,\notag\\
\,&-\,\frac{\upartial[\lambda_\mathrm{s}(k^2+l^2)]}{\upartial T}\,-\,\lambda_\mathrm{s}\frac{\upartial(k^2+l^2)}{\upartial T}\,+\,\frac{\lambda_\mathrm{s}^2}{\sin\theta}\left[\frac{\upartial k}{\upartial\phi}+\frac{\upartial(l\sin\theta)}{\upartial(-\theta)}\right]\,-\,\frac{\upartial(k\sin\theta)}{\upartial T}\notag\\
\,&-\,\frac{1}{\sin\theta}\frac{\upartial[k\mathcal{B}^2\sin^2\theta(k^2+l^2)]}{\upartial\phi}\,-\,\frac{k}{\sin\theta}\frac{\upartial[\mathcal{B}^2\sin^2\theta(k^2+l^2)]}{\upartial\phi}\notag\\
\,&-\,\frac{k^2\mathcal{B}^2\sin^2\theta}{\sin\theta}\left[\frac{\upartial k}{\upartial\phi}+\frac{\upartial(l\sin\theta)}{\upartial(-\theta)}\right]\,-\,k^2l\frac{\ud \mathcal{B}^2}{\ud(-\theta)}\sin^2\theta\,.
\label{EQ_amplitude}
\end{align}\end{linenomath}
It should be noted that this equation is complex at a glance, but conforms to the same rule as the structure that the counterpart in \citet{https://doi.org/10.1002/qj.49709239403} has. At last, the novel conservation law \eqref{EQ_invariant} is obtained from \eqref{DEF_group}, \eqref{DEF_wavenumber_frequency_relation_general}, and \eqref{EQ_amplitude} with \eqref{DEF_dispersion_lambda_general}.\par
%%%%%
In the remainder of this appendix and Appendix \ref{SEC_behaviors}, we shall set about the steady problem in which wave trains with specified values of $k\sin\theta$ and $\lambda_\mathrm{s}$ uniformly continue to be injected on a line of latitude. The discussion of the problem will provide the behaviours of the steady trains near critical latitudes, where $l^2\to\infty$. Its findings are also helpful in learning the behaviours of the trains near turning latitudes, at which $l=0$ (see Appendix \ref{SEC_behaviors}). First of all, consider a region sandwiched between two ray trajectories drawn by the trains injected at two distinct longitudes ($\phi_1$ and $\phi_2$, say). This region may be called a ``ray tube''. It should be noted that, because of the symmetry about the axis of rotation, one of the two trajectories can overlap the other by the rotation $|\phi_1-\phi_2|$ about the axis. The distance between the two trajectories along a line of a colatitude $\theta$ is, therefore, $|\phi_1-\phi_2|\sin\theta$. We then prepare two cross-sections of the tube along two lines of latitude. Let $S$ be the region bounded by the two sections on the tube. For the steady problem with the uniform injection, the surface integral of \eqref{EQ_invariant} over $S$ gives
\begin{equation}
\frac{c_{\mathrm{g},-\theta}}{|\alpha|}\frac{\upartial\mathcal{D}}{\upartial \lambda_\mathrm{s}}|M|^2\sin\theta\,=\,-\frac{\upartial\mathcal{D}}{\upartial l}|M|^2\sin\theta\,=\,\text{const.}\,.
\label{EQ_invariant_steady}
\end{equation}
Then, we obtain $|M|^2\propto[l(\lambda_\mathrm{s}^2-k^2\mathcal{B}^2\sin^2\theta)\sin\theta]^{-1}$, that is, $M=\mathrm{O}(|\theta-\theta_\mathrm{c}|^{-1/4})$ near a critical colatitude $\theta_\mathrm{c}$ since $l=\mathrm{O}(|\theta-\theta_\mathrm{c}|^{-1/2})$ (from \eqref{EQ_local_dispersion_critical}) and $\lambda_\mathrm{s}^2-k^2\mathcal{B}^2\sin^2\theta=\mathrm{O}(|\theta-\theta_\mathrm{c}|)$. This means that steady trains jam into such a latitude and that their amplitudes increase there. It should be noted that the same is true for an isolated packet, as shown in the last paragraph of Section \ref{SEC_discussion}.\par
%%%%%
For constant values of $k\sin\theta$ and $\lambda_\mathrm{s}$, we can get the same expression for $|M|$ as the above in a different way. The solution is again redefined as $\psi_1\equiv\mathrm{Re}[\tilde{\psi}(\theta)\mathrm{e}^{\mathrm{i}\varphi}]$ with $\varphi=|\alpha|^{-1/2}(k\sin\theta)\phi-|\alpha|^{1/2}\lambda_\mathrm{s}\tau$, where we conjecture that $\tilde{\psi}$ can be expressed by the WKBJ (Wentzel--Kramers--Brillouin--Jeffreys) form
\begin{equation}
\tilde{\psi}\,\equiv\,\exp\left[-\mathrm{i}\int^\theta L(\theta_*)\ud\theta_*\right]\,.
\label{DEF_wkbj}
\end{equation}
Consequently, we have $(\ud\tilde{\psi}/\ud(-\theta))=\mathrm{i}L\tilde{\psi}$. Provided that $L=|\alpha|^{-1/2}l+L_0+\mathrm{O}(|\alpha|^{1/2})$, the leading order terms $\mathrm{O}(|\alpha|^0)$ in the ordinary differential equation \eqref{EQ_differential}, which can be written as
\begin{equation}
\begin{aligned}
&|\alpha|\frac{\lambda_\mathrm{s}^2-k^2\mathcal{B}^2\sin^2\theta}{\sin\theta}\frac{\ud}{\ud\theta}\left(\sin\theta\frac{\ud\tilde{\psi}}{\ud\theta}\right)\,-\,|\alpha|k^2\sin^2\theta\frac{\ud\mathcal{B}^2}{\ud\theta}\frac{\ud\tilde{\psi}}{\ud\theta}\\
&\qquad\,-\,\left[k^2(\lambda_\mathrm{s}^2-k^2\mathcal{B}^2\sin^2\theta)+\lambda_\mathrm{s}k\sin\theta-2|\alpha|k^2\sin\theta\mathcal{B}\frac{\ud(\mathcal{B}\cos\theta)}{\ud\theta}\right]\tilde{\psi}\,=\,0\,
\end{aligned}
\label{EQ_differential2}
\end{equation}
if we use $k$ and $\theta$ instead of $m$ and $\mu$, lead to the local dispersion relation \eqref{DEF_local_dispersion}. Its next-to-leading order terms $\mathrm{O}(|\alpha|^{1/2})$ become the equation for $L_0$ in the form
\begin{equation}
\mathrm{i}L_0\,=\,\frac{1}{2}\frac{\ud \ln|l(\lambda_\mathrm{s}^2-k^2\mathcal{B}^2\sin^2\theta)\sin\theta|}{\ud\theta}
\end{equation}
when $l\neq0$ and $\lambda_\mathrm{s}^2\neq k^2\mathcal{B}^2\sin^2\theta$, resulting in
\begin{linenomath}\begin{align}
\tilde{\psi}\,\propto\,\frac{1}{|l(\lambda_\mathrm{s}^2-k^2\mathcal{B}^2\sin^2\theta)\sin\theta|^{1/2}}\exp\left\{-\mathrm{i}\int^\theta\left[|\alpha|^{-1/2}l(\theta_*)+\mathrm{O}(|\alpha|^{1/2})\right]\ud\theta_*\right\}\,.
\label{EQ_wkbj_leading}
\end{align}\end{linenomath}
The pre-exponent of this WKBJ solution is consistent with $|M|$ in \eqref{EQ_invariant_steady}, and $\varphi_\mathrm{G}$ in \eqref{DEF_wavenumber_frequency_general} is equivalent to $\varphi-|\alpha|^{-1/2}\int^\theta l(\theta_*)\ud\theta_*$ as long as we pay attention to the solutions with constant values of $k\sin\theta$ and $\lambda_\mathrm{s}$.

% = = = = = = = = = = = = = = = = = = = = = = = = = = = = = = %

\section{Behaviours of wave packets near a turning latitude}\label{SEC_behaviors}
%%%%%
The last appendix is devoted to an investigation into the behaviours of wave packets near turning latitudes, where $l=0$, when $|\alpha|\ll1$ (see also Section \ref{SEC_discussion} and Appendix \ref{SEC_derivations}). We here write $l^2=(l_\mathrm{t}^2)'(\theta-\theta_\mathrm{t})+\mathrm{O}(|\theta-\theta_\mathrm{t}|^{2})$ in the vicinity of a turning colatitude $\theta=\theta_\mathrm{t}$, where $(l_\mathrm{t}^2)'\equiv(\ud l^2/\ud\theta)|_{\theta=\theta_\mathrm{t}}$ (and we assume that $(l_\mathrm{t}^2)'\neq0$). If a packet advances toward its corresponding turning latitude from a nearby latitude $\theta$, it can then reach there in a finite time
\begin{linenomath}\begin{align}
\int_\theta^{\theta_\mathrm{t}}\frac{|\alpha|}{-c_{\mathrm{g},-\theta}(\theta_*)}\ud\theta_*\,&\simeq\,\pm\left.\frac{2\lambda_\mathrm{s}k^2+k\sin\theta}{2|(l_\mathrm{t}^2)'|^{1/2}(\lambda_\mathrm{s}^2-k^2\mathcal{B}^2\sin^2\theta)}\right|_{\theta=\theta_\mathrm{t}}\int_\theta^{\theta_\mathrm{t}}\frac{\ud\theta_*}{\left\{\operatorname{sgn}\left[(l_\mathrm{t}^2)'\right](\theta_*-\theta_\mathrm{t})\right\}^{1/2}}\notag\\
\,&=\,\mathrm{O}(|\theta-\theta_\mathrm{t}|^{1/2})\,,
\label{EQ_time_turning}
\end{align}\end{linenomath}
and its refraction will occur. The asymptotic expression $M=\mathrm{O}(|\theta-\theta_\mathrm{t}|^{-1/4})$ of its amplitude near the turning latitude is obtained from $l=\mathrm{O}(|\theta-\theta_\mathrm{t}|^{1/2})$ and \eqref{EQ_invariant_steady}. In addition, the valid range \eqref{EQ_wkbj_validity} of the ray theory is given by $|\theta-\theta_\mathrm{t}|\gg\mathrm{O}(|\alpha|^{1/3})$. This tempts us into introducing a new stretched coordinate $y\equiv|\alpha|^{-1/3}\theta$ (with $y_\mathrm{t}\equiv|\alpha|^{-1/3}\theta_\mathrm{t}$). It is to be noted that the scale of $y$ is a little coarser than that of $\varTheta$, since $\theta=|\alpha|^{1/3}y=|\alpha|^{1/2}\varTheta$. The WKBJ solutions \eqref{EQ_wkbj_leading} are then valid in the range $\mathrm{O}(|\alpha|^0)\ll|y-y_\mathrm{t}|$; in particular, the solutions near the turning latitude within the range $\mathrm{O}(|\alpha|^0)\ll|y-y_\mathrm{t}|\ll\mathrm{O}(|\alpha|^{-1/3})$ are written as
\begin{equation}
\tilde{\psi}\,\simeq\,\frac{C}{|\alpha|^{1/12}|(l_\mathrm{t}^2)'|^{1/6}Y^{1/4}|(\lambda_\mathrm{s}^2-k^2\mathcal{B}^2\sin^2\theta)\sin\theta|^{1/2}}\exp\left\{-\mathrm{i}\frac{2}{3}\operatorname{sgn}\left[l(l_\mathrm{t}^2)'\right]\left(Y^{3/2}-Y_0^{3/2}\right)\right\}\,,
\label{EQ_wkbj_leading_turning}
\end{equation}
where $C$ and $Y_0$ are constants and $Y\equiv|(l_\mathrm{t}^2)'|^{-2/3}(l_\mathrm{t}^2)'(y-y_\mathrm{t})$. The signs of $(l_\mathrm{t}^2)'$ and the meridional component $|\alpha|(\lambda_\mathrm{s}/l)$ of the phase velocity determine whether the packet is the incident ($l(l_\mathrm{t}^2)'/\lambda_\mathrm{s}>0$) or reflected waves ($l(l_\mathrm{t}^2)'/\lambda_\mathrm{s}<0$). It should be noted that the scaled stretched coordinate $Y$ is always positive in the oscillatory region $l^2>0$.\par
%%%%%
When $y-y_\mathrm{t}=\mathrm{O}(|\alpha|^0)$, the above approximation breaks down. This is because $\mathrm{i}L_0\simeq(1/2)(\ud\ln|l|/\ud\theta)=\mathrm{O}(|\theta-\theta_\mathrm{t}|^{-1})$ is of the same order of magnitude as $|\alpha|^{-1/2}l$. To match an incident wave with its corresponding reflected one at its corresponding turning latitude $y=y_\mathrm{t}$, we need to return to the original equation \eqref{EQ_differential2}. Without introducing \eqref{DEF_wkbj}, we substitute the Taylor expansion of \eqref{EQ_l2} around the turning latitude in the form
\begin{equation}
-k^2-\frac{\lambda_\mathrm{s}k\sin\theta}{\lambda_\mathrm{s}^2-k^2\mathcal{B}^2\sin^2\theta}\,=\,(l_\mathrm{t}^2)'(\theta-\theta_\mathrm{t})\,+\,\mathrm{O}(|\theta-\theta_\mathrm{t}|^{2})
\end{equation}
into \eqref{EQ_differential2}, and use the variable $Y$ to take the dominant terms at $\theta-\theta_\mathrm{t}=\mathrm{O}(|\alpha|^{1/3})$. We thus obtain
\begin{equation}
\frac{\ud^2\tilde{\psi}}{\ud(-Y)^2}\,=\,(-Y)\tilde{\psi}\,+\,\mathrm{O}(|\alpha|^{1/3})\,,
\end{equation}
which is the Airy differential equation \citep[e.g.][]{abramowitz1964handbook}. Because $\tilde{\psi}$ should vanish as $Y\to-\infty$ (evanescent), the equation has $\tilde{\psi}=C_\mathrm{A}\mathrm{Ai}(-Y)$ as its solution, where $\mathrm{Ai}$ denotes the Airy function of the first kind and $C_\mathrm{A}$ is a constant. We now require that the asymptotic expression for this solution as $Y\to+\infty$ ($\ll\mathrm{O}(|\alpha|^{-1/3})$), which is given by
\begin{equation}
\mathrm{Ai}(-Y)\,\sim\,\frac{1}{\sqrt{\upi}Y^{1/4}}\cos\left(\frac{2}{3}Y^{3/2}-\frac{\upi}{4}\right)\qquad(Y\to+\infty)\,,
\label{EQ_airy}
\end{equation}
should match the sum $\tilde{\psi}_\text{i}+\tilde{\psi}_\text{r}$ of the incident ($C=C_\text{i}$, say) and reflected ($C=C_\text{r}$, say) waves \citep{lighthill1978waves}. It follows that, if $C_\text{i}=C_\text{r}$ (total reflection), we have $(2/3)Y_0^{3/2}=N_0\upi+(\upi/4)$ ($N_0=0,\pm1,\pm2,\ldots$) and 
\begin{equation}
\frac{C_\mathrm{A}}{C_\mathrm{i}}\,=\,\frac{2\sqrt{\upi}(-1)^{N_0}}{|\alpha|^{1/12}|(l_\mathrm{t}^2)'|^{1/6}|(\lambda_\mathrm{s}^2-k^2\mathcal{B}^2\sin^2\theta)\sin\theta|^{1/2}}\,.
\label{EQ_matching}
\end{equation}
These expressions constitute consistent solutions to the incident, reflected and evanescent waves across a turning latitude.\par
\begin{figure}
\begin{center}
	\resizebox*{75mm}{!}{\includegraphics{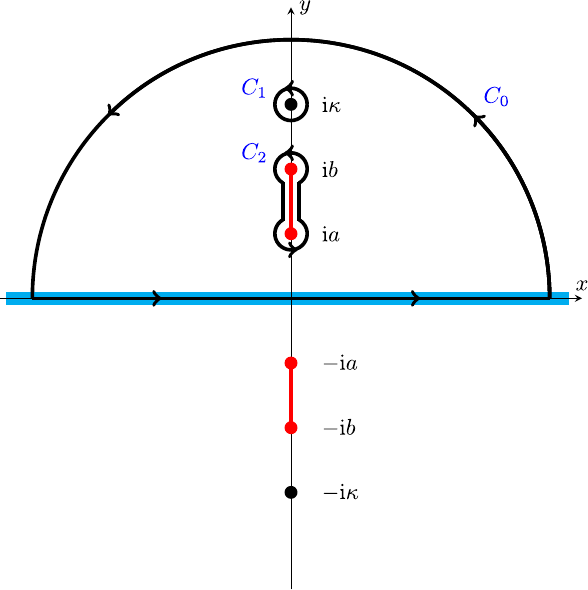}}%
	\caption{Path of integration for the integral \eqref{EQ_elliptic}. The contours $C_0$, $C_1$ and $C_2$ are represented as thick black curves. The black and red circles are poles and branch points, respectively, and the red lines correspond to branch cuts. The integral in \eqref{EQ_quantization_sincos} must be evaluated along the path represented as the cyan line.}
	\label{FIG_contour}
\end{center}
\end{figure}
\begin{figure}
\begin{center}
	\resizebox*{150mm}{!}{\includegraphics{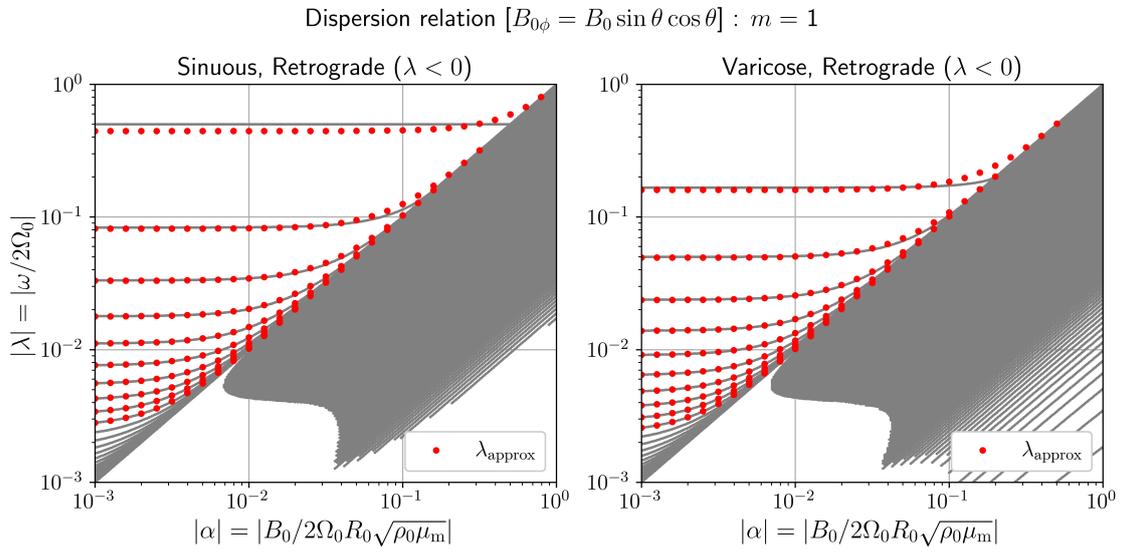}}%
	\caption{Same as Figure \ref{FIG_rossby}, but for the approximate dimensionless angular frequencies $\lambda_\mathrm{approx}$ calculated by the approximate dispersion relation \eqref{EQ_quantization_sincos} instead of \eqref{EQ_dispersion_fast}.}
	\label{FIG_rossby_wkbj}
\end{center}
\end{figure}

%%%%%
We have a by-product of the above study of the matching of solutions. The expression of the constant $Y_0$ of integration, or the phase shift $\upi/4$ in \eqref{EQ_airy}, implies the existence of the quantization condition of waves that have no critical latitudes (which include waves under the Malkus background field $\mathcal{B}=1$, and fast MR waves for the simplest equatorially antisymmetric non-Malkus field $\mathcal{B}=\cos\theta$). The condition is expressed as
\begin{equation}
|\alpha|^{-1/2}\int_{\theta_{\mathrm{t}1}}^{\theta_{\mathrm{t}2}}|l|\ud\theta\,=\,\int_{\theta_{\mathrm{t}1}}^{\theta_{\mathrm{t}2}}\sqrt{-\frac{m\lambda}{\varLambda}-\frac{m^2}{\sin^2\theta}}\ud\theta\,=\,\left(N+\frac{1}{2}\right)\upi\quad(N=0,1,2,\ldots)\,,
\label{EQ_quantization}
\end{equation}
where $l^2>0$ throughout the range $\theta_{\mathrm{t}1}<\theta<\theta_{\mathrm{t}2}$ between two different turning colatitudes $\theta_{\mathrm{t}1}$ and $\theta_{\mathrm{t}2}$. For $\mathcal{B}=1$, we use the substitution $x\equiv\cos\theta/\sqrt{\sin^2\theta-(\lambda^2-m^2\alpha^2)/(-\lambda/m)}$ to calculate the integral, getting
\begin{equation}
\upi\left(\sqrt{-\frac{m\lambda}{\lambda^2-m^2\alpha^2}}-|m|\right)\,=\,\left(N+\frac{1}{2}\right)\upi\,.
\label{EQ_quantization_Malkus}
\end{equation}
Letting $n\equiv |m|+N$, we obtain an approximate formula for the dispersion relation \eqref{EQ_dispersion_Malkus} in the form
\begin{equation}
\lambda\,=\,\frac{-m\pm m\sqrt{1+4\alpha^2[n(n+1)+(1/4)]^2}}{2[n(n+1)+(1/4)]}\,.
\label{EQ_dispersion_quantization_Malkus}
\end{equation}
As expected, the larger the value of $n$ is, the more accurately the above formula approximates the relation \eqref{EQ_dispersion_Malkus}. Next, we proceed to the case where $\mathcal{B}=\cos\theta$. If we perform the substitution $x\equiv\cos\theta/\sqrt{\sin^2\theta-\kappa^{-2}}$, where $\kappa^2\equiv(-\lambda/m-m^2\alpha^2)/(\lambda^2-m^2\alpha^2)$, the condition \eqref{EQ_quantization} becomes
\begin{equation}
(1+m\lambda)\sqrt{-\frac{m\lambda}{\lambda^2-m^2\alpha^2}}\kappa^2\int_{-\infty}^\infty\frac{\ud x}{(x^2+\kappa^2)\sqrt{(x^2+1)(x^2+(-m\lambda)\kappa^2)}}\,=\,\left(N+\frac{1}{2}\right)\upi\,.
\label{EQ_quantization_sincos}
\end{equation}
It should be noted that, in the above equation, either $1<(-m\lambda)\kappa^2<\kappa^2$ or $0<(-m\lambda)\kappa^2<1<\kappa^2$ holds, because we find that $1+m\lambda>0$ during the derivation of \eqref{EQ_quantization_sincos}. To evaluate the integral in \eqref{EQ_quantization_sincos}, we consider the integral of a complex-valued function of $z$ ($\equiv x+\mathrm{i}y$) in the form
\begin{linenomath}\begin{align}
&\oint_{C_0}\frac{\ud z}{(z^2+\kappa^2)\sqrt{(z^2+a^2)(z^2+b^2)}}\notag\\
\,&=\,\oint_{C_1+C_2}\frac{\ud z}{(z^2+\kappa^2)\sqrt{(z^2+a^2)(z^2+b^2)}}\notag\\
\,&=\,2\upi\mathrm{i}\,\mathrm{Res}\left(\frac{1}{(z^2+\kappa^2)\sqrt{(z^2+a^2)(z^2+b^2)}}, \mathrm{i}\kappa\right)\,+\,2\int_a^b\frac{\ud y}{(\kappa^2-y^2)\sqrt{(y^2-a^2)(b^2-y^2)}}\notag\\
\,&=\,-\frac{\upi}{\kappa\sqrt{(\kappa^2-a^2)(\kappa^2-b^2)}}\notag\\
\,&\qquad+\,\frac{2}{\kappa^2b}\int_0^1\frac{\ud v}{\sqrt{(1-v^2)(1-\gamma v^2)}}\,+\,\frac{2a^2}{\kappa^2b(\kappa^2-a^2)}\int_0^1\frac{\ud v}{(1-\chi v^2)\sqrt{(1-v^2)(1-\gamma v^2)}}\,,
\label{EQ_elliptic}
\end{align}\end{linenomath}
where $a\equiv\min(1,\sqrt{-m\lambda}\kappa)$, $b\equiv\max(1,\sqrt{-m\lambda}\kappa)$, $\gamma\equiv1-(a/b)^2$, and $\chi\equiv \gamma\kappa^2/(\kappa^2-a^2)$. In addition, the contours $C_0$, $C_1$ and $C_2$ of integration are shown in Figure \ref{FIG_contour}, and the substitution $y=ab/\sqrt{b^2-(b^2-a^2)v^2}$ has been used in the last equal sign of \eqref{EQ_elliptic}. When the radius of the semicircle bounded by the contour $C_0$ goes to infinity, the left-hand side of \eqref{EQ_elliptic} becomes equal to the integral in \eqref{EQ_quantization_sincos}. Because the integrals in the rightmost side of \eqref{EQ_elliptic} are the complete elliptic integrals of the first and third kinds for the elliptic modulus $\gamma$ and the characteristic $\chi$ \citep[e.g.][]{abramowitz1964handbook}, its calculation requires a numerical method. Figure \ref{FIG_rossby_wkbj} shows the approximate angular frequencies obtained from \eqref{EQ_quantization_sincos}, the calculation of which was performed by the \texttt{scipy.optimize.fsolve} and the \texttt{scipy.integrate.quad} functions of the SciPy library. The condition \eqref{EQ_quantization} will provide the approximate dispersion relation for discrete modes when any basic field is imposed and can be utilised for the check of the numerical calculation of the full eigenvalue problem for such a field.

% = = = = = = = = = = = = = = = = = = = = = = = = = = = = = = %

\end{document}